\documentclass[lettersize,journal]{IEEEtran}

\usepackage{amsmath,amsfonts}
\usepackage{algorithm}
\usepackage{array}
\usepackage[caption=false]{subfig}
\usepackage{textcomp}
\usepackage{stfloats}
\usepackage{url}
\usepackage{verbatim}
\usepackage{graphicx}
\usepackage[noend]{algpseudocode}
\usepackage[section]{placeins}

\usepackage{orcidlink}

\usepackage{booktabs}     
\usepackage{tabu,adjustbox}
\usepackage{multirow,rotating}

\usepackage{csquotes}

\usepackage{amssymb}
\usepackage{nicefrac}

\usepackage{cite}

\begin{document}

%\title{prNet: Solving Fourier Phase Retrieval Problem via Stochastic Refinement}
\title{prNet: Data-Driven Phase Retrieval\\via Stochastic Refinement}

\author{Mehmet Onurcan Kaya~\orcidlink{0009-0006-2606-3992} and Figen S. Oktem~\orcidlink{0000-0002-7882-5120},~\IEEEmembership{Member,~IEEE}
%\thanks{This work is supported by the Scientific and Technological Research Council of Turkey  (TUBITAK) under grant 120E505.}
\thanks{M.~O.~Kaya is with the Department of Applied Mathematics and Computer Science, Technical University of Denmark, Kongens Lyngby, 2800, Denmark (e-mail: monka@dtu.dk). F.~S.~Oktem is with the Department of Electrical Engineering, Middle East Technical University (METU), Cankaya, Ankara 06800, Turkey (e-mail: figeno@metu.edu.tr).}
\thanks{This paper has supplementary downloadable material available at http://ieeexplore.ieee.org., provided by the authors.}
}

%\thanks{Manuscript received April 19, 2021; revised August 16, 2021.}}

% {Shell \MakeLowercase{\textit{et al.}}: prNet: Solving Fourier Phase Retrieval Problem via Stochastic Refinement}
%{Shell \MakeLowercase{\textit{et al.}}: prNet: Data-Driven Phase Retrieval via Stochastic Refinement}


\maketitle

\begin{abstract}

% We present a diffusion-based stochastic refinement framework that addresses this challenge by sampling from the posterior distribution using Langevin dynamics, enabling reconstructions that explicitly navigate the perception–distortion tradeoff. Our method combines stochastic sampling, a learned denoiser, and model-based updates, and is instantiated in three progressively more powerful variants. prNet-Small integrates theoretically grounded Langevin inference with a learnable noise schedule and hybrid HIO-based data-consistency updates. prNet-Large extends this framework with parallel posterior sampling and ensemble averaging to approximate the MMSE estimator, substantially improving distortion metrics. prNet-Large-Adv further enhances perceptual quality through an adversarially trained aggregation network. We additionally introduce test-time augmentation for phase retrieval by exploiting inherent symmetries in the Fourier magnitude operator. Extensive experiments demonstrate that our models consistently achieve state-of-the-art reconstruction quality across multiple benchmarks, improving both fidelity and perceptual realism.

Phase retrieval is an ill-posed inverse problem in which existing methods often struggle to balance reconstruction fidelity and perceptual quality. 
We propose a novel Langevin-based framework for phase retrieval that bridges model-driven and data-driven approaches by combining stochastic posterior sampling, learned denoising, and model-based updates to better navigate the perception-distortion tradeoff. 
The framework consists of three progressively more advanced variants.
The first, prNet-Small, provides a theoretically grounded and lightweight pipeline that integrates Langevin dynamics with learned noise and denoising processes and alternating projection-based updates. 
Building upon this foundation, prNet-Large exploits parallel sampling of diverse reconstructions and aggregates them to approximate the MMSE solution, significantly improving distortion metrics. Finally, prNet-Large-Adversarial enhances perceptual quality by replacing simple averaging with an adversarially trained aggregation network while preserving reconstruction fidelity. In addition, we incorporate test time augmentation to further improve the performance. 
Extensive experiments demonstrate that the proposed framework consistently outperforms classical, deep learning-based, and diffusion-based baselines across multiple benchmarks 
in both reconstruction fidelity and perceptual quality.
The source code and trained models are available at \href{https://github.com/METU-SPACE-Lab/prNet-for-Phase-Retrieval}{https://github.com/METU-SPACE-Lab/prNet-for-Phase-Retrieval}.

% yielding reconstructions that explicitly balance distortion and perceptual quality. Unlike conventional approaches that prioritize pixel-wise accuracy, our methods navigate the perception-distortion tradeoff through a principled combination of stochastic sampling, learned denoising, and model-based updates. The framework comprises three variants of increasing complexity, integrating theoretically grounded Langevin inference, adaptive noise schedule learning, parallel reconstruction sampling, and warm-start initialization from classical solvers. 

\end{abstract}


%%%%%%%%%%%%%%%%%%%%%%%%%%%%%%%%%%%%%%%%%%%%%%%%%%%%%%%%%%%%%%%%%%%%%%%%%%%%%%%%%%%%%%%%%%%%%%%%%%
\section{Introduction}
% \label{sec:intro}
% The measurement model:
% \begin{equation}
% 	\mathbf{y^2} =   \mathbf{\vert Ax \vert^2 + w}, \quad \quad \mathbf{w} \sim \mathcal{N}(\mathbf{0}, \alpha^2 \text{diag}(\vert \mathbf{Ax} \vert^2)) 
%     \label{eq:measurementmodel}
% 	\end{equation}

\IEEEPARstart{P}{hase} retrieval (PR)
%Phase retrieval (PR)
is a fundamental inverse problem encountered in many scientific and engineering disciplines, where the goal is to reconstruct a signal using only intensity measurements such as Fourier intensities. This problem is critical in various applications including microscopy, holography, crystallography, and coherent diffraction imaging~\cite{Dong_2023}. 
Mathematically, the PR problem involves reconstructing an unknown signal $\mathbf{x} \in \mathbb{C}^{n}$ from its noisy intensity measurements expressed as
\begin{equation}
%\mathbf{y^2} = \mathbf{\vert Ax \vert^2 + w}
 	\mathbf{y^2} =   \mathbf{\vert Ax \vert^2 + w}, \quad \quad \mathbf{w} \sim \mathcal{N}(\mathbf{0}, \alpha^2 \text{diag}(\vert \mathbf{Ax} \vert^2)) 
\label{eq:measurementmodel}
\end{equation}
where $\mathbf{A} \in \mathbb{C}^{m \times n}$ is a known measurement operator, $\mathbf{y^2} \in \mathbb{R}^{m}$ denotes intensity measurements, and $\mathbf{w}$ represents measurement noise which is often Poisson-distributed but is commonly approximated as Gaussian with a noise level parameter $\alpha$ in many practical settings~\cite{pmlr-v80-metzler18a}. An important special case is Fourier PR, where $\mathbf{A}$ corresponds to the discrete Fourier transform matrix.

The primary challenge in PR lies in the loss of phase information, making the problem highly non-linear and ill-posed. Classical approaches to PR are predominantly based on alternating projection methods, which iteratively enforce known constraints in both the spatial and measurement domains. Among the earliest of these are the Gerchberg-Saxton (GS)~\cite{gs1978} and Error Reduction (ER) algorithms~\cite{fienup1978reconstruction}, which apply hard projections for each available constraint. While simple and computationally efficient, these algorithms are highly sensitive to initialization and often stagnate at suboptimal solutions. 
An important variant is the Hybrid Input-Output (HIO) algorithm~\cite{fienup1982comparison}, which introduces a feedback mechanism in the object domain. Rather than fully projecting onto the constraint set, HIO applies a feedback-driven update rule in the region where constraints are violated. This prevents the algorithm from stagnating at fixed points of the simpler ER algorithm, making HIO one of the most widely used methods in the field.
Nevertheless, all of these classical approaches remain vulnerable to noise, and suffer from artifacts and stagnation to local minima, particularly in high-dimensional and low-SNR regimes~\cite{marchesini2007invited}. To mitigate these issues, more advanced techniques have been developed, including methods based on semidefinite programming and sparsity regularization. However, these approaches often introduce a significant computational burden and rely on strong prior assumptions, limiting their applicability in practical settings \cite{waldspurger2015phase, jaganathan2013sparse}.

In recent years, deep learning has emerged as a powerful tool for solving various inverse problems in imaging, including phase retrieval \cite{lopeztaipa, Wang_2024}. Data-driven approaches based on deep neural networks (DNNs) have initially demonstrated remarkable success in directly reconstructing images from measurements or refining initial estimates from classical methods \cite{Jin2016DeepCN,Isil:19}. Moreover, plug-and-play (PnP) and unrolled methods have integrated deep priors learned from data into iterative model-based optimization frameworks \cite{Romano2016TheLE,pmlr-v80-metzler18a,Isil:20,isil2024deep,chang2021LPR}. However, such existing deep learning solutions for phase retrieval often suffer from limited performance due to domain shift, which occurs when the training data and real-world test data follow different distributions \cite{zhang2024wrong}. Moreover, these methods face challenges in perceptual quality, as they primarily rely on minimum mean squared error (MMSE) or maximum a posteriori (MAP) estimation. Such estimators tend to produce overly smooth outputs with reduced perceptual quality due to the perception-distortion tradeoff \cite{Blau2017ThePT}.

% To address these limitations, generative models—particularly score-matching and diffusion-based approaches—have gained traction for their ability to sample from the posterior distribution via Langevin dynamics. Unlike deterministic estimators, generative models yield diverse, high-quality samples that align more closely with the natural image manifold. This property is especially valuable in phase retrieval, where the inherent ill-posedness of the problem admits multiple valid solutions. The diffusion framework, in particular, offers a principled way to incorporate measurement constraints while leveraging learned priors, as evidenced by Tweedie’s formula, which connects score-based sampling to denoising processes \cite{Luo2022UnderstandingDM}.

% Recent work has explored the application of diffusion models to phase retrieval. For instance, DDRM-PR \cite{zhang2024wrong} employs pretrained diffusion models to solve the problem but relies on generic denoisers that do not explicitly learn the measurement model, limiting adaptability to specific imaging conditions. Similarly, DOLPH \cite{Shoushtari2022DOLPHDM} integrates diffusion priors with nonconvex data-fidelity terms but uses a fixed noise schedule and suboptimal data-consistency steps, which can restrict performance. These limitations highlight the need for a more flexible and adaptive framework that jointly optimizes the denoising process and measurement consistency.
To address these limitations, deep generative models, particularly score-based and diffusion-based approaches, have gained increasing attention in the inverse problem literature due to their ability to model and sample from complex posterior distributions \cite{Kawar2021SNIPSSN}.
These methods are closely connected to score-based Langevin dynamics, providing a principled framework for posterior sampling. In particular, Tweedie’s formula~\cite{Luo2022UnderstandingDM} establishes a relationship between denoising and score estimation, which underlies many iterative diffusion-based sampling methods.

Unlike point estimators such as MMSE or MAP, which summarize the posterior distribution with a single estimate, diffusion models can generate plausible reconstructions through sampling guided by a learned score function.
This capability is especially advantageous for ill-posed inverse problems such as phase retrieval, where inherent ambiguities allow multiple signals to be consistent with the same observation, making a single point estimate insufficient to represent the underlying solution space.

Recent works have increasingly applied the diffusion models to various phase retrieval problems, aiming to leverage their generative capabilities as learned priors. For instance, DDRM-PR~\cite{kaya2025ddrm} utilizes the unsupervised posterior sampling framework of denoising diffusion restoration models for phase retrieval. % to solve Fourier phase retrieval problem, 
However, its reliance on pretrained unconditional diffusion models that are not specifically tailored to the underlying physical forward model limits its reconstruction performance. 
%In fact, the authors themselves note that task-specific diffusion training could further improve reconstruction quality for phase retrieval problems.
%DDRM-PR exploits the efficient and unsupervised posterior sampling framework of denoising diffusion restoration models (DDRMs) for Fourier phase retrieval. This approach combines the model-based alternating-projection methods with the DDRM framework to utilize pretrained unconditional diffusion priors. 
Similarly, DOLPH~\cite{Shoushtari2022DOLPHDM} employs pretrained diffusion models %for for coded diffraction pattern (CDP) settings 
and incorporates gradient-based correction into the reverse diffusion process to achieve consistency with the measurements.
%modifies the reverse diffusion process with a gradient correction step to achieve that candidate solutions remain consistent with the measurements. 
%DOLPH alternates data-consistency updates with the sampling steps of the diffusion model to achieve that candidate solutions remain consistent with the measurements. 
In addition to relying on a pretrained model, the method also employs a fixed diffusion noise schedule and relatively simple data-consistency updates, which may limit its reconstruction performance beyond the coded diffraction pattern scenarios considered in the paper.

Beyond these phase retrieval-specific approaches, several diffusion-based Bayesian posterior-sampling strategies have been developed to approximate the posterior score function $\nabla_x \log p (\mathbf{x} |\mathbf{y})$. For example, Diffusion Posterior Sampling (DPS)~\cite{chung2023dps} approximates the analytically intractable likelihood score via a Tweedie-style expectation to efficiently handle general noisy inverse problems, while PnP-DM~\cite{wu2024principled} integrates pretrained diffusion priors into a principled plug-and-play sampling scheme. More recently, DAPS~\cite{zhang2025improving} decouples the diffusion noise schedule from the measurement update through an annealed sampling procedure, showing competitive performance across a range of inverse problems.

Specialized architectures have also emerged to address the specific constraints of phase retrieval. DeepECpr~\cite{shastri2025fast} provides a fast and robust model-driven alternative by combining deep denoising with an expectation-consistent approximation; however, it remains focused on computing a single approximate MMSE estimate and does not explicitly address the perception-distortion tradeoff. In the context of Fourier phase retrieval specifically, DiffFPR~\cite{li2024difffpr} introduces a diffusion prior tailored for oversampled Fourier measurements, while RED-PhaseCut~\cite{cha2024red} couples a diffusion-based variational sampler with a PhaseCut-inspired data-consistency regularization step.  I2I-PR~\cite{kaya2026i2i} heuristically adapts
image-to-image diffusion models to Fourier
phase retrieval by combining HIO-based warm-starts with the fixed noise schedule of the InDI framework~\cite{delbracio2023inversion}.

Despite these advancements, many existing methods treat the generative prior and the noise schedule as independent components, often lacking the flexibility to adapt to the complexities of the PR problem. These shortcomings highlight the need for a framework that jointly learns the generative prior and the diffusion noise schedule in a problem-aware manner to achieve robust performance across diverse PR scenarios.

%Despite these advances, a common feature of existing methods is that the diffusion prior (score network) and the noise schedule are treated as fixed, independent components. In other words, the generative model is pretrained and the noise-annealing schedule is designed separately from the measurement model. This can limit flexibility when adapting to complex, nonlinear phase-retrieval constraints. These limitations motivate more adaptive frameworks that jointly learn both the diffusion prior and the noise schedule in a problem-aware manner, to better conform to the physics of phase retrieval and ensure robust performance across varied PR scenarios.

We propose a novel Langevin-based framework for phase retrieval that bridges model-driven and data-driven approaches.
Our framework consists of three progressively more advanced methods: 
(1) prNet-Small, a theoretically grounded and lightweight pipeline that integrates Langevin dynamics with learned noise and denoising processes and model-driven HIO updates; (2) prNet-Large, which exploits parallel sampling of diverse reconstructions to approximate the MMSE solution via ensemble averaging, significantly improving distortion metrics; and (3) prNet-Large-Adversarial, extending prNet-Large with an adversarially trained aggregation network to enhance perceptual quality while preserving reconstruction fidelity.
Additionally, we leverage test time augmentation (TTA) for enhanced phase retrieval by exploiting the inherent symmetries and structure of the problem.

Our proposed methods show superior performance compared to classical and state-of-the-art techniques.
Moreover, our framework demonstrates promise for developing reliable stochastic solvers for nonlinear inverse problems, which could have broader implications beyond PR. Some preliminary results of
this research were presented in \cite{kaya2025_mlsp}.

The subsequent sections of this paper are organized as follows: Section \ref{sec:relatedworksprnet} reviews related research that informed the development of our approach. Our developed approach is detailed in Section \ref{sec:methodprnet}, followed by a comparative performance analysis against classical and state-of-the-art methods in Section \ref{sec:resultsprnet}. Lastly, Section \ref{sec:conclusionprnet} summarizes our findings and outlines future research directions.

%%%%%%%%%%%%%%%%%%%%%%%%%%%%%%%%%%%%%%%%%%%%%%%%%%%%%%%%%%%%%%%%%%%%%%%%%%%%%%%%%%%%%%%%%%%%%%%%%%

\section{Related Work}
\label{sec:relatedworksprnet}

% \subsection{Classical Iterative Projection Techniques for Phase Retrieval}
\subsection{Iterative Projection Techniques for Phase Retrieval}

Iterative projection techniques have become fundamental tools for phase retrieval. One of the earliest and most well-known algorithms is the classical Gerchberg-Saxton (GS) algorithm~\cite{gs1978}, which iteratively applies magnitude constraints in both the spatial and measurement domains to reconstruct an unknown signal. An enhancement of the GS algorithm is the Error Reduction (ER) algorithm, which incorporates additional spatial domain constraints beyond just magnitude~\cite{fienup1978reconstruction}. A particularly significant and widely used method among alternating projection techniques is the Hybrid Input-Output (HIO) algorithm~\cite{fienup1982comparison}, which builds upon the principles of the ER algorithm.

In the HIO method, measurement fidelity constraints and various spatial domain constraints (such as support, non-negativity, and real-valuedness) are iteratively applied, similar to the ER algorithm. However, the key distinction is that HIO does not force the iterates to strictly satisfy the constraints at every step. Instead, it uses the iterates to progressively guide the algorithm towards a solution that meets the constraints. In its most general form, the HIO iterations can be mathematically expressed as follows~\cite{fienup1982comparison,netrapalli2013phase,chandra2019phasepack}:
\begin{equation}\label{eq:prnetttthio2}
\begin{aligned}
\mathbf{x}_{k+1}[n] = \left\{ \begin{array}{rcl}
\mathbf{x}_k'[n] & \text{for} & n \notin \gamma \\
\mathbf{x}_{k}[n] - \beta \mathbf{x}_k'[n] & \text{for} & n \in \gamma \\
\end{array}\right.
\end{aligned}
\end{equation}
where
\begin{equation}\label{eq:prnetttthio1}
\begin{aligned}
\mathbf{x}_k' = \mathbf{A}^{\dagger}\left\{\mathbf{y} \odot \frac{\mathbf{A} \mathbf{x}_k}{\vert \mathbf{A} \mathbf{x}_k \vert}\right\}.
\end{aligned}
\end{equation}
In these equations, $\mathbf{x}_k \in \mathbb{R}^{m}$ represents the reconstruction at the $k^{th}$ iteration, $\mathbf{A^{\dagger}}$ denotes the pseudoinverse of the forward system matrix, $\odot$ signifies element-wise multiplication, $\beta$ is a constant parameter (commonly set to 0.9), and $\gamma$ is the set of indices $n$ where $\mathbf{x}_k'[n]$ fails to meet the spatial domain constraints~\cite{fienup1982comparison}.

Despite the lack of a comprehensive theoretical understanding of the HIO method's convergence behavior, it has been empirically observed to converge to acceptable solutions in a wide array of applications. However, the reconstructions produced by HIO can sometimes contain artifacts and errors. These issues are often attributed to the algorithm getting trapped in local minima or to the amplification of noise within the solution~\cite{shechtman2015phase, marchesini2007invited}. To address these limitations, numerous variations and enhancements of the HIO method have been proposed, aiming to improve its reconstruction performance and reliability~\cite{stefanoqianpty, Maiden:17,Isil:20,isil2024deep}.

\subsection{Deep Learning for Inverse Problems}

Deep learning-based reconstruction techniques have emerged as a compelling alternative to traditional analytical methods. These approaches demonstrate the potential to achieve high reconstruction quality and computational efficiency across various imaging problems, including phase retrieval \cite{lopeztaipa, Wang_2024}. 
The integration of deep learning into phase retrieval represents a significant advancement, offering new solutions to longstanding challenges. Deep learning priors are particularly useful for phase retrieval because they can effectively capture complex structures and patterns in data, which are difficult to represent with traditional analytical techniques. By learning from large datasets, deep learning models can provide robust priors that guide the phase retrieval process to reduce the impact of noise and improve convergence to accurate solutions \cite{Wang_2024}.

The current landscape of deep learning-based reconstruction in the literature can be broadly categorized into four main classes: 1) learning-based direct inversion, 2) plug-and-play (PnP) regularization, 3) learned iterative reconstruction based on unrolling, and 4) generative methods.

Learning-based direct inversion methods aim to bypass iterative reconstruction altogether by directly mapping measurements to the desired image using a deep neural network (DNN). This approach trains the DNN to learn the inverse function of the forward model solely on the basis of the training data.  While achieving state-of-the-art performance for simpler inverse problems such as denoising \cite{El_Helou_2020}, these methods face challenges with complex observation models, significant discrepancies between observations and the target image, or limited training data availability. Such end-to-end direct inversion schemes also exist for the phase retrieval problem \cite{Nishizaki2020AnalysisON, Uelwer2019PhaseRU, Shevkunov2021DeepCN}. However, due to the nature of the phase retrieval problem, they generally do not perform well compared to other approaches \cite{zhang2024wrong}.

To address these limitations, a common strategy involves applying an efficient analytical approximation of the forward model to generate an initial reconstruction. This initial estimate then serves as a ``warm start" for a subsequent DNN refinement step. This hybrid approach, which combines neural networks with analytical methods, has demonstrably succeeded in various real-valued 2D reconstruction problems, including deconvolution, super-resolution, tomography, and phase retrieval \cite{Jin2016DeepCN,Isil:19,Nishizaki2020AnalysisON}. A key advantage of learning-based direct inversion methods lies in their low computational complexity due to their feed-forward (non-iterative) nature, making them suitable for real-time imaging applications.

In contrast to learning-based direct inversion, PnP regularization, and unrolled learning methods embrace iterative strategies. Their core principle lies in replacing hand-crafted analytical priors with data-driven deep priors within model-based reconstruction frameworks.
%Plug-and-play methods first train a deep prior on dedicated datasets and then leverage it as a regularizer for an iterative model-based inversion algorithm \cite{6737048, Romano2016TheLE}.
PnP methods leverage a pretrained generic denoiser as a deep prior, integrating it as a regularizer within an iterative model-based inversion framework \cite{6737048, Romano2016TheLE}.
Maximum A Posteriori (MAP) problem given Gaussian noise assumption can be written as an optimization problem in the form of $\max_\mathbf{x} -\| \mathbf{y} - \mathcal{A} (\mathbf{x}) \|^2 + \mathcal{R}(\mathbf{x}) $ which can be split into data-fidelity and regularization steps. Thus, this framework allows us to solve various inverse problems by leveraging the impressive capabilities of existing denoising models in the regularization steps while model-based algorithms can be used jointly in the data-fidelity steps.
Such PnP methods are widely used in the current phase retrieval literature \cite{Isil:20, isil2024deep, Wang2020WhenDD,chang2021LPR}.
While achieving superior image quality, flexibility, and generalizability compared to direct inversion methods, they typically require higher memory usage and computational complexity due to their iterative nature. This complexity stems from the need to compute the forward operator and its adjoint at each iteration.

Unrolled learning takes iterative methods utilizing proximal operators or deep priors, such as those employed in PnP approaches, and transforms them into end-to-end trainable networks. This representation allows the algorithm to be concatenated as a series of layers, running a finite number of times as it passes through the network. This unrolling aims to further improve reconstruction quality \cite{Monga2019AlgorithmUI}. However, similar to PnP methods, unrolled iterative learning generally suffers from high computational demands. Furthermore, unlike direct inversion and PnP methods, unrolled approaches necessitate the computation of both forward and adjoint operators during training, leading to a significant increase in training time and complexity. This can make them impractical for large-scale reconstruction problems. Despite these limitations, unrolled learning has shown success in phase retrieval \cite{Naimipour2020UPRAM,Deng2019PhysicsED,Naimipour2020UnfoldedAF,Wang2020PhaseRW, Liu2023PRISTANetDI}.
%Unrolled learning takes iterative methods utilizing proximal operators or deep priors, such as those employed in plug-and-play approaches, and transforms them into end-to-end trainable networks. This representation allows the algorithm to be concatenated as a series of layers, executing it a finite number of times as it passes through the network. It aims to improve reconstruction quality further \cite{Aggarwal2017MoDLMD, Monga2019AlgorithmUI}. However, similar to plug-and-play methods, unrolled iterative learning generally suffers from high computational demands, particularly when dealing with large sensing matrices during forward operator and adjoint evaluations. Furthermore, unlike direct inversion and plug-and-play methods, unrolled approaches necessitate the computation of both forward and adjoint operators during training, leading to a significant increase in training time and complexity. Despite these limitations, unrolled learning has shown success in phase retrieval \cite{Naimipour2020UPRAM,Deng2019PhysicsED,Naimipour2020UnfoldedAF,Wang2020PhaseRW, Liu2023PRISTANetDI}.

A recent PR-specific instance is deepECpr~\cite{shastri2025fast}, which combines a deep denoiser with an expectation-consistent approximation step to obtain fast and robust reconstructions. Unlike the stochastic posterior-sampling framework we develop, deepECpr targets a single minimum-mean-square-error estimate and does not address the perception-distortion tradeoff explicitly.

\subsection{Generative Models for Inverse Problems}

All of the aforementioned deep learning methods focus on Maximum A Posteriori or Minimum Mean Squared Error (MMSE) estimation. 
As theoretically shown in \cite{Blau2017ThePT} and empirically observed in \cite{8099502}, these estimates may deviate significantly from the natural image manifold, leading to reconstructions with overly smooth features. Interestingly, the work by Işıl et al. \cite{Isil:19} attributes this smoothing behavior to an unavoidable inherent limitation of DNNs in the context of phase retrieval. 
However, as long as reconstruction algorithms prioritize minimizing distortion metrics, such as mean squared error, we can only expect limited improvements in perceptual quality.

To achieve reconstructions that are visually accurate to human observers, a shift in our strategy for solving inverse problems is necessary. Instead of focusing solely on the conditional mean of the posterior distribution, we should aim to sample directly from this posterior distribution $p (\mathbf{x}|\mathbf{y})$. This allows us to generate images that are more likely to belong to the true underlying distribution of natural images.

In cases of severe information loss, the image reconstruction problem becomes ill-posed, meaning that there can be multiple valid solutions that explain the observed measurements. This challenge is particularly relevant in phase retrieval, where intrinsic system symmetries can map different input images to the same output, which affects network performance \cite{Tayal2020InversePD}. The MMSE solution attempts to average these potential solutions, resulting in smoothed images lacking the fine details often present in real-world scenes. Given the existence of multiple valid solutions, a successful approach should incorporate stochasticity, as ill-posed problems inherently have multiple viable solutions for the same data. Generative models provide an ideal framework for this purpose, allowing us to sample from the posterior distribution and generate diverse yet plausible reconstructions.

Generative models, which include techniques such as Generative Adversarial Networks, Variational Autoencoders (VAE), flow-based approaches, and diffusion models, have demonstrated impressive performance in diverse inverse problem tasks \cite{Dimakis_2022, 10004774}. By learning to generate samples from the posterior distribution, generative models can produce reconstructions that better capture the variability and richness of natural images. Notably, generative models have also been successfully applied to phase retrieval \cite{Uelwer2019PhaseRU, Gladrow2019DigitalPH, Shoushtari2022DOLPHDM}. Uelwer et al. \cite{Uelwer2019PhaseRU} demonstrated that conditional generative adversarial networks (cGANs) can optimize phase retrieval processes by incorporating measurement knowledge, thus achieving superior performance compared to traditional methods. Similarly, Gladrow et al. \cite{Gladrow2019DigitalPH} utilized deep conditional generative models, such as cGAN and conditional VAE,
to solve the inverse problem of digital holography, showcasing the potential of data-driven approaches in handling optical aberrations. Shoushtari et al. \cite{Shoushtari2022DOLPHDM} introduced DOLPH, a diffusion model-based architecture, which effectively integrates image priors with nonconvex data-fidelity terms, providing robust and high-quality solutions for phase retrieval. These studies collectively highlight the versatility and robustness of generative models in enhancing phase retrieval outcomes.

\subsection{Diffusion Models for Inverse Problems}

Diffusion models, a subclass of generative models, have recently gained prominence for their effectiveness in high-dimensional data generation and reconstruction tasks. These models work by simulating a diffusion process that transforms simple, noise-like data into complex structures over time. The process is guided by learned score functions, which estimate the gradients of the data distribution at each step to gradually denoise the data and refine the generated outputs.

The significance of diffusion models lies in their theoretical foundation and practical success. Historically, these models draw inspiration from non-equilibrium thermodynamics and stochastic processes. The influential works on diffusion models have demonstrated their capability to generate high-quality, diverse samples, rivalling or surpassing other generative models such as GANs and VAEs. The iterative nature of diffusion models allows them to incrementally refine solutions \cite{chan2024tutorial, 10004774, Uelwer2019PhaseRU, Gladrow2019DigitalPH, Shoushtari2022DOLPHDM}, making them particularly suitable for tasks requiring high precision, such as phase retrieval.

In the context of phase retrieval, diffusion models provide a powerful framework for incorporating deep learning priors \cite{kaya2025ddrm}. The iterative denoising process aligns well with the need to progressively refine phase estimates from initial noisy guesses.
Diffusion models, when trained on large-scale image datasets, learn rich statistical priors that capture the structure of natural images. These learned priors can be leveraged to more effectively guide the phase retrieval process, leading to significantly improved reconstruction accuracy and visual fidelity.
%By training on large image datasets, diffusion models learn to capture the underlying statistical properties of the data, which can then be leveraged to guide the phase retrieval process towards more accurate reconstructions.

One of the key advantages of using diffusion models for phase retrieval is their robustness to noise and initialization. Traditional phase retrieval algorithms often suffer from convergence to local minima and sensitivity to the initial guess. Diffusion models, with their probabilistic and iterative nature, can mitigate these issues by providing a systematic approach to explore the solution space and progressively enhance the quality of the reconstructions \cite{kaya2025ddrm, Shoushtari2022DOLPHDM}.

Moreover, the flexibility of diffusion models allows their adaptation to various types of data and measurement settings. Whether dealing with coded diffraction patterns, multi-plane intensity measurements, or different wavelengths, diffusion models can be trained to incorporate these variations, enabling a unified framework for phase retrieval across diverse applications \cite{kaya2025ddrm, Shoushtari2022DOLPHDM}.

\subsection{Posterior Sampling via Score/Diffusion-Based Models}

Unconditional diffusion/score-based models are known for their ability to generate high-quality samples from a prior distribution using the score function $\nabla_x \log p (\mathbf{x})$ via Langevin dynamics. It is worth mentioning that since score-based and diffusion-based interpretations are equivalent due to Tweedie's formula \cite{Luo2022UnderstandingDM, chan2024tutorial, li2023diffusion}, we can focus solely on the score-based approach here.

Although directly learning the score function is an option, most work utilizes a deep denoiser instead. This substitution is based on the relationship given by \cite{miyasawa1961empirical}
\begin{equation}
\nabla_{\mathbf{x}_t} \log p\left(\mathbf{x}_t\right)=\frac{\text{Denoiser}\left(\mathbf{x}_t, \sigma_t \right)-\mathbf{x}_t}{\sigma_t^2}
\end{equation}
where $\mathbf{x}_t \triangleq \mathbf{x} + \mathbf{v}$ with $\mathbf{v} \sim \mathcal{N}(\mathbf{0}, \sigma_t^2 \mathbf{I})$.

Several strategies have been explored to extend this score-based approach for sampling from a posterior distribution $p (\mathbf{x} |\mathbf{y})$, leveraging the posterior score function $\nabla_x \log p (\mathbf{x} |\mathbf{y})$ within Langevin dynamics. Here, we can discuss four common methods for approximating the posterior score function for inverse problems: 1) conditioning via initialization, 2) conditional denoiser, 3) hard projection, and 4) Bayesian approach.

%The ``conditioning via initialization" approaches, such as SDEdit \cite{meng2022sdedit}, initialize Langevin dynamics sampling with a plausible estimate obtained from a simpler classical method, but it does not modify the score function itself. While this method is simple to implement, it lacks a guarantee of consistency with the observations. Consequently, the resulting outputs may be inconsistent with the actual measurements.

The ``conditioning via initialization" approaches,
such as SDEdit \cite{meng2022sdedit}, % SDEdit classical inverse problemler için değil ama bu tarz basit bir metodu inverse problemler için bulamadım, bulabildiğim en yakın citation bu
employ a ``warm start" strategy, initializing Langevin dynamics sampling from a heuristic or classical estimate instead of pure noise. While computationally efficient, requiring no modifications to the unconditional pretrained diffusion model's score function, this approach inherits a fundamental limitation: the initialization does not enforce measurement consistency. Since sampling begins at an intermediate denoised state (skipping early noise-heavy diffusion steps) yet still relies on an unconditional score estimator, the generated samples may diverge from the true data manifold or violate physical constraints. Thus, despite its simplicity, this paradigm lacks theoretical guarantees of fidelity to observations, often producing artifacts or biased reconstructions.

In ``conditional denoiser" techniques, such as SR3 \cite{saharia2022image} or Palette \cite{saharia2022palette}, they give $\mathbf{y}$ to denoiser as $\nabla_{\mathbf{x}_t} \log p\left(\mathbf{x}_t | \mathbf{y}\right)=\frac{\text{Denoiser}\left(\mathbf{x}_t, \mathbf{y}, \sigma_t \right)-\mathbf{x}_t}{\sigma_t^2}$ and rely fully on the learning process. This approach is still simple, but for many inverse problems, the influence of the estimation can be very challenging to learn as it requires learning the complex measurement model. Also, there is no theoretical guarantee for conditioning.

The ``hard projection" methods, such as ReSample \cite{song2024solving}, utilize a regular denoiser followed by a projection step to match with $\mathbf{y}$, more mathematically, $
\hat{\mathbf{x}} = \arg \min_\mathbf{z} \frac{1}{2} \| \mathbf{z} - \text{Denoiser}(\mathbf{x}, \sigma_t ) \|^2 \text{ s.t. } \mathbf{y} = \mathcal{A}(\mathbf{z})
$. Although relatively simple to implement, this approach might not be applicable to all inverse problems. Additionally, it can suffer from inaccuracies as the projection step might not achieve perfect conditioning on the measurement.

The Bayesian approaches leverage Bayes' rule to derive the posterior score function as $\nabla_{\mathbf{x} } \log p\left(\mathbf{x}_{t-1} | \mathbf{y}\right) = \nabla_{\mathbf{x}} \log p\left(\mathbf{y} | \mathbf{x}_{t-1}\right) + \nabla_{\mathbf{x}} \log p\left(\mathbf{x}_{t-1}\right)
$. Offering a mathematically well-founded approach for posterior sampling, this method has been successfully applied to linear inverse problems in the SNIPS method \cite{Kawar2021SNIPSSN}.

Therefore, one promising approach for achieving high perceptual quality reconstructions is to employ a posterior score-based sampler, as demonstrated by Kawar et al. \cite{Kawar2021SNIPSSN}. This strategy offers a multitude of potential solutions for attaining perfect perceptual quality, albeit potentially at the expense of distortion metrics.

Closely related Bayesian posterior-sampling methods include DPS~\cite{chung2023dps}, which approximates the likelihood score $\nabla_{\mathbf{x}} \log p(\mathbf{y} | \mathbf{x}_t)$ via a Tweedie-style expectation under the unconditional denoiser, and PnP-DM~\cite{wu2024principled}, which integrates pretrained diffusion priors into a principled plug-and-play sampling scheme. DAPS~\cite{zhang2025improving} decouples the diffusion noise schedule from the measurement update through an annealed sampling procedure and has shown competitive results on a range of inverse problems. These methods all belong to the broader family of Bayesian posterior-sampling approaches discussed above and differ primarily in how they tractably approximate $\nabla_{\mathbf{x}} \log p(\mathbf{y} | \mathbf{x}_t)$ and in how the prior denoiser is composed with the data-consistency step. The Bayesian PnP-Langevin framework of Laumont~et~al.~\cite{laumont2022bayesian}, which formalizes the convergence of Langevin sampling with a learned denoiser via Tweedie's identity, provides theoretical grounding for this family of methods.

Moreover, several recent works have used specialized diffusion models for phase retrieval. DiffFPR~\cite{li2024difffpr} introduces a diffusion prior for oversampled Fourier PR and serves as one of our principal baselines in this work. RED-PhaseCut~\cite{cha2024red} couples a diffusion-based variational sampler with a PhaseCut-style data-consistency step. DiffPhase~\cite{peer2023diffphase} targets STFT phase retrieval for speech signals; although the forward model and signal statistics differ from the image setting considered here, the use of an unconditional diffusion model as a generative prior is methodologically related. Moreover, I2I-PR~\cite{kaya2026i2i} adapts
image-to-image diffusion models %to Fourier phase retrieval 
by heuristically combining multiple HIO-based warm-starts under the fixed noise schedule of the InDI framework~\cite{delbracio2023inversion}.
%Most closely related to our work, I2I-PR~\cite{kaya2026i2i} adapts the InDI framework~\cite{delbracio2023inversion} to Fourier PR, conditioning the denoiser on multiple warm-starts and interleaving HIO data-consistency updates. 
In contrast, the proposed prNet framework is theoretically grounded, with update rules analytically derived from Langevin posterior sampling, while training on intermediate reconstructions enables both the noise and denoising schedules to be learned rather than fixed. % and learned noise and denoising schedules.

\subsection{Wasserstein Adversarial Loss}

Generative Adversarial Networks (GANs) have shown strong performance in generating realistic images and have been applied to inverse problems for producing high-quality outputs \cite{10004774}. These models aim to generate diverse images that both satisfy measurement constraints and match the distribution of clean examples. In phase retrieval, GAN-based approaches have been explored \cite{Uelwer2019PhaseRU, Hand2018PhaseRU}, though they often assume noiseless measurements, a limitation in practical settings. %\cite{kipeng}.

Beyond generative modeling, adversarial loss can help counteract over-smoothing caused by distortion-based optimization. A common formulation adds an adversarial term to the distortion loss \cite{Blau2017ThePT}:
\begin{equation}
\ell_{\text {total }}=\ell_{\text {distortion }}+\lambda \ell_{\text {adv }},
\end{equation}
where $\ell_{\text {adv}}$ is the standard GAN loss.

However, standard adversarial loss, based on Jensen-Shannon (JS) divergence, often leads to training instability, mode collapse, and poor sample quality when real and generated distributions diverge. To address this, Wasserstein GAN (WGAN) replaces JS divergence with the Wasserstein (Earth Mover’s) distance, offering a smoother, more meaningful measure of distributional difference. This results in more stable training and improved generation quality \cite{Arjovsky2017WassersteinG}.

%An improved version of WGAN, known as WGAN with Gradient Penalty (WGAN-GP), further enhances the training stability and sample quality by imposing a gradient penalty on the discriminator. This penalty term penalizes the norm of the gradients of the discriminator with respect to its inputs, thereby enforcing the Lipschitz continuity. By constraining the Lipschitz constant of the discriminator, WGAN-GP mitigates the risk of mode collapse and training instability while promoting smoother convergence. The algorithm for WGAN-GP involves alternately updating the discriminator and generator parameters while incorporating the gradient penalty term into the loss function. This regularization technique not only improves the robustness of the discriminator/critic but also facilitates a more effective training of the generator, resulting in higher-quality generated samples \cite{Gulrajani2017ImprovedTO}. Overall, by leveraging the Wasserstein distance and integrating gradient penalty regularization, WGAN-GP offers a more reliable and effective framework for training GANs, particularly in applications such as inverse problems where stability and sample quality are paramount.

\subsection{Test Time Augmentation}

Test time augmentation (TTA) is a powerful technique in deep learning that leverages data properties to enhance performance without an additional training requirement. It involves creating slightly modified versions of the test images (flips, rotations, crops) and feeding them through the trained model. The predictions from these augmented versions are then combined (typically by averaging) to produce a final prediction \cite{shorten2019survey}. This approach acts as a form of ensembling, effectively increasing the training data by leveraging the inherent equivariance properties of the model and the data distribution \cite{Kimura2024UnderstandingTA, shanmugam2021better}.

TTA is particularly beneficial when models struggle with small input variations. In image classification, for instance, flipping an image might not significantly alter the content, but the model could potentially misclassify the flipped version. By combining predictions from both versions, TTA achieves a more robust and generalizable performance. This strategy has demonstrably improved accuracy and robustness across various deep learning domains, including image classification \cite{shanmugam2021better}, object detection \cite{casado2020ensemble}, and image segmentation \cite{wang2019aleatoric}.

% In image reconstruction tasks, TTA can capitalize on the algorithm's invariances. By processing different versions of the test image, TTA integrates additional information at test time, ultimately enhancing the reconstructed output quality. It also mitigates the model's vulnerability to spatial transformations and noise patterns in test data that might have been underrepresented or absent during training.

Closely related ideas have also appeared in image restoration. Geometric self-ensembling was introduced in \cite{timofte2016seven}, in which transformed inputs are passed through a network and the inverse-transformed outputs are averaged to improve super-resolution quality. This idea was adapted within DPIR through a periodical geometric self-ensemble \cite{zhang2021plug}, applying a single transformation per iteration of the plug-and-play loop rather than averaging at the output. More recently, equivariance was studied formally in the PnP context in \cite{terris2024equivariant}, where it was shown that replacing the denoiser with its Reynolds average over a group of transformations, or a Monte Carlo approximation thereof, symmetrizes the Jacobian, reduces the Lipschitz constant, and stabilizes PnP
%, RED,
and Langevin algorithms without retraining. These works suggest viewing TTA in image reconstruction not merely as a post-hoc ensembling heuristic but as a principled way of enforcing approximate equivariance to known invariances of the forward model.

%%%%%%%%%%%%%%%%%%%%%%%%%%%%%%%%%%%%%%%%%%%%%%%%%%%%%%%%%%%%%%%%%%%%%%%%%%%%%%%%%%%%%%%%%%%%%%%%%%
\section{Developed Methods}
\label{sec:methodprnet}

\subsection{prNet-Small}

%The core of our method is the Langevin dynamics algorithm, which can be used to generate samples from a given posterior probability distribution, $p\left(\mathbf{x}_{t} | \mathbf{y}\right)$, based on the score function of the posterior distribution and a noise term. This is represented by the following equation:
We rely on Langevin dynamics to sample from the posterior distribution \( p\left(\mathbf{x}_{t} | \mathbf{y} \right) \), using both the score function of the posterior and Gaussian noise injection. Hence our approach is based on the following update step~\cite{Kawar2021SNIPSSN}:
\begin{equation}
\mathbf{x}_{t+1} \leftarrow \mathbf{x}_{t}+\gamma \nabla_{\mathbf{x}} \log p\left(\mathbf{x}_{t} | \mathbf{y}\right)+\sqrt{2 \gamma} \mathbf{v}_t, \quad 1 \leq t \leq T,
\label{eq:langevindyanmics}
\end{equation}
where \( \gamma \) denotes the step size, \( \mathbf{v}_t \sim \mathcal{N}(\mathbf{0}, \mathbf{I}) \) is standard Gaussian noise, and \( T \) is the total number of iterations.
%Here, $\gamma$ is the learning rate, $\mathbf{v}_t$ represents standard white Gaussian noise, and $T$ is the number of iterations \cite{Kawar2021SNIPSSN}. Applying Bayes' rule yields:
Using Bayes' rule for the score function $\nabla_{\mathbf{x}} \log p\left(\mathbf{x}_{t} | \mathbf{y}\right)$ of the posterior yields:
\begin{equation}
\mathbf{x}_{t+1} \leftarrow \mathbf{x}_{t}+ \gamma\nabla_{\mathbf{x}} \log p\left(\mathbf{x}_{t}\right) +\gamma \nabla_{\mathbf{x}} \log p\left(\mathbf{y} | \mathbf{x}_{t}\right) +\sqrt{2 \gamma} \mathbf{v}_t
\label{eq:langevindyanmics2}
\end{equation}

We use a learned denoiser model $\mathcal{D}_{\mathbf{\theta}}$ with parameters $\mathbf{\theta}$ to approximate the unconditional score function $\nabla_{\mathbf{x}} \log p\left(\mathbf{x}_t\right)$, assuming degradation with $\mathbf{x}_t \triangleq \mathbf{x} + \mathbf{v}$ where $\mathbf{v} \sim \mathcal{N}(\mathbf{0}, \sigma_t^2 \mathbf{I})$ 
%$\mathcal{N}(\mathbf{0}, \sigma_t^2 \mathbf{I})$
\cite{miyasawa1961empirical}:
\begin{equation}\nabla_{\mathbf{x}} \log p\left(\mathbf{x}_t\right)=\frac{\mathcal{D}_{\mathbf{\theta}}\left(\mathbf{x}_t, t\right)-\mathbf{x}_t}{\sigma_t^2}.\end{equation}  
% We define the synthetic annealing noise model as 
% $$
% \mathbf{x}_{t} = \mathbf{x}+\mathbf{v}, \quad\quad \mathbf{v} \sim \mathcal{N}(\mathbf{0}, \sigma_t^2 \mathbf{I})
% $$
Substituting this into Eq.~\eqref{eq:langevindyanmics2}, we obtain:
\begin{equation}
\begin{aligned}
\mathbf{x}_{t+1} \leftarrow & \left(1-\frac{\gamma}{\sigma_t^2}\right) \mathbf{x}_{t} + \frac{\gamma}{\sigma_t^2} \mathcal{D}_{\mathbf{\theta}}(\mathbf{x}_{t}, t) \\
&+ \gamma \nabla_{\mathbf{x}} \log p\left(\mathbf{y} | \mathbf{x}_{t}\right) + \sqrt{2 \gamma} \mathbf{v}_t.
\end{aligned}
\label{eq:denoisersubslangevin}
\end{equation}
%
%For successful conditioning of the generation process on the observation, the formulation of the likelihood-related term specific to the inverse problem $\nabla_{\mathbf{x}} \log p\left(\mathbf{y} | \mathbf{x}_{t}\right)$ is crucial. Now, we will derive this term. 
%
%Using the change of variables formula with $\mathbf{y} \odot \mathbf{y} = \mathbf{y^2}$ transformation, Eq.~\eqref{eq:measurementmodel} gives:

To derive the log-likelihood gradient term \( \nabla_{\mathbf{x}} \log p\left(\mathbf{y} | \mathbf{x}_{t} \right) \) for phase retrieval, we consider the following simplified forward model: % in Eq.~\eqref{eq:measurementmodel} as follows: 
\begin{equation}
	\mathbf{y} = \mathbf{\vert Ax \vert} + \mathbf{w}, \quad \quad \mathbf{w} \sim \mathcal{N}\left( \mathbf{0}, \left(\frac{\alpha}{2}\right)^2 \mathbf{I} \right),
    \label{eq:eq10}
\end{equation}
This approximation is justified by the delta method under general conditions, such as sufficiently high SNR.
Using this model together with the defined diffusion process $\mathbf{x}_t = \mathbf{x} + \mathbf{v}$, the likelihood term $p\left(\mathbf{y} | \mathbf{x}_{t}\right)$ can be obtained. Through linearization with first-order Taylor series expansion, this can be approximated by a multivariate Gaussian distribution with mean vector $\mathbf{\vert Ax}_t\vert$ and covariance matrix $((\frac{\alpha}{2})^2 + \sigma_t^2)\mathbf{I}$ for a unitary measurement matrix \(\mathbf{A}\). 
%%%FIGEN: Makalede bununla ilgili adımları göstermeliyiz. Genel olarak basit toplama, çıkarma, çarpma yaptığımız adımlar hariç tüm adımları göstermeliyiz metodun derivasyonuyla ilgili.
Then the log-likelihood gradient term is given by
%%%%%%%%%%%%%%%%%%%%%%%%%%%%%%%%%%%%%%%%%%%%%%%%
% BU HALA APPROXIMATION, OBTAINED DEMEYELİM.
% Bu forma varmak için
% 1. Independence btw. x_t and y gerekiyor
% 2. \sigma_t^2 küçük olmalı, large SNR gerekiyor
% Burada çok ara stepi atladım. Çoğu diffusion for inverse problems paperı da atlıyor.
% İki modelimiz var:
% y = |Ax| + w where w ~ N(0, (alpha/2)^2 I)
% x_t = x + v where v ~ N(0, sigma_t^2 I)
% Ve biz p(y|x_t) bulmak istiyoruz.
% x GT görüntü. x_t mevcut adımdaki tahminimiz, additive Gaussian ile hatasını modelledik çünkü düzgün eğitirsek önceki adımlardan perfect tahmin geliyor ve bu adıma gelmeden üstüne biz gürültü v ekliyoruz gibi düşünüyoruz.
% p(y | x_t)
% = p(y - |Ax_t| | x_t)
% = p(|Ax| + w - |A(x+v)| | x_t)
% = p(|Ax| + w - |Ax| - J(x)*v | x_t) (first order Taylor expansion around x)
% = p(w - J(x)*v | x_t)
% Burada w - J(x)*v ikisinin de zero-mean Gaussian olduğunu ve varyanslarını biliyoruz. Ama x_t nasıl etkiliyor bilmiyoruz. Onu nasıl kullanacağız?
% İşte, independence assumptionı bunun için gerekli x_t independent olmalı ki 
% p(w - J(x)*v | x_t) = p(w - Jv)
% diyebilelim
% x_t aslında bunlardan etkileniyor independent değil. Ama bu assumption common in the literature, OK.
% Sonra buradan (alpha/2)^2  I + sigma_t^2 * J(x) * J(x)^H olarak varyans geliyor.
% J(x) yani |Ax| fonksiyonunun gt. olan x'teki gradyanını
%%%%%%%%%%%%%%%%%%%%%%%%%%%%%%%%%%%%%%%%%%%%%%%%%
\begin{equation}
\nabla_{\mathbf{x} } \log p\left(\mathbf{y} | \mathbf{x}_{t}\right)
= -\frac{\nicefrac{1}{2}}{(\frac{\alpha}{2})^2 + \sigma_t^2} \nabla_{\mathbf{x} } \| \mathbf{y} - \mathbf{\vert Ax}_t \vert  \|^2, 
    \label{eq:eq11}
\end{equation}
with the following (sub-)gradient expression:
%that has full column rank and orthonormal columns:
%for a full column rank $\mathbf{A}$:
\begin{equation}
 %-\frac{2}{\alpha^2}
2 \left(
\mathbf{{x}}_t -
\mathbf{A}^{\dagger} \left(
\frac{\mathbf{ A{x} }_t}{\mathbf{\vert A{x}}_t \vert }
 \odot \mathbf{y}  
 \right)
 \right)
\in
%\nabla_{\mathbf{{x}} } \log p\left(\mathbf{y} | \mathbf{{x}}_{t}\right).
\nabla_{\mathbf{x} } \| \mathbf{y} - \mathbf{\vert Ax}_t \vert  \|^2.
    \label{eq:eq12}
\end{equation}

%Employing gradient lookahead, where $\mathbf{\tilde{x}}_{t} = \mathcal{D}_{\mathbf{\theta}}(\mathbf{x}_t, t)$, involves using the gradient after the denoising step rather than the gradient at $\mathbf{x}_t$. Thus, the Langevin dynamics update in Eq.~\eqref{eq:denoisersubslangevin} is given by:
We employ gradient lookahead; in other words, we evaluate the log-likelihood gradient after the denoising step. % rather than at \(\mathbf{x}_t\). 
Hence we use \(\mathbf{\tilde{x}}_{t} = \mathcal{D}_{\mathbf{\theta}}(\mathbf{x}_t, t)\),  the denoised estimate, in place of \(\mathbf{x}_t\) in the sub-gradient expression. Substituting the final log-likelihood gradient term into Eq.~\eqref{eq:denoisersubslangevin} yields the following Langevin dynamics update:
\begin{equation}
\begin{aligned}    
\mathbf{x}_{t+1} \leftarrow&
%\frac{4\gamma}{\alpha^2}
\frac{\gamma}{(\frac{\alpha}{2})^2 + \sigma_t^2}
\mathbf{A}^{\dagger} \left(
\frac{\mathbf{ A }\mathbf{\tilde{x}}_{t}}{\mathbf{\vert A}\mathbf{\tilde{x}}_{t} \vert }
 \odot \mathbf{y}  
 \right) 
+
\left(1-\frac{\gamma}{\sigma_t^2}\right) \mathbf{x}_{t}
\\
&+ \left( \frac{\gamma}{\sigma_t^2} -
%\frac{4\gamma}{\alpha^2}
\frac{\gamma}{(\frac{\alpha}{2})^2 + \sigma_t^2}
\right) \mathbf{\tilde{x}}_{t} 
 + \sqrt{2 \gamma} \mathbf{v}_t .
\end{aligned}
    \label{eq:eq13}
\end{equation}
By defining $\lambda_t = \frac{\gamma}{(\frac{\alpha}{2})^2 + \sigma_t^2}$, this can be expressed in a more compact form as follows: 
\begin{equation}
\begin{aligned}    
\mathbf{x}_{t+1} \leftarrow&  
\mathbf{A}^{\dagger} \left(
\frac{\mathbf{ A }\mathbf{\tilde{x}}_{t}}{\mathbf{\vert A}\mathbf{\tilde{x}}_{t} \vert }
 \odot \left(
 %\frac{4\gamma}{\alpha^2}
 %\frac{\gamma}{(\frac{\alpha}{2})^2 + \sigma_t^2}
 \lambda_t
 \mathbf{y}  + \left( \frac{\gamma}{\sigma_t^2} -
 %\frac{4\gamma}{\alpha^2}
 %\frac{\gamma}{(\frac{\alpha}{2})^2 + \sigma_t^2}
 \lambda_t
 \right) \mathbf{\vert A}\mathbf{\tilde{x}}_{t} \vert \right)
 \right) \\
&+ \left(1-\frac{\gamma}{\sigma_t^2}\right) \mathbf{x}_{t} + \sqrt{2 \gamma} \mathbf{v}_t .
\end{aligned}
\label{eq:eq14}
\end{equation}
Here
%Instead of using fixed $\gamma$, we can learn $\lambda_t = \frac{4\gamma}{\alpha^2}$ by making the 
$\lambda_t$ is a time-dependent learnable parameter that controls the strength of the measurement update. %and time-dependent measurement weight parameter used for the %optimal measurement update. % weights during the training process.
%We generalize this by introducing a learnable, timestep-dependent parameter \( \lambda_t := \frac{4\lambda}{\alpha^2} \), yielding:
For a simpler expression, we set $\gamma = \sigma_t^2$
%\frac{4\sigma_t^2}{\alpha^2}$
to arrive at the following update equation:
%\begin{equation}
%\mathbf{x}_{t+1} \leftarrow \mathbf{A}^{\dagger} \left(
%\frac{\mathbf{ A }\mathbf{\tilde{x}}_{t}}{\mathbf{\vert A}\mathbf{\tilde{x}}_{t} \vert }
% \odot \left( \frac{4\gamma}{\alpha^2} \mathbf{y}  + \left( 1 - \frac{4\gamma}{\alpha^2} \right) \mathbf{\vert A}\mathbf{\tilde{x}}_{t} \vert \right)
% \right) + 
% \sqrt{2 \gamma} \mathbf{v}_t 
%\end{equation}
\begin{equation}
\begin{aligned}    
\mathbf{x}_{t+1} \leftarrow& \mathbf{A}^{\dagger} \left(
\frac{\mathbf{ A }\mathbf{\tilde{x}}_{t}}{\mathbf{\vert A}\mathbf{\tilde{x}}_{t} \vert }
 \odot \left( \lambda_t \mathbf{y}  + \left( 1 - \lambda_t \right) \mathbf{\vert A}\mathbf{\tilde{x}}_{t} \vert \right)
 \right)\\
 &+ \alpha \sqrt\frac{\lambda_t}{2 (1-\lambda_t)} \mathbf{v}_t .
\end{aligned}
\label{eq:eq15}
\end{equation}
where $\lambda_t \in (0,1)$.

Note that the first term above corresponds to a single measurement-space projection step of an alternating projection algorithm such as ER, where the initial estimate is denoised $\mathbf{\tilde{x}}_{t}$ and the input data is the updated measurement $\lambda_t \mathbf{y}  + \left( 1 - \lambda_t \right) \mathbf{\vert A}\mathbf{\tilde{x}}_{t} \vert$.
However, (sub-)gradient methods and alternating projection algorithms such as ER are known to be less effective for phase retrieval. When additional object-domain constraints, such as real-valuedness or non-negativity, are available, this step can instead be replaced with the Hybrid Input-Output (HIO) algorithm, which typically provides improved empirical performance.
This important modification~\cite{Isil:20,isil2024deep} is often not incorporated in prior diffusion-based PR methods such as \cite{Shoushtari2022DOLPHDM}, which typically rely on simpler update rules that may lead to inferior reconstruction performance. %convergence behavior.
%Notably, prior diffusion-based approaches to phase retrieval, such as \cite{Shoushtari2022DOLPHDM}, do not account for this consideration.
%Moreover, HIO also allows to incorporate the available object-domain constraints such as real-valuedness and non-negativity, without hardly enforcing them.
%the expression \(\nabla_{\mathbf{x}} \log p\left(\mathbf{y} \mid \mathbf{x}_{t}\right)\) is incomplete, as it neglects additional prior information about \(\mathbf{x}\) beyond its Fourier magnitude, specifically, spatial properties such as realness and non-negativity. To approximate the effect of incorporating these spatial constraints, i.e., simulating \(\nabla_{\mathbf{x}} \log p\left(\mathbf{y}, \text{spatial constraints} \mid \mathbf{x}_{t}\right)\), we avoid hard enforcement and instead integrate the constraints into the HIO feedback steps to improve convergence.
We also
empirically 
observe that
performing multiple HIO iterations, rather than a single update, further improves reconstruction performance.
%performing multiple HIO iterations, rather than a single update, leads to improved reconstruction performance.
With this final modification to the data consistency term, our update becomes:
\begin{equation}
\begin{aligned}
\mathbf{x}_{t+1} \leftarrow&  \text{HIO}(\mathcal{D}_{\mathbf{\theta}}(\mathbf{x}_{t}, t) ; \lambda_t \mathbf{y} + (1 - \lambda_t) |\mathbf{A} \mathcal{D}_{\mathbf{\theta}}(\mathbf{x}_{t}, t) |) \\
 &+ \alpha \sqrt\frac{\lambda_t}{2 (1-\lambda_t)} \mathbf{v}_t .
\end{aligned}
\end{equation}
It is worth noting that when no object-domain constraints are available, HIO can be replaced with other alternating projection-based solvers.

To further improve the performance, %conditioning on the observations, 
we adopt a warm-start strategy rather than initializing the process with a pure noise image. Adopting such a strategy requires deviating from the standard diffusion sampling process, which typically begins from noise. %a purely noisy input. 
Specifically, we begin with a plausible estimate obtained using classical methods such as HIO, through the initialization stage proposed in \cite{pmlr-v80-metzler18a}. This warm-start
%, as explained in Section \ref{section:initialization},
simplifies the learning task by allowing the diffusion model to focus on refining a rough estimate rather than generating one from scratch. Given that classical phase retrieval algorithms are computationally efficient and can already produce fairly accurate reconstructions, it is more efficient to leverage these intermediate solutions rather than wasting denoiser model capacity on early-stage reconstruction. %Instead this utilizes the denoiser’s capacity to focus on high-fidelity refinement. 
Such ``image-to-image'' rather than ``noise-to-image'' paradigm has also been recently exploited for various inverse problems in imaging~\cite{delbracio2023inversion, whang2022deblurring, 
Bansal2022ColdDI, jaiswal2023physics}.

Together with the warm-start strategy, the proposed pipeline is summarized in Algorithm~\ref{alg:capsmall}. Detailed derivations for Eqs.~\eqref{eq:eq10}--\eqref{eq:eq12} are provided in the supplementary document.

\begin{algorithm}
\caption{Proposed algorithm: prNet-Small}\label{alg:capsmall}
\textbf{Input: $\mathbf{y}, \alpha$}

\textbf{Hyperparameters: $ T, K, \beta$}

\textbf{Learned parameters:} Denoiser weights $\theta$, measurement update weights $\boldsymbol{\lambda} \in (0,1)^T$ (initialized as a logarithmically decreasing vector)

%\textbf{Output: $\mathbf{z}_T^{(0)}$}

%\textbf{Output: x′T\mathbf{x_T'}}
\begin{algorithmic}[1]
\State $\mathbf{x}_0' \gets $ Initialization Stage($\mathbf{y}$)
\For{$ i = 1 $ to $T$}
    \State $\mathbf{x}_i \gets \mathcal{D}_{\mathbf{\theta}}(\mathbf{x}_{i-1}'$, $i-1)$
    
    \If{$i = T$} 
        \State \Return $\mathbf{x}_i$
    \EndIf

    \State $ {\mathbf{y}_i}' \gets \lambda_i \mathbf{y} + (1-\lambda_i) |\mathbf{A}\mathbf{x}_i| $
    
    \State $\mathbf{z}_i^{(0)} \gets \mathbf{x}_i $

	\For{$k=1$ to $K$}
    \State ${\mathbf{z}_i^{(k)}}' \gets \mathbf{A^{\dagger}} \left( {\mathbf{y}_i}' \odot \frac{\mathbf{Az}_i^{(k-1)}}{|\mathbf{Az}_i^{(k-1)}|} \right)  $ 

\State $\gamma \gets$ \parbox[t]{\dimexpr0.8\columnwidth-\algorithmicindent}{indices where ${\mathbf{z}_i^{(k)}}'$ violates spatial constraints (e.g., support and non-negativity)}
    
\State ${\mathbf{z}_i^{(k)}}[n] \gets 
\begin{cases} 
  {\mathbf{z}_{i}^{(k)}}'[n] & \text{, }  n \notin \gamma \\
  
    \mathbf{z}_{i}^{(k-1)}[n] - \beta {\mathbf{z}_{i}^{(k)}}'[n]
  
  & \text{, }  n \in \gamma
\end{cases}$

    \EndFor

\State $\boldsymbol{\epsilon} \gets \mathcal{N}(\mathbf{0},\mathbf{I})$

\State $\mathbf{x}_i' \gets \mathbf{z}_i^{(K)} + \alpha \sqrt\frac{\lambda_i}{2 (1-\lambda_i)}  \mathbf{\epsilon} $ 
\EndFor
\end{algorithmic}
\end{algorithm}

\subsection{prNet-Large}

Because our initialization stage incorporates stochasticity, different outputs are obtained for the same measurement across different runs. Some of these outputs reconstruct certain regions of the image better than others. This observation motivates leveraging multiple initialization results within our prNet-Large pipeline.

The prNet-Large pipeline enhances reconstruction quality through multiple parallel reconstructions. Specifically, the initialization stage produces \( k \) diverse estimates. In the main loop, each of them is passed through a denoiser to yield \( k/2 \) refined outputs. These outputs, along with those from the data consistency step, are concatenated and perturbed with Gaussian noise before the next iteration. Thus, each step of the main loop receives \( k \) inputs and produces \( k \) outputs.

In the final stage of the prNet-Large pipeline, we compute the average of these \( k \) outputs. Given that our method approximates samples from the posterior \( p(\mathbf{x} | \mathbf{y}) \), averaging them yields an estimate of the MMSE solution. Since the MMSE estimator minimizes expected distortion, this averaging can improve distortion-based metrics.

\subsection{prNet-Large-Adversarial}

\begin{figure*}[tb!]
% \centering\includegraphics[width=\linewidth, page=1]{figures/newlargepipeline.pdf}
\centering\includegraphics[width=\linewidth, page=1]{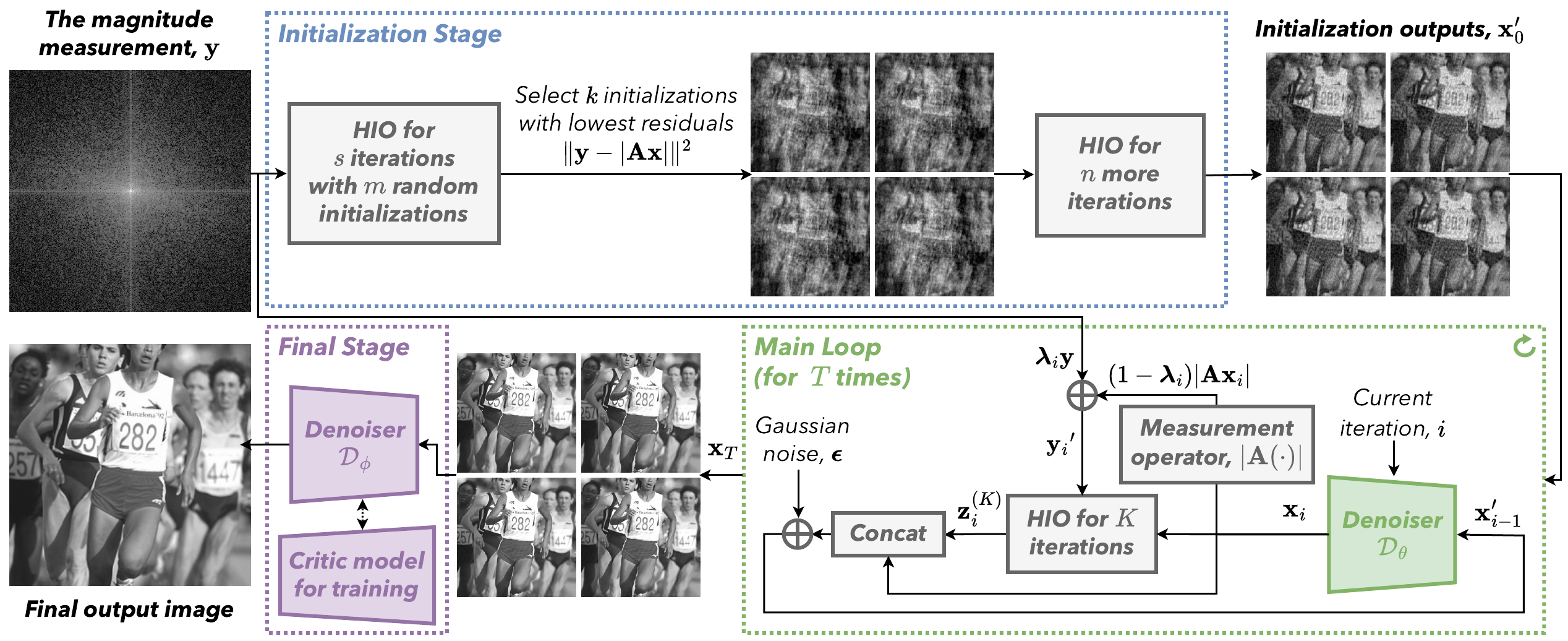}
\caption{The overall pipeline of prNet-Large-Adversarial.
Our approach begins with the Initialization Stage, where \( m \) random initializations are refined using HIO for \( s \) steps. The top \( k \) candidates with the lowest residuals are selected and further refined with \( n \) additional HIO iterations to produce initial estimates \( \mathbf{x}_0' \). In the Main Loop, each estimate is stochastically perturbed with Gaussian noise and iteratively refined using a combination of classical HIO updates and a learned denoiser \( D_\theta \). This process is repeated for \( T \) iterations. In the Final Stage, the refined outputs are passed through a learned denoiser \( D_\phi \), trained adversarially via a critic model to enhance perceptual quality. Compared to prNet-Large, which uses simple averaging in the Final Stage, prNet-Large-Adversarial incorporates a learned denoiser for aggregation. Additionally, prNet-Large-Adversarial refines multiple reconstructions in parallel, while prNet-Small operates on a single initialization throughout.}
\label{fig:prnettttoveralllarge}
\end{figure*}

The prNet-Large-Adversarial pipeline includes an additional refinement stage to combine the multiple reconstructions from the main loop, as illustrated in Fig.~\ref{fig:prnettttoveralllarge}. Rather than directly averaging the outputs from the main loop, a final denoiser \( \mathcal{D}_{\boldsymbol{\phi}} \) with parameters \( \boldsymbol{\phi} \) takes the multiple reconstructions as input and produces a refined output. This step compensates for the perceptual degradation often introduced by naive averaging, allowing us to improve both distortion and perceptual quality simultaneously.

During training, the final denoiser is optimized using both a distortion-based reconstruction loss and an adversarial loss based on the improved Wasserstein GAN with gradient penalty (WGAN-GP)~\cite{Gulrajani2017ImprovedTO} to enhance perceptual quality. Specifically, the denoiser output is evaluated by a critic network, which encourages the refined reconstruction to be indistinguishable from real images while preserving reconstruction fidelity. The adversarial loss mitigates the over-smoothing commonly observed in denoisers trained solely with pixel-wise losses such as MSE, leading to perceptually more realistic reconstructions. The critic network is used only during training, while inference requires only the learned denoiser \( \mathcal{D}_{\boldsymbol{\phi}} \).

\subsection{Initialization Stage}
\label{section:initialization}

Due to the inherent nonlinearity and non-convexity of the phase retrieval problem, reconstruction algorithms are highly susceptible to the initial guess. In order to address this challenge and enhance the robustness of our method, 
this initialization procedure runs the HIO algorithm for a small number of $s$ iterations for $m$ different random phase initializations, as in prDeep~\cite{pmlr-v80-metzler18a}. This initial exploration aims to identify promising regions in the search space and is highly parallelizable.
% After selecting the reconstruction with the lowest residual ${\left\Vert \mathbf{y-\vert Ax \vert} \right\Vert}^2_2$, this reconstruction is then further refined using HIO for a larger number of $n$ iterations, as in \cite{pmlr-v80-metzler18a}.
% This multi-restart strategy of prDeep, this initialization procedure runs the HIO algorithm for a small number of $s$ iterations for $m$ different random phase initializations.
For prNet-Small, the reconstruction with the lowest residual ${\left\Vert \mathbf{y-\vert Ax \vert} \right\Vert}^2_2$ is selected and further refined using HIO for a larger number of $n$ iterations. For prNet-Large and prNet-Large-Adversarial, the top $k$ candidates with the lowest residuals are instead retained and each is independently refined for $n$ iterations, producing $k$ diverse warm-starts that feed the parallel main loop.

\subsection{Denoiser Model}

\begin{figure}[b]
\centering\includegraphics[width=\columnwidth, page=1]{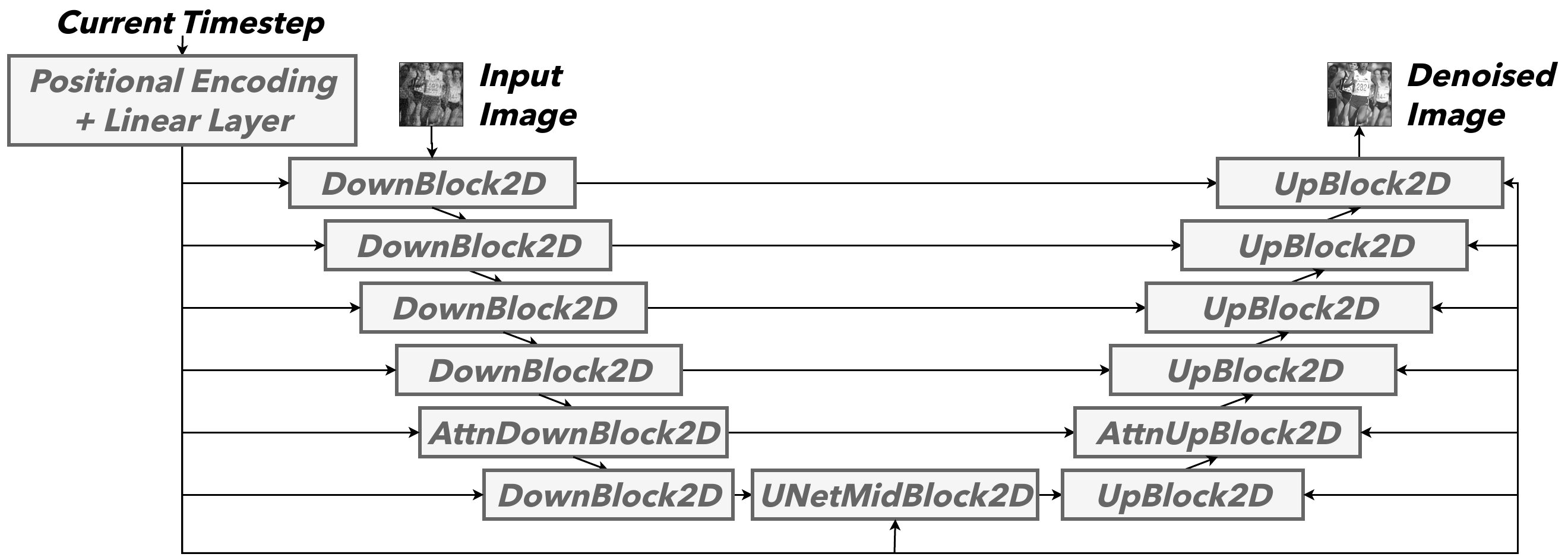}
\caption{Architecture of the UNet denoiser with timestep input. For prNet-Small, the input and output are single images, as illustrated. In the main loop denoiser of prNet-Large, the network processes multiple input images and produces multiple output images.}
\label{fig:prnettttunetarch}
\end{figure}

As the denoising component of our pipelines, we employed a customized UNet architecture \cite{ronneberger2015u} depicted in Fig. \ref{fig:prnettttunetarch}, a well-established framework renowned for its efficacy in image restoration tasks \cite{elad2023image}. Notably, this implementation of UNet incorporates timestep information as an additional input that is intricately linked to the noise level of the input image. The denoiser operates by estimating the residual, thus facilitating the refinement of the reconstructed image by focusing only on the discrepancy between the noisy input and the desired clean output.
Our customization of the UNet architecture includes blocks that utilize attention mechanisms. These mechanisms enable the network to selectively focus on relevant parts of the input image, enhancing its ability to capture intricate details and effectively suppress noise. This incorporation of attention mechanisms is crucial for improving denoising performance, particularly in scenarios where noise levels vary across different regions of the image.

\subsection{Progressive Training Process}

During training, the denoiser model receives the output from the previous iteration and produces an estimate of the clean image. Our training loss is a standard MSE loss between this estimate and the ground truth image. Additionally, since there are other learnable parameters following the denoising block (e.g., $\boldsymbol{\lambda}$ in the data consistency layers), our loss function also includes a term corresponding to the reconstruction error at the output of the data consistency block.

We adopt a progressive strategy that evolves over training epochs, enabling effective learning across all iterations. Initially, we focus on training the early iterations of the main loop to learn the initial stages of reconstruction. Within each epoch, we gradually increase the mean of the random timesteps used during training, allowing the model to adapt to increasingly complex temporal structures. As training progresses to later epochs, the focus shifts toward optimizing the final iterations. This progressive schedule mirrors the concept of algorithm unrolling, where outputs from earlier iterations inform the training of subsequent ones, thereby improving both training efficiency and coherence.

%%%%%%%%%
% LOSS FORMULATIONS:

To formalize the progressive training objective described above, we define the training loss for prNet as follows:
\begin{equation}
\min_{\theta, \boldsymbol{\lambda}}
 \mathbb{E}_{i \sim p(i),\, \mathbf{x} \sim p(\mathbf{x})} \left[ \mu_1 \left\|
 \mathcal{D}_\theta(\mathbf{x}_i', i) - \mathbf{x} \right\|^2
 %\mathcal{D}_{\mathbf{\theta}}(\mathbf{x}_{i-1}', i-1) - \mathbf{x}  \right\|^2
 + \mu_2 \left\|
 \mathbf{z}_{i+1}^{(K)} - \mathbf{x} \right\|^2 \right]
\end{equation}
where \( \mathbf{x} \) denotes the ground truth image sampled from the data distribution \( p(\mathbf{x}) \), and \( \mathbf{x}_i' \) is the input to the denoiser at iteration \( i \) generated by the previous iterations using the current version of the denoiser model and noise process, as described in Algorithm~\ref{alg:capsmall}. The denoiser %output 
\( \mathcal{D}_\theta(\mathbf{x}_i', i) \) is trained to approximate the clean image, with \( \mu_1 \) controlling the weight of the denoising loss given in the first term. The second term enforces fidelity at the output of the data consistency block, where \( \mathbf{z}_{i+1}^{(K)} \) is the output after \( K \) HIO updates, %and spatial constraint corrections, 
as defined in Algorithm~\ref{alg:capsmall}. The scalar \( \mu_2 \) controls the relative importance of this term.
% TODO: more details...
The sampling distribution \( p(i) \) is the probability density function of the chosen iteration, whose mean increases linearly with the training epoch, thus progressively focusing more on later %refinement steps 
iterations as the training proceeds. Since the input \( \mathbf{x}_i \) depends on the outputs of previous iterations, we adopt a stage-wise training strategy that initially focuses on earlier steps, allowing their denoisers to stabilize before focusing on later iterations in subsequent epochs.

In contrast to methods that assume a fixed noise schedule and train denoisers accordingly, our approach explicitly leverages the actual outputs of previous iterations during training. This use of exact, rather than approximated, inputs simplifies the learning process but may increase overall training time. A key advantage of our framework is its flexibility in learning the denoising schedule. Unlike approaches that rely on a fixed, predefined diffusion process, our pipeline allows both noise and denoising schedule to be learned during training. This flexibility enables the model to discover an optimal denoising strategy tailored to the reconstruction task, potentially leading to improved performance.

To train the final denoiser \( \mathcal{D}_{\boldsymbol{\phi}} \) in the prNet-Large-Adversarial pipeline, we augment the reconstruction objective with an 
%we incorporate an additional 
improved Wasserstein GAN loss with gradient penalty (WGAN-GP)~\cite{Gulrajani2017ImprovedTO}. %into the training objective.
%This term addresses perceptual quality alongside distortion metrics, helping to balance the perception-distortion tradeoff. Denoisers trained solely with distortion-based losses, such as MSE, often produce overly smooth outputs that are easily distinguishable by a critic model. The WGAN loss penalizes such outputs, promoting more realistic reconstructions. Since both the Langevin dynamics framework used in the main loop and this adversarial term used for the final denoiser explicitly tackle the perception-distortion tradeoff, our overall training scheme provides a comprehensive framework aligned with the principles discussed in \cite{Blau2017ThePT}.
Hence the final output from the learned denoiser \( \mathcal{D}_\phi \) is evaluated by a critic network using WGAN-GP.
The resulting training objective includes an additional adversarial term
\begin{equation}
\mu_{\text{adv}} \cdot \mathcal{L}_{\text{WGAN}}\!\left(\mathcal{D}_{\boldsymbol{\phi}}(\mathbf{x}_T^{(1)}, \dots, \mathbf{x}_T^{(k)})\right),
\end{equation}
where \( \mathbf{x}_T^{(1)}, \dots, \mathbf{x}_T^{(k)} \) denote the \( k \) reconstructions obtained from the final iteration of the prNet-Large pipeline, and \( \mu_{\text{adv}} \) is a hyperparameter controlling the tradeoff between distortion-based reconstruction and perceptual quality.

Since both the Langevin dynamics framework used in the main loop and this adversarial term used for the final denoiser are designed to improve the perception-distortion tradeoff, the overall training scheme provides a framework aligned with the principles discussed in~\cite{Blau2017ThePT}.

%For prNet-Large-Adversarial, an additional adversarial loss is introduced to enhance perceptual quality. 
%
%The resulting training objective has an extra term \(\mu_{\text{adv}} \cdot \mathcal{L}_{\text{WGAN}}(\mathcal{D}_\phi(\mathbf{x}_T^{(1)}, \dots, \mathbf{x}_T^{(k)}))\) where \( \mathbf{x}_T^{(1)}, \dots, \mathbf{x}_T^{(k)} \) are the \( k \) reconstructions from the final iteration of the prNet-Large pipeline, and \( \mu_{\text{adv}} \) is a hyperparameter balancing perceptual realism against distortion-based reconstruction. The WGAN loss encourages the final denoiser \( \mathcal{D}_\phi \) to generate samples indistinguishable from real images, thereby mitigating the oversmoothing typically introduced by MSE-based training.

\subsection{Test Time Augmentation}

We can leverage inherent invariances of the measurement operator 
to improve reconstruction quality through test time augmentation (TTA). Many measurement operators exhibit invariance under certain image transformations such as flipping or rotation. For example, in the case of Fourier magnitude, the magnitude spectrum of a flipped image remains identical to that of the original image.
%Since our measurement model is invariant to flipping, meaning that the Fourier magnitude of a flipped image is identical to that of the original, we can leverage this mathematical property for test time augmentation.
This property enables us to apply corresponding transformations during test time to enrich our reconstruction process.

\begin{figure}[t!]
%\centering\includegraphics[width=\columnwidth, page=1]{figures/tta_ew.pdf}
\centering\includegraphics[width=\columnwidth, page=1]{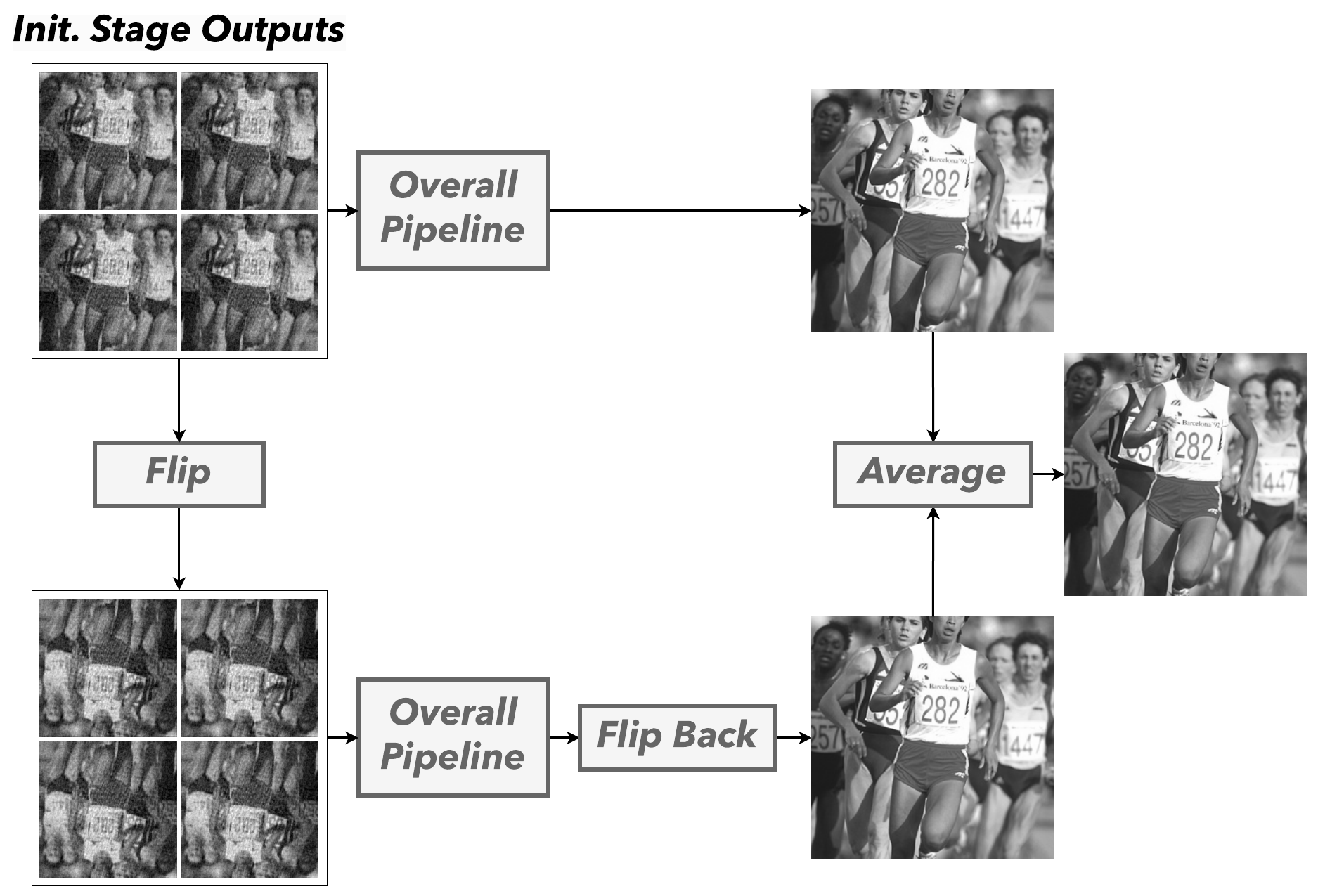}
\caption{Test time augmentation (TTA): We execute the full pipeline on both the original initialization outputs and their flipped versions, then average the results to produce the final output.}
\label{fig:prnettttttapipeline}
\end{figure}

As depicted in Fig. \ref{fig:prnettttttapipeline}, following the robust initialization stage, we can apply flipping to these initialization outputs and execute our pipeline for the flipped versions of the images for the Fourier PR problem. Subsequently, combining the flipped outputs with the original outputs allows us to obtain a more refined estimate. Test time augmentation is widely applicable across various deep learning domains and can also be beneficial for enhancing the performance of image reconstruction tasks.

\begin{figure}[tb]
%\centering\includegraphics[width=0.9\columnwidth, page=1]{figures/ttad4.pdf}
\centering\includegraphics[width=\columnwidth, page=1]{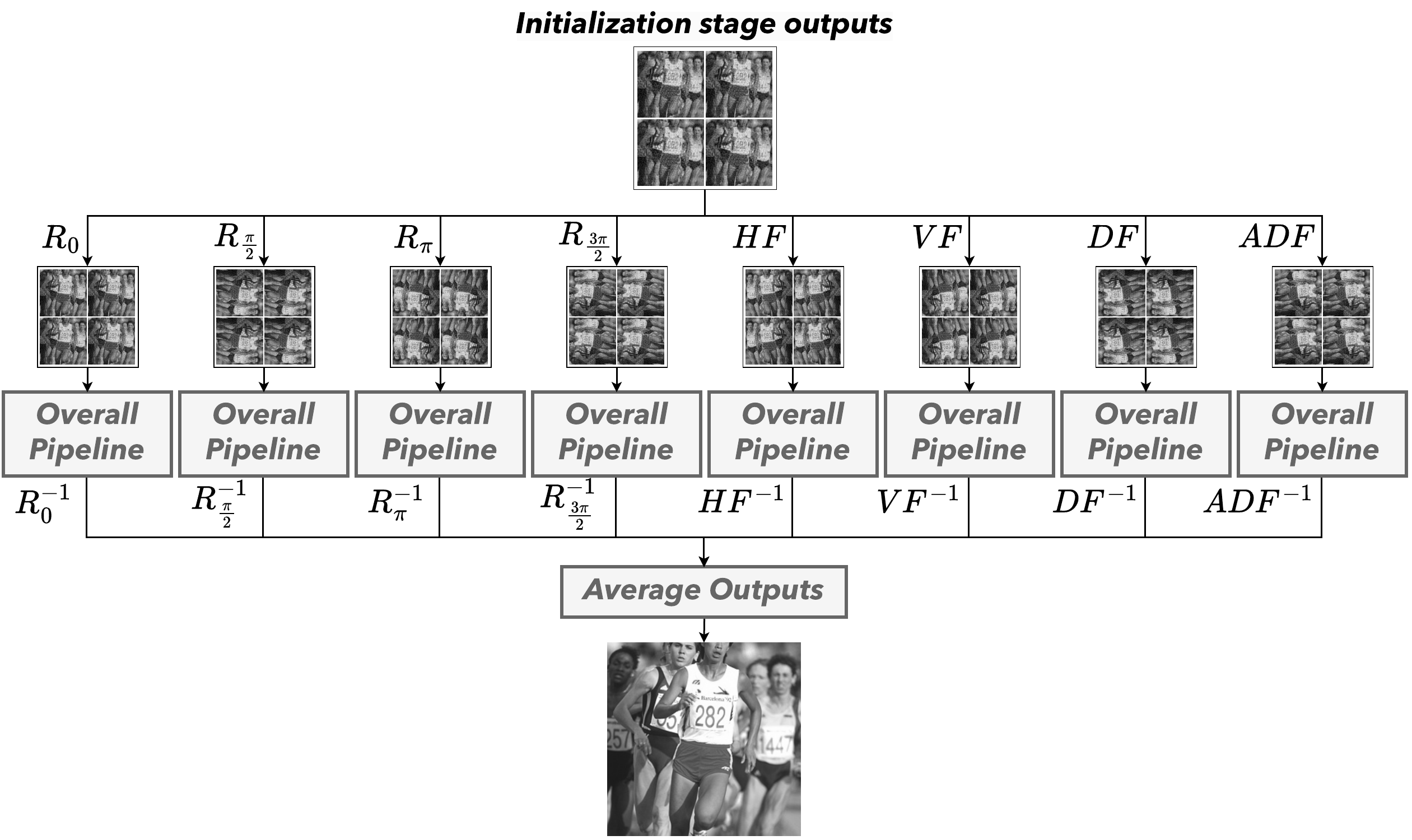}
\caption{Test time augmentation using dihedral group $D_4$
 (TTA $D_4$).}
\label{fig:prnettttttapipeline_d4}
\end{figure}

%%%
A more advanced Test Time Augmentation technique called TTA \(D_4\), as illustrated in Fig. \ref{fig:prnettttttapipeline_d4}, leverages the properties of the \(D_4\) dihedral group, which includes all symmetries of a square, such as rotations and reflections. This method enhances the initial TTA by applying each transformation from the \(D_4\) group to the outputs from the robust initialization stage, covering rotations ($R_0$, $R_{\pi/2}$, $R_{\pi}$, $R_{3\pi/2}$) and reflections (Horizontal Flip $HF$, Vertical Flip $VF$, Diagonal Flip $DF$, and Anti-Diagonal Flip $ADF$). Formally, we process the initialization outputs $\{ \hat{\mathbf{x}}_{\text{init}}^{\textit{(m)}} \}_{m = 1}^k$ with a transform $\mathcal{T}$ to generate new sets of initializations $\{ \mathcal{T}( \hat{\mathbf{x}}_{\text{init}}^{\textit{(m)}} ) \}_{m = 1}^k$. We also know the effects of these transformations in the Fourier domain, thus, we also apply the corresponding transformation in the Fourier domain to the observation $\mathbf{y}$. These transformed initializations are then iteratively refined, producing different final outputs. The combined final result is obtained by averaging over all \(D_4\) transformations, expressed as:
\begin{equation}
\hat{\mathbf{x}}_{\text{final}}^{\text{(combined)}} = \frac{1}{|D_4|} \sum_{\mathcal{T} \in D_4} \mathcal{T}^{-1} (\hat{\mathbf{x}}_{\text{final}}^{\mathcal{T}})
\label{eq:ttaugd4}
\end{equation}

where $|D_4|=8$ is the order of the \(D_4\) dihedral group.

By incorporating all transformations from the \(D_4\) dihedral group, this advanced TTA technique maximizes the use of symmetry properties and available data, significantly enhancing the robustness and quality of image reconstructions. This approach is particularly effective in image reconstruction tasks, where the enriched data from augmentation helps mitigate overfitting and improves generalization performance.

The TTA $D_4$ strategy admits a clean interpretation as Reynolds averaging applied at the pipeline level. Let $\Phi$ denote the composition of the main loop and the final stage, i.e., the mapping from an initialization $\hat{\mathbf{x}}_{\text{init}}$ and measurement $\mathbf{y}$ to a final reconstruction $\hat{\mathbf{x}}_{\text{final}}$. Since the dihedral group $D_4$ leaves the Fourier magnitude $|\mathbf{F}\mathbf{x}|$ invariant up to known transformations of $\mathbf{y}$, Eq.~\eqref{eq:ttaugd4} approximates an equivariant reconstruction operator $\Phi_{D_4}(\mathbf{y}) = \frac{1}{|D_4|} \sum_{\mathcal{T} \in D_4} \mathcal{T}^{-1}\bigl(\Phi(\mathcal{T}\mathbf{y})\bigr)$. This is closely related to the equivariant PnP framework of~\cite{terris2024equivariant}, which enforces equivariance at the denoiser level via either the full Reynolds average or a stochastic per-iteration sample, and to the periodical geometric self-ensemble used within DPIR~\cite{zhang2021plug}, which cycles through transformations across iterations. Our formulation differs in two respects. First, the group action is applied at the pipeline level rather than per denoiser call, thus each transformed input is independently refined by the full main loop before averaging. Importantly, we apply $D_4$ transformations only after the initialization stage rather than rerunning HIO from $m$ random initializations for every group element; the initialization stage is itself stochastic and already produces diverse seeds $\{\hat{\mathbf{x}}_{\text{init}}^{(m)}\}_{m=1}^k$, thus re-running it under each transformation would inflate compute without adding diversity beyond what the existing $m$ random restarts provide. Second, the diversity required for ensembling at the main-loop level is therefore already supplied upstream by these HIO restarts, and applying $D_4$ to the selected seeds complements rather than duplicates this diversity by injecting symmetry-aligned variation that the random restarts alone do not span.

\section{Experimental Results}
 \label{sec:resultsprnet}
 
To evaluate the performance of our methods, we conduct numerical simulations using a large image dataset. Our experiments focus on the challenging Fourier phase retrieval problem, which involves recovering an image from the magnitude of its oversampled Fourier transform. We assess generalization capability and computational efficiency, and compare reconstruction quality against both classical and state-of-the-art phase retrieval algorithms.

\subsection{Experimental Setup}

Noisy Fourier intensity measurements are simulated according to Eq.~\eqref{eq:measurementmodel} at the noise level $\alpha=3$.
% To ensure the uniqueness of the solution (up to trivial ambiguities),
% we employ an oversampled discrete Fourier transform matrix $\mathbf{A}=\mathbf{F}$ with an oversampling ratio of $m = 4n$~\cite{hayes1982}.
We employ an oversampled discrete Fourier transform, modeled as $\mathbf{A} = \mathbf{F}\mathbf{O}$, where $\mathbf{O} \in \mathbb{R}^{m \times n}$ is a zero-padding operator that converts the $\sqrt{n}\times \sqrt{n}$ image into an image of size $\sqrt{m}\times \sqrt{m}$, and $\mathbf{F} \in \mathbb{C}^{m \times m}$ is the unitary $\sqrt{m}\times \sqrt{m}$ point discrete Fourier transform matrix. We assume the images are real-valued and compactly supported, consistent with typical Fourier phase retrieval setups.
Oversampling provides additional measurement redundancy, which helps to constrain the solution space and improve the stability of the inversion process. 
We use an oversampling ratio of $m = 4n$, which guarantees uniqueness of the solution up to trivial ambiguities~\cite{hayes1982}. The pseudoinverse is given by $\mathbf{A}^\dagger = \mathbf{O}^\top \mathbf{F}^{\mathrm{H}}$, i.e., the inverse DFT followed by truncation to the original image support of $\sqrt{n}\times \sqrt{n}$.

The training set consists of $44,000$ grayscale natural images, including the dataset used in \cite{Isil:19, isil2024deep} as well as randomly selected images from ImageNet~\cite{deng2009imagenet,zhang2017learning}. Only natural images are used for training. All images have $256 \times 256$ pixels. The test set, identical to that used in \cite{Isil:19, isil2024deep}, contains 230 natural and 6 unnatural grayscale images used in \cite{pmlr-v80-metzler18a}.

A customized UNet architecture is used for the denoiser models, while a simple ResNet18 network~\cite{He2015DeepRL} is employed as the critic model for the prNet-Large-Adversarial pipeline. For the case with $\alpha=3$ noise level, training is performed using AdamW optimizer with decoupled weight decay regularization~\cite{Loshchilov2017DecoupledWD} along with cosine annealing and linear warmup~\cite{Loshchilov2016SGDRSG}.
% The total training times for prNet-Small, prNet-Large, and prNet-Large-Adversarial are approximately four days (for 90 epochs), five days (for 40 epochs), and one day (for 25 epochs), respectively, using a single NVIDIA A100 80GB GPU. 
The total training times for prNet-Small and prNet-Large are respectively around four days (90 epochs) and five days (40 epochs) on a single NVIDIA A100 80GB GPU. prNet-Large-Adversarial is obtained by attaching the final aggregation denoiser $\mathcal{D}_\phi$ and its critic on top of an already-trained prNet-Large model and training only these additional components for an additional day (25 epochs); the underlying prNet-Large parameters are frozen during this training.

In the initialization stage of prNet-Small, the HIO method was initially run with $m=50$ different random starting points for $s=50$ iterations each. The reconstruction with the lowest residual error was selected for an additional HIO run of $n=1000$ iterations. The resulting reconstruction was then used as input for the main loop containing the denoiser update, measurement update, HIO iterations, and random perturbation. In this iterative phase, consisting of $T=18$ blocks, the number of HIO iterations is chosen as  $K=5$.

The selected hyperparameters for the prNet-Large pipeline differ from the prNet-Small pipeline only in the initialization stage. In the prNet-Large initialization stage, $k=10$ multiple outputs are generated from the best $k=10$ initializations with the lowest residuals among the $m=100$ different random initializations.

\begin{figure}[t!]
%\centering\includegraphics[width=9cm]{figures/histogram_alpha3_large_small_new.pdf}
\centering\includegraphics[width=\columnwidth]{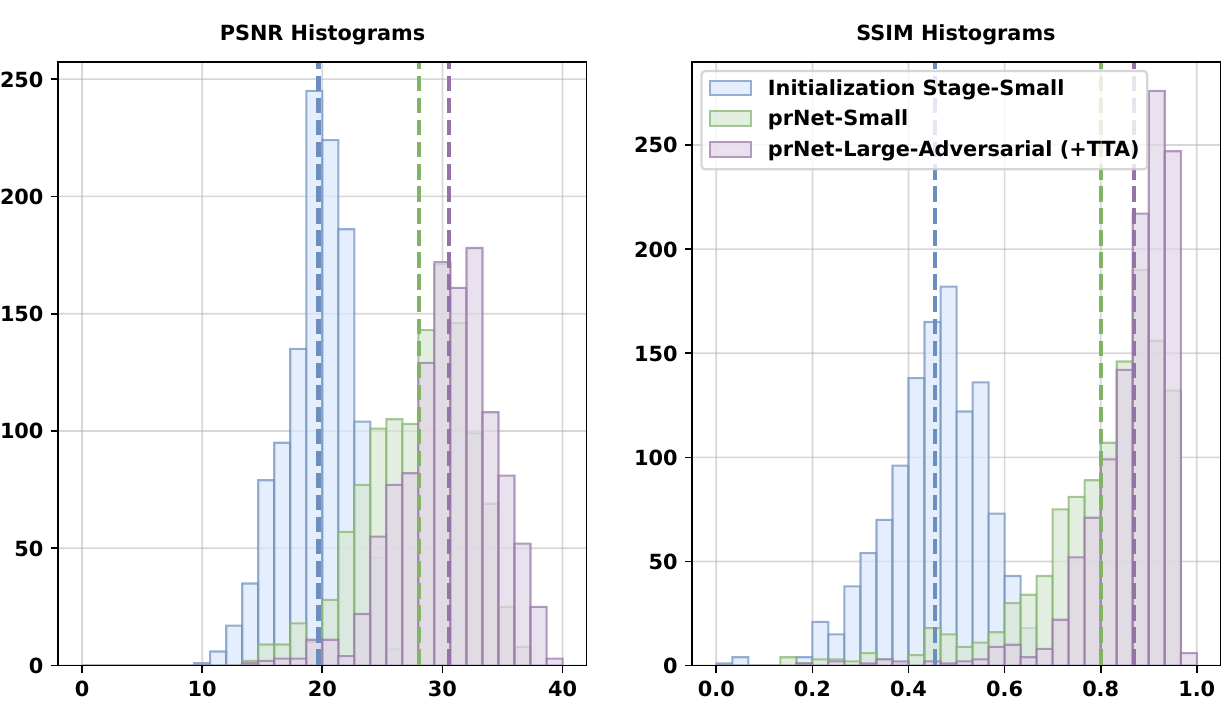}
\caption{The histograms of PSNR (left column) and SSIM (right column) for the reconstructions produced by various methods across 236 test images and 5 Monte Carlo runs for the $\alpha=3$ scenario. Vertical dashed lines indicate the mean PSNR and SSIM values. %Overlapping histograms for each column are shown at the bottom.
}
\label{fig:prnetttthistograms}
\end{figure}

\renewcommand{\arraystretch}{0.99}  % a little more vertical padding
\setlength{\tabcolsep}{10pt}        % a little more horizontal padding
\begin{table*}[htb!]
  \centering
  \caption{
  %Average reconstruction performances for 236 test images across 5 Monte Carlo runs.
  Average reconstruction performance over 236 test images (natural and unnatural) across 5 Monte Carlo runs. All results are obtained using the same model trained on natural images at noise level $\alpha = 3$, and evaluated at multiple noise levels ($\alpha = 2, 3, 4$) to assess generalization across image domains and robustness to varying noise conditions. For each plug-and-play diffusion-based baseline, we report two variants: 
    $^{\dagger}$ using the hyperparameters reported in the original paper, and 
    $^{\ddagger}$ using hyperparameters tuned via Bayesian optimization on our validation set.
  }
  \begin{adjustbox}{width=\textwidth}
    \begin{tabu}{c l c c c c c c c}
      \toprule
      \textbf{} 
        & \textbf{} 
        & \textbf{}
        & \multicolumn{3}{c}{\textbf{Average PSNR (dB) $\uparrow$}} 
        & \multicolumn{3}{c}{\textbf{Average SSIM $\uparrow$}} 
        \\
      \cmidrule(lr){4-6} \cmidrule(lr){7-9}
      \textbf{Noise Level} 
        & \textbf{Method} 
        & \textbf{Runtime (s) $\downarrow$}
        & Overall & Natural & Unnatural 
        & Overall & Natural & Unnatural  \\
      \midrule
        \multirow{16}{*}{\rotatebox{90}{\shortstack{$\alpha=2$\\(Avg. SNR: 33.24dB)}}}
        & HIO~\cite{fienup1982comparison}          & 0.25 & 19.79 & 19.73 & 21.92 & 0.50 & 0.50 & 0.49  \\
        & prDeep~\cite{pmlr-v80-metzler18a}       & 59.32 & 23.45 & 23.49 & 21.72 & 0.65 & 0.66 & 0.58  \\
        & DIR~\cite{Isil:19}                      & 21.59 & 23.61 & 23.60 & 24.02 & 0.72 & 0.72 & 0.73  \\
        & PnP-HIO~\cite{Isil:20}                   & 24.11 & 24.87 & 24.86 & 25.56 & 0.74 & 0.74 & 0.74  \\
        & I2I-PR~\cite{kaya2026i2i} %($T=4,$ no aggregation)
        & 1.10 & 28.59 & 28.65 & 26.39 & 0.79 & 0.80 & 0.67\\
        & DiffFPR$^{\dagger}$~\cite{li2024difffpr}                   & 63.49 & 19.35 & 19.40 & 17.61 & 0.45 & 0.45 & 0.35  \\
        & DiffFPR$^{\ddagger}$~\cite{li2024difffpr}                   & 64.92 & 25.10 & 25.16 & 23.12 & 0.65 & 0.65 & 0.54  \\
        & DAPS$^{\dagger}$~\cite{zhang2025improving}                   & 94.70 & 18.86 & 18.89 & 18.04 & 0.44 & 0.44 & 0.48 \\
        & DAPS$^{\ddagger}$~\cite{zhang2025improving}                   & 348.73 & 25.25 & 25.31 & 22.77 & 0.72 & 0.72 & 0.66  \\
        & LatentDAPS$^{\dagger}$~\cite{zhang2025improving}                   & 85.67 & 17.42 & 17.40 & 18.05 & 0.35 & 0.35 & 0.42 \\
        & LatentDAPS$^{\ddagger}$~\cite{zhang2025improving}                   & 78.41 & 27.99 & 27.90 & 31.38 & 0.79 & 0.79 & 0.80  \vspace{0.3em}\\
        \cline{2-9}\\[-0.9em]
        & Initialization Stage                   & 0.45 & 20.64 & 20.55 & 24.25 & 0.53 & 0.53 & 0.57  \\
        & prNet‑Small                            & 0.98 & 29.60 & 29.67 & 26.81 & 0.85 & 0.85 & 0.78  \\
        & prNet‑Large                            & 1.46 & 32.24 & 32.28 & 30.77 & 0.90 & 0.90 & 0.89  \\
        & prNet‑Large‑Adversarial                        & 1.50 & 32.38 & 32.42 & 30.77 & 0.90 & 0.90 & 0.89  \\
        & prNet‑Large‑Adversarial (+TTA) &1.80& 32.66&32.69&31.25&0.91&0.91&0.89\\
        & prNet‑Large‑Adversarial (+TTA $D_4$) &3.12& 32.92&32.94&31.88&0.91&0.91&0.88\\

      \midrule
      \multirow{16}{*}{\rotatebox{90}{\shortstack{$\alpha=3$\\(Avg. SNR: 31.53dB)}}}
        & HIO~\cite{fienup1982comparison}            & 0.27 & 18.92 & 18.89 & 20.34 & 0.43 & 0.43 & 0.43 \\
        & prDeep~\cite{pmlr-v80-metzler18a}         & 59.41 & 22.06 & 22.09 & 20.91 & 0.59 & 0.59 & 0.54 \\
        & DIR~\cite{Isil:19}                        & 21.72 & 22.87 & 22.85 & 23.50 & 0.68 & 0.68 & 0.71 \\
        & PnP-HIO~\cite{Isil:20}                     & 24.35 & 23.92 & 23.92 & 23.98 & 0.70 & 0.70 & 0.69 \\
        & I2I-PR~\cite{kaya2026i2i} %($T=4,$ no aggregation)
        & 1.11 & 26.78 & 26.85 & 24.18 & 0.73 & 0.73 & 0.61\\
        & DiffFPR$^{\dagger}$~\cite{li2024difffpr}                   & 63.50 & 19.17 & 19.19 & 18.30 & 0.44 & 0.44 & 0.39 \\
        & DiffFPR$^{\ddagger}$~\cite{li2024difffpr}                   & 64.54 & 23.67 & 23.71 & 22.14 & 0.58 & 0.58 & 0.51  \\
        & DAPS$^{\dagger}$~\cite{zhang2025improving}                   & 94.66 & 18.72 & 18.74 & 18.01 & 0.43 & 0.43 & 0.46  \\
        & DAPS$^{\ddagger}$~\cite{zhang2025improving}                   & 347.33 & 25.58 & 25.68 & 21.69 & 0.72 & 0.73 & 0.59  \\
        & LatentDAPS$^{\dagger}$~\cite{zhang2025improving}                   & 85.72 & 17.41 & 17.40 & 17.98 & 0.35 & 0.35 & 0.42  \\
        & LatentDAPS$^{\ddagger}$~\cite{zhang2025improving}                   & 78.67 & 26.93 & 26.96 & 25.90 & 0.76 & 0.76 & 0.72   \vspace{0.3em}\\
        \cline{2-9}\\[-0.9em]
        & Initialization Stage                     & 0.47 & 19.73 & 19.68 & 21.65 & 0.46 & 0.46 & 0.45 \\
        & prNet‑Small                              & 1.03 & 28.08 & 28.13 & 25.93 & 0.80 & 0.81 & 0.70 \\
        & prNet‑Large                              & 1.48 & 30.17 & 30.24 & 27.79 & 0.86 & 0.86 & 0.78 \\
        & prNet‑Large‑Adversarial                          & 1.52 & 30.22 & 30.28 & 27.79 & 0.86 & 0.87 & 0.79 \\
        & prNet‑Large‑Adversarial (+TTA) &1.82& 30.52&30.57&27.88&0.87&0.87&0.79\\
        & prNet‑Large‑Adversarial (+TTA $D_4$) &3.13& 30.73&30.81&27.76&0.87&0.87&0.76\\

      \midrule
      \multirow{16}{*}{\rotatebox{90}{\shortstack{$\alpha=4$\\(Avg. SNR: 30.24dB)}}}
        & HIO~\cite{fienup1982comparison}           & 0.28 & 18.52 & 18.48 & 19.80 & 0.39 & 0.39 & 0.40  \\
        & prDeep~\cite{pmlr-v80-metzler18a}        & 59.68 & 20.69 & 20.70 & 20.38 & 0.53 & 0.53 & 0.51  \\
        & DIR~\cite{Isil:19}                       & 21.95 & 21.80 & 21.77 & 22.79 & 0.62 & 0.62 & 0.69  \\
        & PnP-HIO~\cite{Isil:20}                    & 24.43 & 22.41 & 22.39 & 23.09 & 0.63 & 0.63 & 0.65  \\
        & I2I-PR~\cite{kaya2026i2i} %($T=4,$ no aggregation)
        & 1.10 & 25.43 & 25.46 & 24.23 & 0.66 & 0.66 & 0.59\\
        & DiffFPR$^{\dagger}$~\cite{li2024difffpr}                   & 63.52 & 18.94 & 18.92 & 19.73 & 0.42 & 0.42 & 0.43  \\
        & DiffFPR$^{\ddagger}$~\cite{li2024difffpr}                   & 64.83 & 22.37 & 22.38 & 21.90 & 0.51 & 0.51 & 0.49  \\
        & DAPS$^{\dagger}$~\cite{zhang2025improving}                   & 94.48 & 18.90 & 18.91 & 18.42 & 0.44 & 0.44 & 0.49  \\
        & DAPS$^{\ddagger}$~\cite{zhang2025improving}                   & 348.57 & 24.93 & 25.04 & 20.65 & 0.71 & 0.71 & 0.59  \\
        & LatentDAPS$^{\dagger}$~\cite{zhang2025improving}                   & 85.94 & 17.44 & 17.43 & 17.87 & 0.35 & 0.35 & 0.41  \\
        & LatentDAPS$^{\ddagger}$~\cite{zhang2025improving}                   & 78.43 & 26.07 & 26.04 & 27.01 & 0.72 & 0.72 & 0.69   \vspace{0.3em}\\
        \cline{2-9}\\[-0.9em]
        & Initialization Stage                    & 0.48 & 19.12 & 19.08 & 20.63 & 0.41 & 0.41 & 0.41  \\
        & prNet‑Small                             & 1.02 & 26.69 & 26.75 & 24.48 & 0.76 & 0.76 & 0.67  \\
        & prNet‑Large                             & 1.48 & 28.29 & 28.35 & 25.98 & 0.81 & 0.81 & 0.72  \\
        & prNet‑Large‑Adversarial                         & 1.52 & 28.29 & 28.35 & 25.91 & 0.81 & 0.81 & 0.72  \\
        & prNet‑Large‑Adversarial (+TTA)&1.81 & 28.56&28.61&26.49&0.82&0.82&0.73\\
        & prNet‑Large‑Adversarial (+TTA $D_4$)&3.12& 28.74&28.80&26.50&0.83&0.83&0.73\\
      \bottomrule
    \end{tabu}
  \end{adjustbox}
  \label{table:prnettttresultsqua}
\end{table*}

% \setlength{\belowcaptionskip}{1pt}  % Reduces space between figure and caption
% \begin{figure*}[t!]
%     \centering
%     \begin{subfigure}[b]{0.18\textwidth}
%         \caption{Ground truth}
%         \includegraphics[width=\columnwidth]{figures/122/122_0_gt_0.000_0.000_31.605.pdf}
%     \end{subfigure}
%     \hfill
%     \begin{subfigure}[b]{0.18\textwidth}
%         \caption{prDeep,\\PSNR: $23.44$, SSIM: $0.66$}%0.46
%         \includegraphics[width=\columnwidth]{figures/cagatayao2019/209_3_1_prdeep}
%     \end{subfigure}
%     \hfill
%     \begin{subfigure}[b]{0.18\textwidth}
%         \caption{DIR,\\PSNR: $20.37$, SSIM: $0.54$}%0.40
%         \includegraphics[width=\columnwidth]{figures/cagatayao2019/209_3_1_unet2}
%     \end{subfigure}
%     \hfill
%     \begin{subfigure}[b]{0.18\textwidth}
%         \caption{prNet-Small,\\PSNR: $27.12$, SSIM: $0.88$}
%         \includegraphics[width=\columnwidth]{figures/122/122_0_small_main_loop_27.176_0.881_31.605.pdf}
%     \end{subfigure}
%     \hfill
%     \begin{subfigure}[b]{0.18\textwidth}
%         \caption{prNet-Large,\\PSNR: $30.87$, SSIM: $0.93$}
%         \includegraphics[width=\columnwidth]{figures/122/122_0_large_main_loop_adversarial_30.872_0.924_31.605.pdf}
%     \end{subfigure}
    
%     \caption{The outputs of various algorithms for the ``Cameraman'' test image subjected to $\alpha=3$ noise (SNR: 31.61 dB).}
%     \label{fig:prnettttcameraman}
% \end{figure*}

\subsection{Comparison with Other Methods}

The reconstructions of the developed approach are compared with the true images using the peak signal-to-noise ratio (PSNR) and structural similarity index (SSIM). To comparatively evaluate the performance, we consider several state-of-the-art baseline methods, including the classical HIO algorithm \cite{fienup1982comparison}, DIR—a deep-learning based refinement approach \cite{Isil:19}, plug-and play methods with deep priors such as prDeep \cite{pmlr-v80-metzler18a} and PnP-HIO \cite{Isil:20, isil2024deep}, and recent diffusion-based approaches including I2I-PR~\cite{kaya2026i2i}, DiffFPR~\cite{li2024difffpr}, DAPS~\cite{zhang2025improving}, and LatentDAPS~\cite{zhang2025improving}.

For the diffusion-based baselines DiffFPR, DAPS, and LatentDAPS, we report results for two variants: (i) with the hyperparameters provided in the original papers for noisy Fourier PR, and (ii) with the hyperparameters tuned on our validation set via Bayesian optimization for the case with $\alpha=3$ noise level. For hyperparameter optimization, we use the tree-structured Parzen estimator (TPE) sampler~\cite{NIPS2011_86e8f7ab}, with PSNR as the optimization objective. To ensure a comparable computational budget across methods, the optimization procedure is run for approximately five days for each baseline, matching the training time of our own pipeline. This allows us to additionally compare each method under similar resource budget.

\begin{figure}[]
    \centering

    % First row
    \subfloat[][\parbox{0.31\columnwidth}{\centering (a) Ground truth}\label{fig:gt}]{%
        \includegraphics[width=0.31\columnwidth]{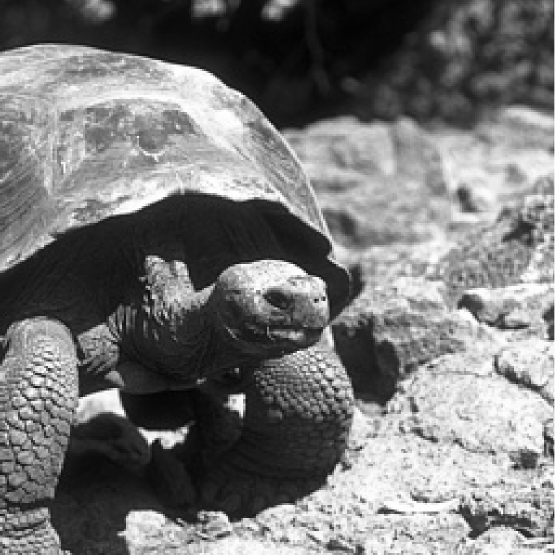}
    }
    \hfill
    \subfloat[][\parbox{0.31\columnwidth}{\centering (b) prDeep~\cite{pmlr-v80-metzler18a}\\PSNR:25.35,SSIM:0.71}\label{fig:prdeep}]{%
        \includegraphics[width=0.31\columnwidth]{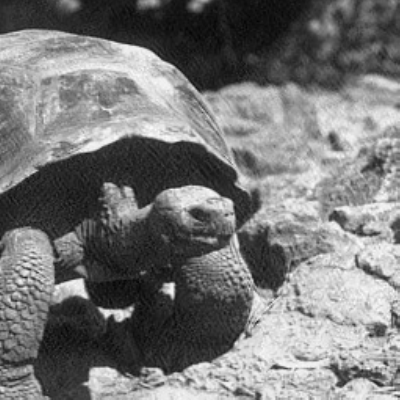}
    }
    \hfill
    \subfloat[][\parbox{0.31\columnwidth}{\centering (c) DIR~\cite{Isil:19}\\PSNR:26.49,SSIM:0.73}\label{fig:dir}]{%
        \includegraphics[width=0.31\columnwidth]{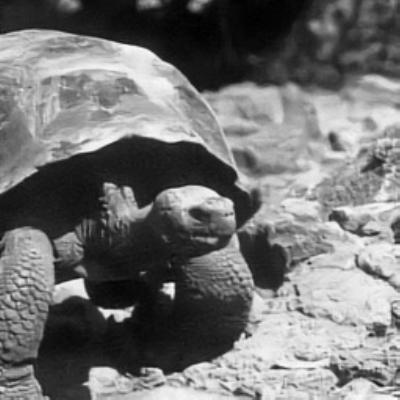}
    }

    % First row
    \subfloat[][\parbox{0.31\columnwidth}{\centering (d) DiffFPR$^{\ddagger}$~\cite{li2024difffpr}\\PSNR:25.06,SSIM:0.71}]{%
        \includegraphics[width=0.31\columnwidth]{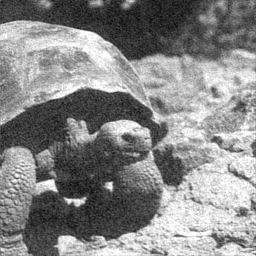}
    }
    \hfill
    \subfloat[][\parbox{0.31\columnwidth}{\centering (e) DAPS$^{\ddagger}$~\cite{zhang2025improving}\\PSNR:14.10,SSIM:0.70}\label{fig:prdeep}]{%
        \includegraphics[width=0.31\columnwidth]{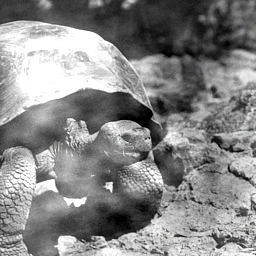}
    }
    \hfill
    \subfloat[][\parbox{0.31\columnwidth}{\centering (f) LatentDAPS$^{\ddagger}$~\cite{zhang2025improving}\\PSNR:16.73,SSIM:0.76}\label{fig:dir}]{%
        \includegraphics[width=0.31\columnwidth]{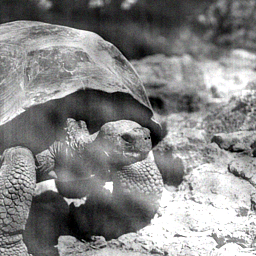}
    }
    
    % Second row
    \subfloat[][\parbox{0.31\columnwidth}{\centering (g) PnP-HIO~\cite{Isil:20}\\PSNR:27.87,SSIM:0.77}\label{fig:mbwddp}]{%
        \includegraphics[width=0.31\columnwidth]{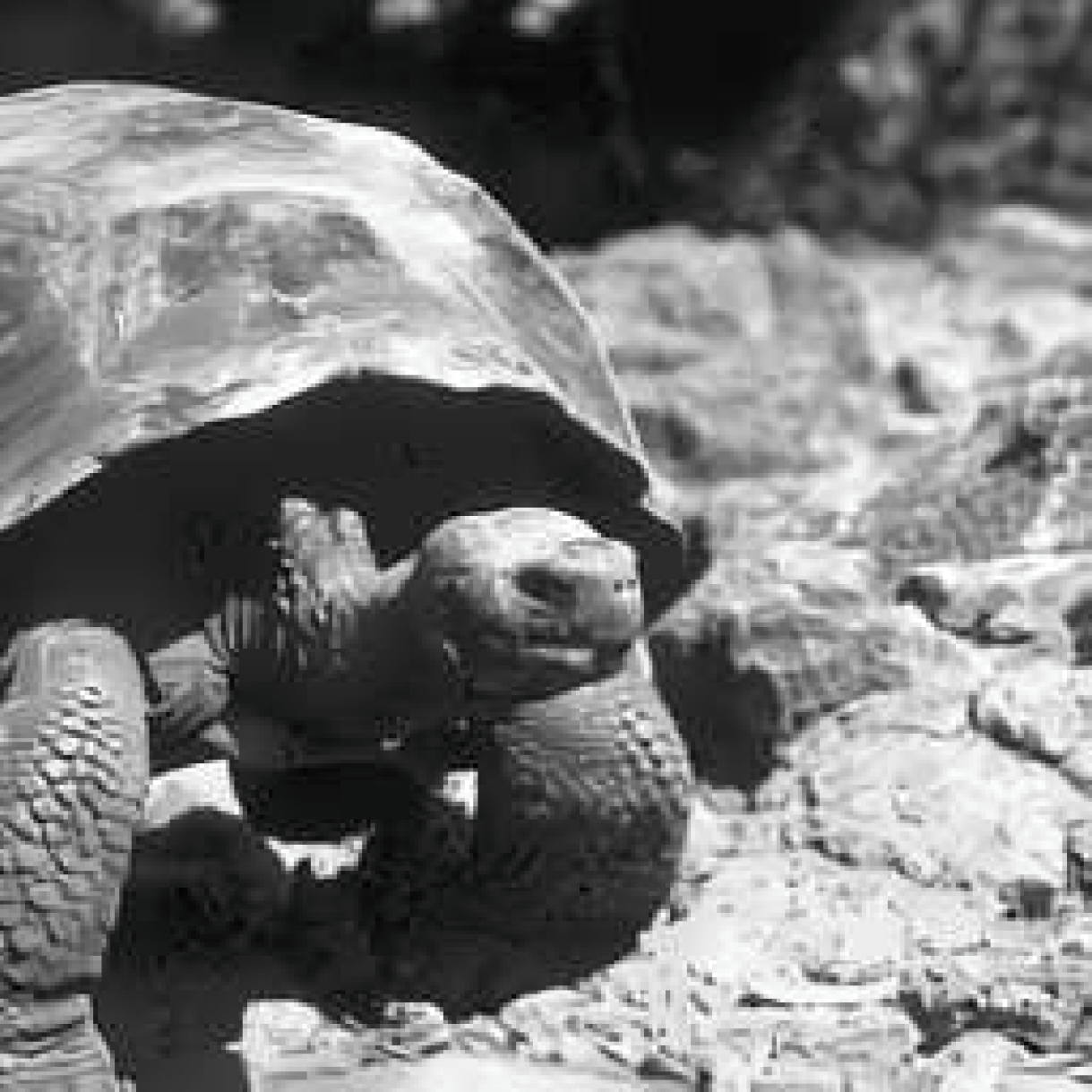}
    }
    \hfill
    \subfloat[][\parbox{0.31\columnwidth}{\centering (h) prNet-Small\\PSNR:29.74,SSIM:0.91}\label{fig:small}]{%
        \includegraphics[width=0.31\columnwidth]{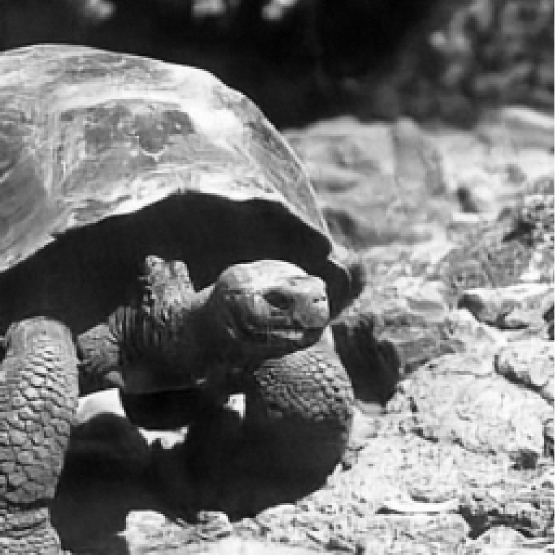}
    }
    \hfill
    \subfloat[][\parbox{0.31\columnwidth}{\centering (i) prNet-Large-Adversarial (+TTA)\\PSNR:31.31,SSIM:0.93}\label{fig:large}]{%
        \includegraphics[width=0.31\columnwidth]{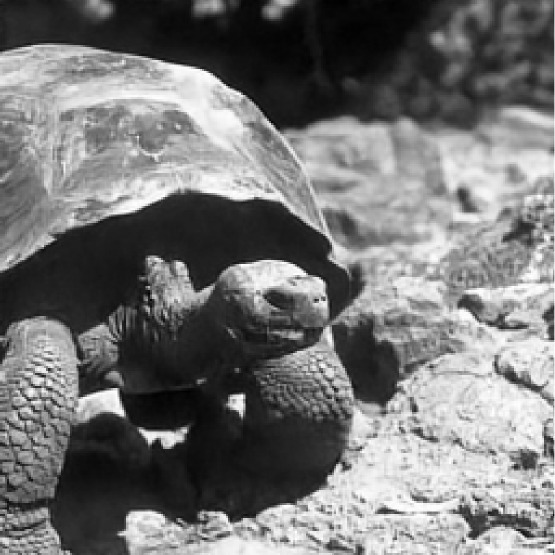}
    }

    \caption{The outputs of various algorithms for the ``Turtle" test image subjected to $\alpha=3$ noise (SNR=31.89dB).}
    \label{fig:prnettttturtle}
\end{figure}

\begin{figure}[]
    \centering

    % First row
    \subfloat[][\parbox{0.31\columnwidth}{\centering (a) Ground truth}]{%
        \begin{minipage}[t]{0.31\columnwidth}
            \begin{tikzpicture}
                \node[anchor=south west,inner sep=0] (image) at (0,0) {
                    \includegraphics[width=\linewidth]{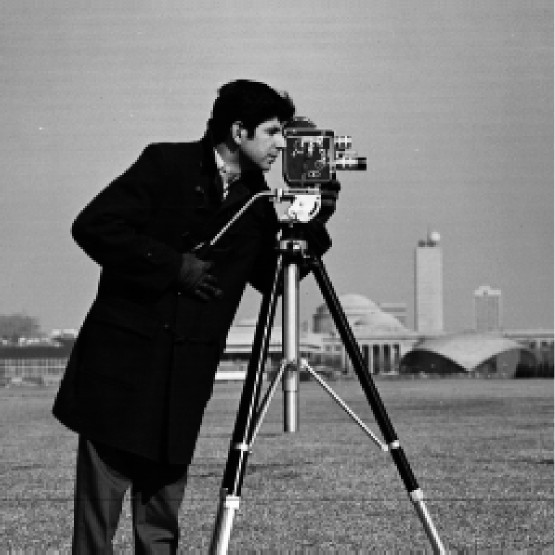}
                };
                \begin{scope}[x={(image.south east)}, y={(image.north west)}]
                    \draw[white, dashed, very thick] (0.35,0.56) rectangle (0.70,0.85);
                \end{scope}
            \end{tikzpicture}
            \vspace{2pt}
            \begin{adjustbox}{center,minipage=2.85\linewidth,trim={0.35\width, 1.1\height, 0.3\width, 0.3\height},clip}
                \includegraphics[width=\linewidth]{figures/122/122_0_gt_0.000_0.000_31.605.pdf}
            \end{adjustbox}
        \end{minipage}
    }
    \hfill
    \subfloat[][\parbox{0.31\columnwidth}{\centering (b) prDeep~\cite{pmlr-v80-metzler18a}\\PSNR:23.44,SSIM:0.66}]{%
        \begin{minipage}[t]{0.31\columnwidth}
            \includegraphics[width=\linewidth]{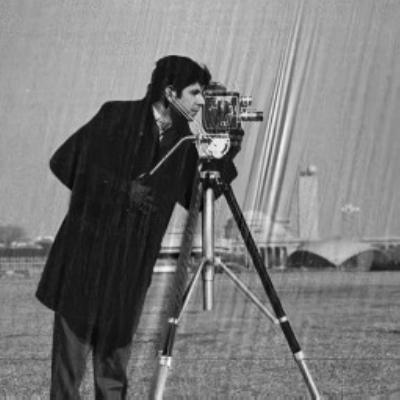}
            \vspace{2pt}
            \begin{adjustbox}{center,minipage=2.85\linewidth,trim={0.35\width, 1.1\height, 0.3\width, 0.3\height},clip}
                \includegraphics[width=\linewidth]{figures/cagatayao2019/209_3_1_prdeep}
            \end{adjustbox}
        \end{minipage}
    }
    \hfill
    \subfloat[][\parbox{0.31\columnwidth}{\centering (c) I2I-PR~\cite{kaya2026i2i} %DIR~\cite{Isil:19}
    \\PSNR:27.39,SSIM:0.70}]{%
        \begin{minipage}[t]{0.31\columnwidth}
            \includegraphics[width=\linewidth]{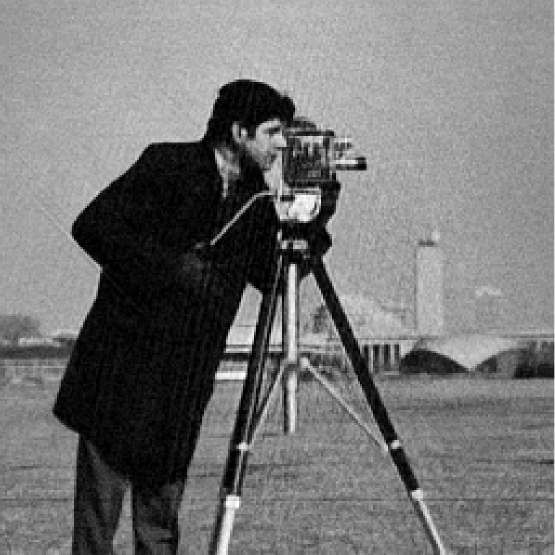}
            \vspace{2pt}
            \begin{adjustbox}{center,minipage=2.85\linewidth,trim={0.35\width, 1.1\height, 0.3\width, 0.3\height},clip}
                \includegraphics[width=\linewidth]{figures/indipr/122_1_output_27.388_0.700_31.900.pdf}
            \end{adjustbox}
        \end{minipage}
    }

    % Second row
    \subfloat[][\parbox{0.31\columnwidth}{\centering (d) DiffFPR$^{\ddagger}$~\cite{li2024difffpr}\\PSNR:21.44,SSIM:0.49}]{%
        \begin{minipage}[t]{0.31\columnwidth}
            \includegraphics[width=\linewidth]{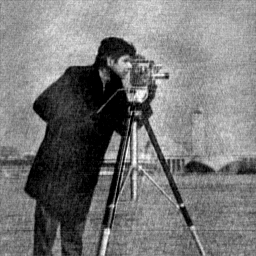}
            \vspace{2pt}
            \begin{adjustbox}{center,minipage=2.85\linewidth,trim={0.35\width, 1.1\height, 0.3\width, 0.3\height},clip}
                \includegraphics[width=\linewidth]{figures/new_diffbaselines_qualitative/cameraman/difffpr.png}
            \end{adjustbox}
        \end{minipage}
    }
    \hfill
    \subfloat[][\parbox{0.31\columnwidth}{\centering (e) DAPS$^{\ddagger}$~\cite{zhang2025improving}\\PSNR:13.49,SSIM:0.42}]{%
        \begin{minipage}[t]{0.31\columnwidth}
            \includegraphics[width=\linewidth]{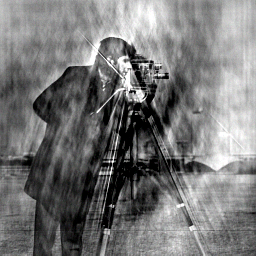}
            \vspace{2pt}
            \begin{adjustbox}{center,minipage=2.85\linewidth,trim={0.35\width, 1.1\height, 0.3\width, 0.3\height},clip}
                \includegraphics[width=\linewidth]{figures/new_diffbaselines_qualitative/cameraman/pixelDAPS.png}
            \end{adjustbox}
        \end{minipage}
    }
    \hfill
    \subfloat[][\parbox{0.31\columnwidth}{\centering (f) LatentDAPS$^{\ddagger}$~\cite{zhang2025improving}\\PSNR:12.78,SSIM:0.30}]{%
        \begin{minipage}[t]{0.31\columnwidth}
            \includegraphics[width=\linewidth]{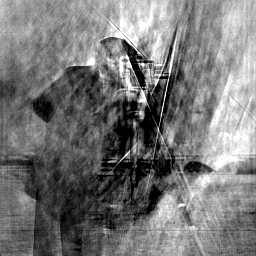}
            \vspace{2pt}
            \begin{adjustbox}{center,minipage=2.85\linewidth,trim={0.35\width, 1.1\height, 0.3\width, 0.3\height},clip}
                \includegraphics[width=\linewidth]{figures/new_diffbaselines_qualitative/cameraman/latentDAPS.png}
            \end{adjustbox}
        \end{minipage}
    }

    % Second row
    \subfloat[][\parbox{0.31\columnwidth}{\centering (g) prNet-Small\\PSNR:27.12,SSIM:0.88}]{%
        \begin{minipage}[t]{0.31\columnwidth}
            \includegraphics[width=\linewidth]{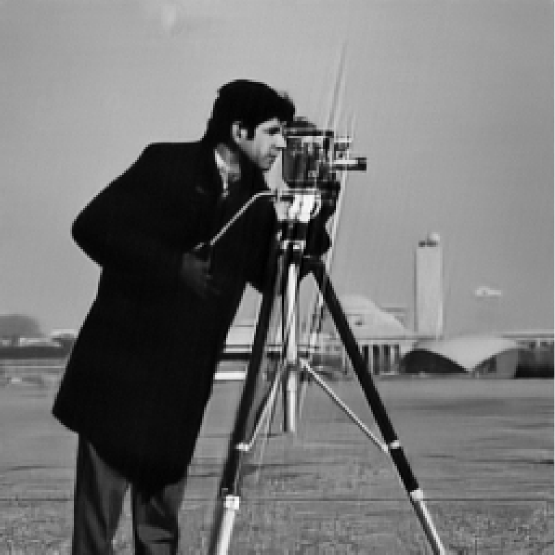}
            \vspace{2pt}
            \begin{adjustbox}{center,minipage=2.85\linewidth,trim={0.35\width, 1.1\height, 0.3\width, 0.3\height},clip}
                \includegraphics[width=\linewidth]{figures/122/122_0_small_main_loop_27.176_0.881_31.605.pdf}
            \end{adjustbox}
        \end{minipage}
    }
    \hfill
    \subfloat[][\parbox{0.31\columnwidth}{\centering (h) prNet-Large\\PSNR:30.87,SSIM:0.93}]{%
        \begin{minipage}[t]{0.31\columnwidth}
            \includegraphics[width=\linewidth]{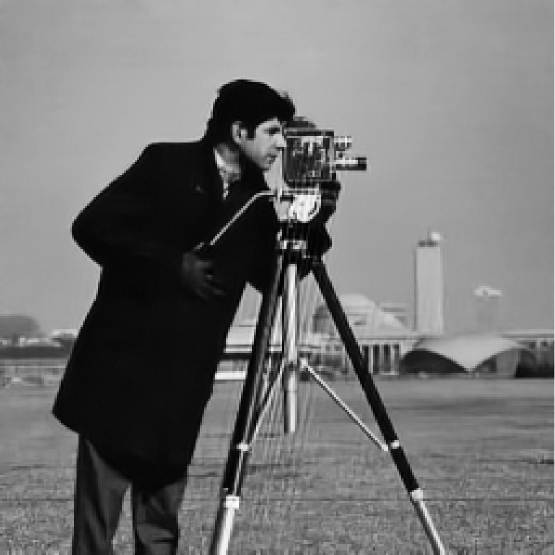}
            \vspace{2pt}
            \begin{adjustbox}{center,minipage=2.85\linewidth,trim={0.35\width, 1.1\height, 0.3\width, 0.3\height},clip}
                \includegraphics[width=\linewidth]{figures/122/122_0_large_main_loop_adversarial_30.872_0.924_31.605.pdf}
            \end{adjustbox}
        \end{minipage}
    }
    \hfill
    \subfloat[][\parbox{0.31\columnwidth}{\centering (i) prNet-Large-Adversarial (+TTA)\\PSNR:31.23,SSIM:0.93}]{%
        \begin{minipage}[t]{0.31\columnwidth}
            \includegraphics[width=\linewidth]{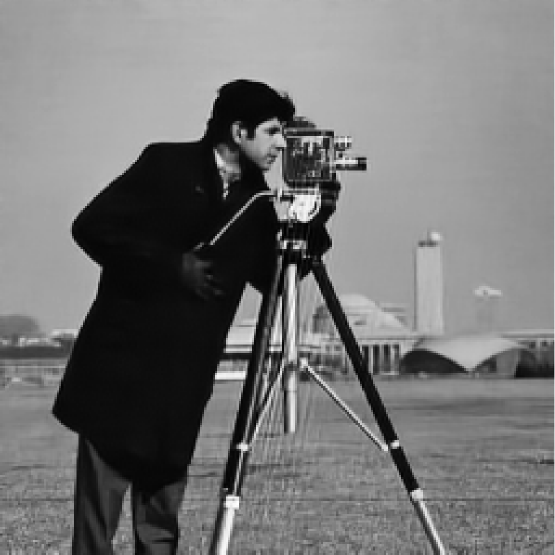}
            \vspace{2pt}
            \begin{adjustbox}{center,minipage=2.85\linewidth,trim={0.35\width, 1.1\height, 0.3\width, 0.3\height},clip}
                \includegraphics[width=\linewidth]{figures/122/122_0_large_main_loop_adversarial_tta_31.225_0.927_31.605.pdf}
            \end{adjustbox}
        \end{minipage}
    }

    \caption{The outputs of various algorithms for the ``Cameraman" test image subjected to $\alpha=3$ noise (SNR=31.61dB).}
    \label{fig:prnettttcameraman}
\end{figure}

Table \ref{table:prnettttresultsqua} presents the average reconstruction performance of different approaches for $236$ test images over $5$ Monte Carlo runs under varying levels of Poisson noise ($\alpha = 2, 3, 4$). The average SNR values are also reported in Table~\ref{table:prnettttresultsqua}, where SNR is calculated as $10 \log\left(\Vert |\mathbf{Fx}|^2 \Vert_2 / \Vert \mathbf{y}^2 - |\mathbf{Fx}|^2 \Vert_2\right)$. The developed methods consistently surpass other methods in both PSNR and SSIM metrics across all noise levels, while only necessitating a marginal increase in runtime compared to HIO and the initialization stage. 
The superiority of our methods can also be seen visually in the sample reconstructions presented in Figs. \ref{fig:prnettttturtle} and \ref{fig:prnettttcameraman}.

With our methods, reconstruction artifacts can be effectively removed while preserving fine image details. In contrast to the other methods, our approach generally avoids introducing artifacts and mitigates the over-smoothing commonly associated with distortion-based optimization~\cite{Isil:19}, thanks to the incorporation of adversarial loss. As a result, it provides higher reconstruction quality while minimizing distortions and enhancing the perceptual quality. Moreover, our methods add only a marginal runtime overhead on top of the initialization stage while achieving substantially higher reconstruction quality than the learning-based baselines with comparable or lower runtime. 
%Moreover, our methods add only a marginal runtime overhead on top of the initialization stage, achieving substantially higher reconstruction quality than the baselines at a fraction of their runtime.

% Moreover, our methods exhibit computational efficiency comparable to the initialization stage, demonstrating superior reconstruction quality and computational efficiency.

\begin{figure}[]
    \centering
    \subfloat[][\parbox{0.31\columnwidth}{\centering (a) Ground truth}]{%
        \includegraphics[width=0.31\columnwidth]{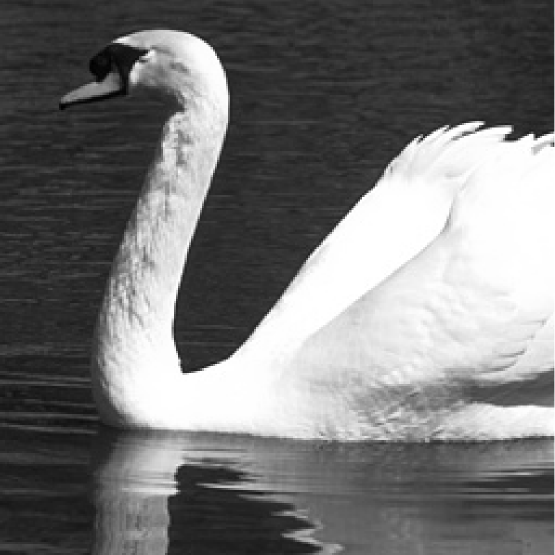}
    }
    \hfill
    \subfloat[][\parbox{0.31\columnwidth}{\centering (b) Initialization stage\\PSNR:14.28,SSIM:0.39}]{%
        \includegraphics[width=0.31\columnwidth]{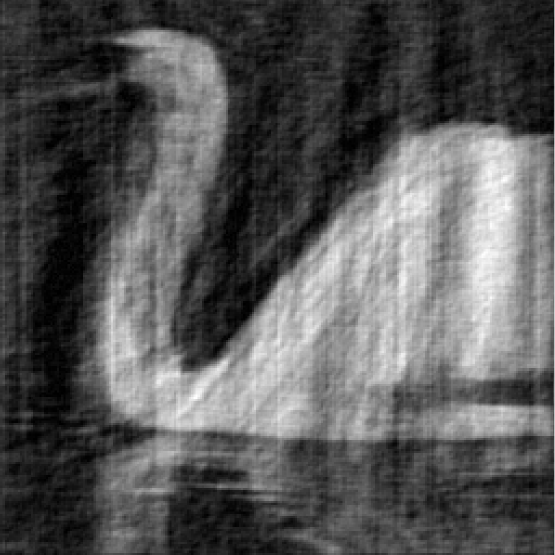}
    }
    \hfill
    \subfloat[][\parbox{0.31\columnwidth}{\centering (c) prNet-Small\\PSNR:24.97,SSIM:0.84}]{%
        \includegraphics[width=0.31\columnwidth]{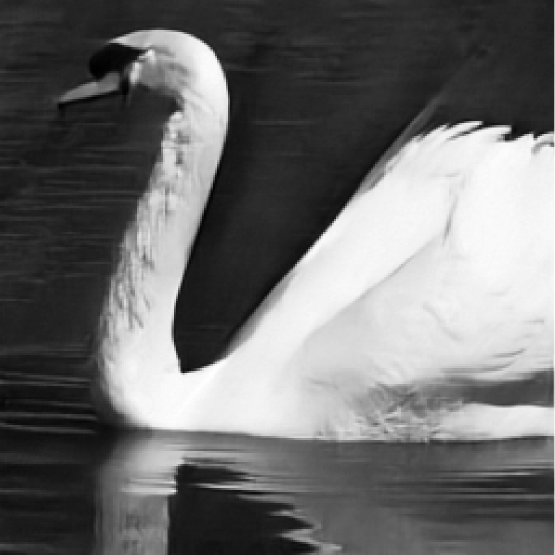}
    }

    \subfloat[][\parbox{0.31\columnwidth}{\centering (d) Main loop-Large\\PSNR:34.48,SSIM:0.93}]{%
        \includegraphics[width=0.31\columnwidth]{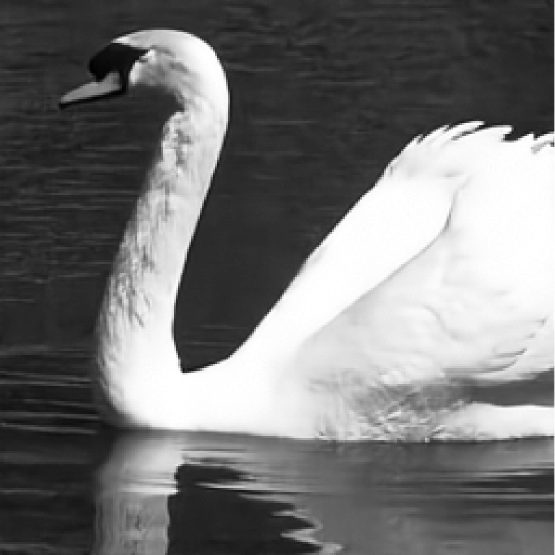}
    }
    \hfill
    \subfloat[][\parbox{0.31\columnwidth}{\centering (e) prNet-Large\\PSNR:35.02,SSIM:0.93}]{%
        \includegraphics[width=0.31\columnwidth]{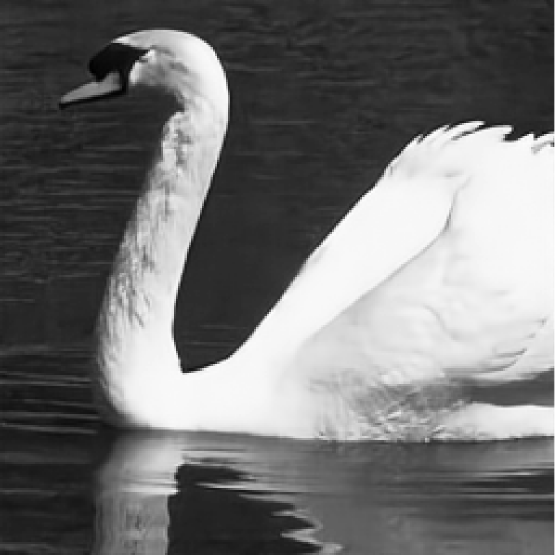}
    }
    \hfill
    \subfloat[][\parbox{0.31\columnwidth}{\centering (f) prNet-Large-Adversarial (+TTA)\\PSNR:35.21,SSIM:0.93}]{%
        \includegraphics[width=0.31\columnwidth]{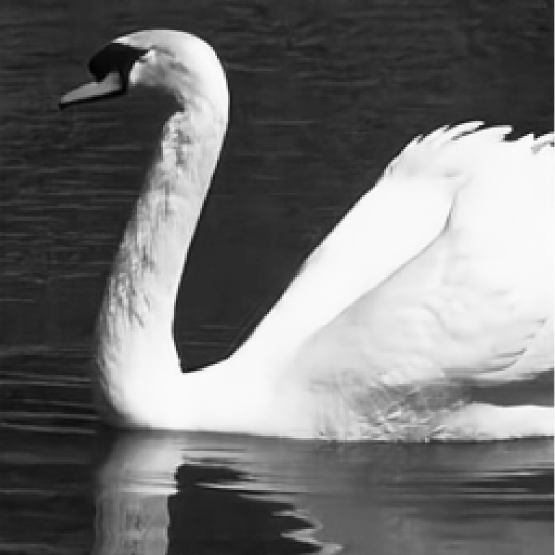}
    }
    \caption{Intermediate reconstruction results for the ``Swan" test image at a noise level of $\alpha=3$ (SNR=31.69dB).}

    \label{fig:prnettttintermediate}
\end{figure}

Intermediate reconstructions for a natural image in the test dataset are shown in Fig. \ref{fig:prnettttintermediate} to illustrate the effect of each stage of the proposed pipeline. Additional results and error maps are also provided in the supplementary document.
As seen, the initialization stage produces a coarse estimate that already captures the global structure but is corrupted by reconstruction artifacts. % and residual noise. 
The main loop of prNet-Large progressively refines this estimate, removing artifacts while recovering fine details that are absent in the initial coarse estimate. Averaging of the multiple reconstruction samples in prNet-Large yields a further small improvement in distortion metrics, and the final adversarially trained denoiser in prNet-Large-Adversarial sharpens the result with further texture-level corrections. In fact, the residual error in textured and edge regions shrinks from prNet-Large to the prNet-Large-Adversarial (+TTA) variant. This progress illustrates how each stage of the pipeline contributes complementary refinements rather than redundant updates.

For the diffusion-based baselines DiffFPR, DAPS, and LatentDAPS, the two reported variants reveal a notable property of these methods. Although they are presented as PnP approaches that pair a pretrained diffusion prior with a generic data-consistency step, the configurations released by the authors transfer poorly to the setting considered here, yielding PSNR/SSIM values similar to the classical HIO baseline. Reaching satisfactory performance requires a substantial per-task tuning effort, after which the same methods become competitive with our distortion-oriented baselines. In this sense, the practical use of these methods on a new forward operator is closer to a problem-specific tuning procedure than to a true plug-and-play deployment. We therefore report both variants to provide their performance both before and after task-specific tuning.

A natural counterpart to the discussion above concerns the hyperparameters of our own pipeline. Our method involves a number of manually set hyperparameters spanning initialization ($m$, $s$, $n$, $k$), iteration ($T$, $K$, $\beta$), training (progressive schedule, $\mu_\mathrm{adv}$), and test time augmentation. The reported values reflect choices that worked well in our preliminary experiments without optimizing over a fine grid. Unlike the diffusion-based baselines above, where Bayesian optimization can be applied directly because each evaluation involves only inference with a pretrained model, a comparable sweep is impractical here: each setting of the iteration- and training-related hyperparameters requires retraining the full pipeline at a cost of four to five days on a single A100, making a systematic search infeasible within our available computational resources.

For practical guidance, we found the initialization parameters ($m$, $s$, $n$, $k$) the easiest to tune since they affect only the warm-start cost and %can be adjusted post hoc; 
can be adjusted without retraining the whole network.
Larger $m$ and $n$ trade runtime for a more reliable starting point, while $k$ controls the diversity available to prNet-Large. The iteration count $T$ and inner HIO iteration number $K$ trade runtime for fidelity at fixed model capacity, with the values reported here lying near the knee of this tradeoff. The adversarial weight $\mu_\mathrm{adv}$ controls the perception--distortion balance and should be tuned to the user's downstream priority. 
Even with these default hyperparameter settings, our method substantially outperforms all baselines, including the PnP-based diffusion baselines under their own tuned configurations, demonstrating robust performance without requiring extensive hyperparameter optimization.
%Even with these defaults, our method substantially outperforms all baselines, including the diffusion baselines under their own tuned configurations, suggesting that our setting is not unduly sensitive to fine-grained tuning.

\subsection{Generalization Capability and Robustness}

\begin{figure}[th]
    \centering

    % First row
    \subfloat[][\parbox{0.31\columnwidth}{\centering (a) Ground truth}]{%
        \begin{minipage}[t]{0.31\columnwidth}
            \begin{tikzpicture}
                \node[anchor=south west,inner sep=0] (image) at (0,0) {
                    \includegraphics[width=\linewidth]{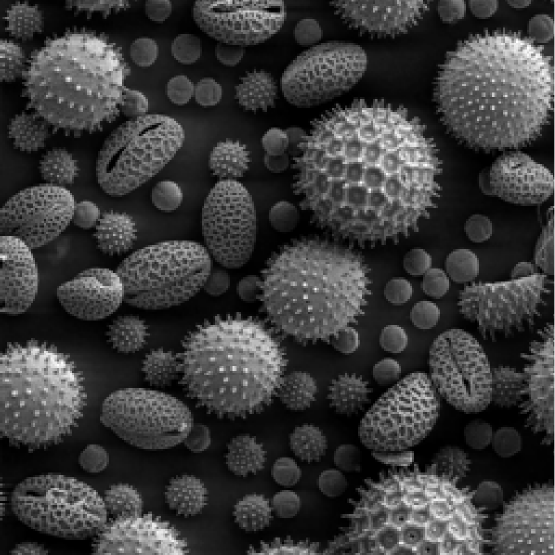}
                };
                \begin{scope}[x={(image.south east)}, y={(image.north west)}]
                    \draw[white, dashed, very thick] (0.35,0.35) rectangle (0.65,0.64);
                \end{scope}
            \end{tikzpicture}
            \vspace{2pt}
            \begin{adjustbox}{center,minipage=3.33\linewidth,trim={.35\width,.35\height,.25\width,.25\height},clip}
                \includegraphics[width=\linewidth]{figures/117/117_0_gt_0.000_0.000_28.102.pdf}
            \end{adjustbox}
        \end{minipage}
    }
    \hfill
    \subfloat[][\parbox{0.31\columnwidth}{\centering (b) prDeep~\cite{pmlr-v80-metzler18a}\\PSNR:19.37,SSIM:0.47}]{%
        \begin{minipage}[t]{0.31\columnwidth}
            \includegraphics[width=\linewidth]{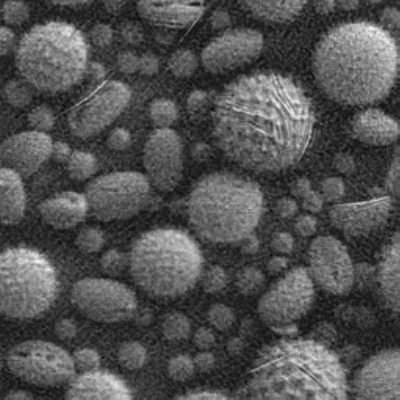}
            \vspace{2pt}
            \begin{adjustbox}{center,minipage=3.33\linewidth,trim={.35\width,.35\height,.25\width,.25\height},clip}
                \includegraphics[width=\linewidth]{figures/cagatayao2019/204_3_1_prdeep}
            \end{adjustbox}
        \end{minipage}
    }
    \hfill
    \subfloat[][\parbox{0.31\columnwidth}{\centering (c) I2I-PR~\cite{kaya2026i2i} %DIR~\cite{Isil:19}
    \\PSNR:28.95,SSIM:0.86}]{%
        \begin{minipage}[t]{0.31\columnwidth}
            \includegraphics[width=\linewidth]{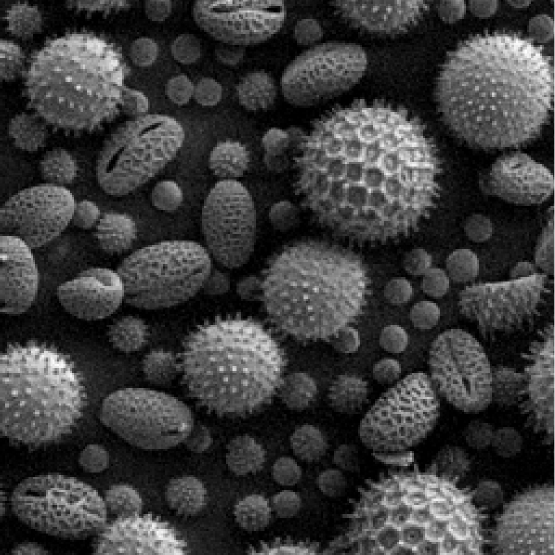}
            \vspace{2pt}
            \begin{adjustbox}{center,minipage=3.33\linewidth,trim={.35\width,.35\height,.25\width,.25\height},clip}
                \includegraphics[width=\linewidth]{figures/indipr/117_0_output_28.954_0.861_27.692.pdf}
            \end{adjustbox}
        \end{minipage}
    }

    % Second row
    \subfloat[][\parbox{0.31\columnwidth}{\centering (d) DiffFPR$^{\ddagger}$~\cite{li2024difffpr}\\PSNR:24.92,SSIM:0.73}]{%
        \begin{minipage}[t]{0.31\columnwidth}
            \includegraphics[width=\linewidth]{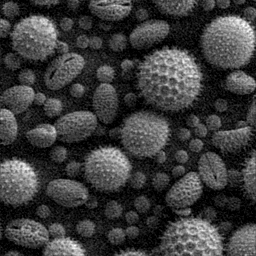}
            \vspace{2pt}
            \begin{adjustbox}{center,minipage=3.33\linewidth,trim={.35\width,.35\height,.25\width,.25\height},clip}
                \includegraphics[width=\linewidth]{figures/new_diffbaselines_qualitative/pollen/difffpr.png}
            \end{adjustbox}
        \end{minipage}
    }
    \hfill
    \subfloat[][\parbox{0.31\columnwidth}{\centering (e) DAPS$^{\ddagger}$~\cite{zhang2025improving}\\PSNR:24.10,SSIM:0.85}]{%
        \begin{minipage}[t]{0.31\columnwidth}
            \includegraphics[width=\linewidth]{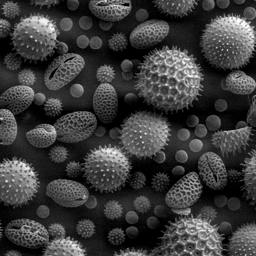}
            \vspace{2pt}
            \begin{adjustbox}{center,minipage=3.33\linewidth,trim={.35\width,.35\height,.25\width,.25\height},clip}
                \includegraphics[width=\linewidth]{figures/new_diffbaselines_qualitative/pollen/pixelDAPS.png}
            \end{adjustbox}
        \end{minipage}
    }
    \hfill
    \subfloat[][\parbox{0.31\columnwidth}{\centering (f) LatentDAPS$^{\ddagger}$~\cite{zhang2025improving}\\PSNR:21.53,SSIM:0.80}]{%
        \begin{minipage}[t]{0.31\columnwidth}
            \includegraphics[width=\linewidth]{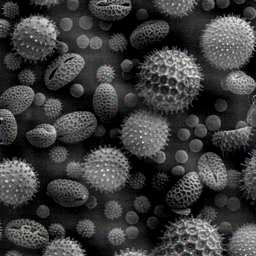}
            \vspace{2pt}
            \begin{adjustbox}{center,minipage=3.33\linewidth,trim={.35\width,.35\height,.25\width,.25\height},clip}
                \includegraphics[width=\linewidth]{figures/new_diffbaselines_qualitative/pollen/latentDAPS.png}
            \end{adjustbox}
        \end{minipage}
    }

    % Second row
    \subfloat[][\parbox{0.31\columnwidth}{\centering (g) PnP-HIO~\cite{Isil:20}\\PSNR:26.28,SSIM:0.71}]{%
        \begin{minipage}[t]{0.31\columnwidth}
            \includegraphics[width=\linewidth]{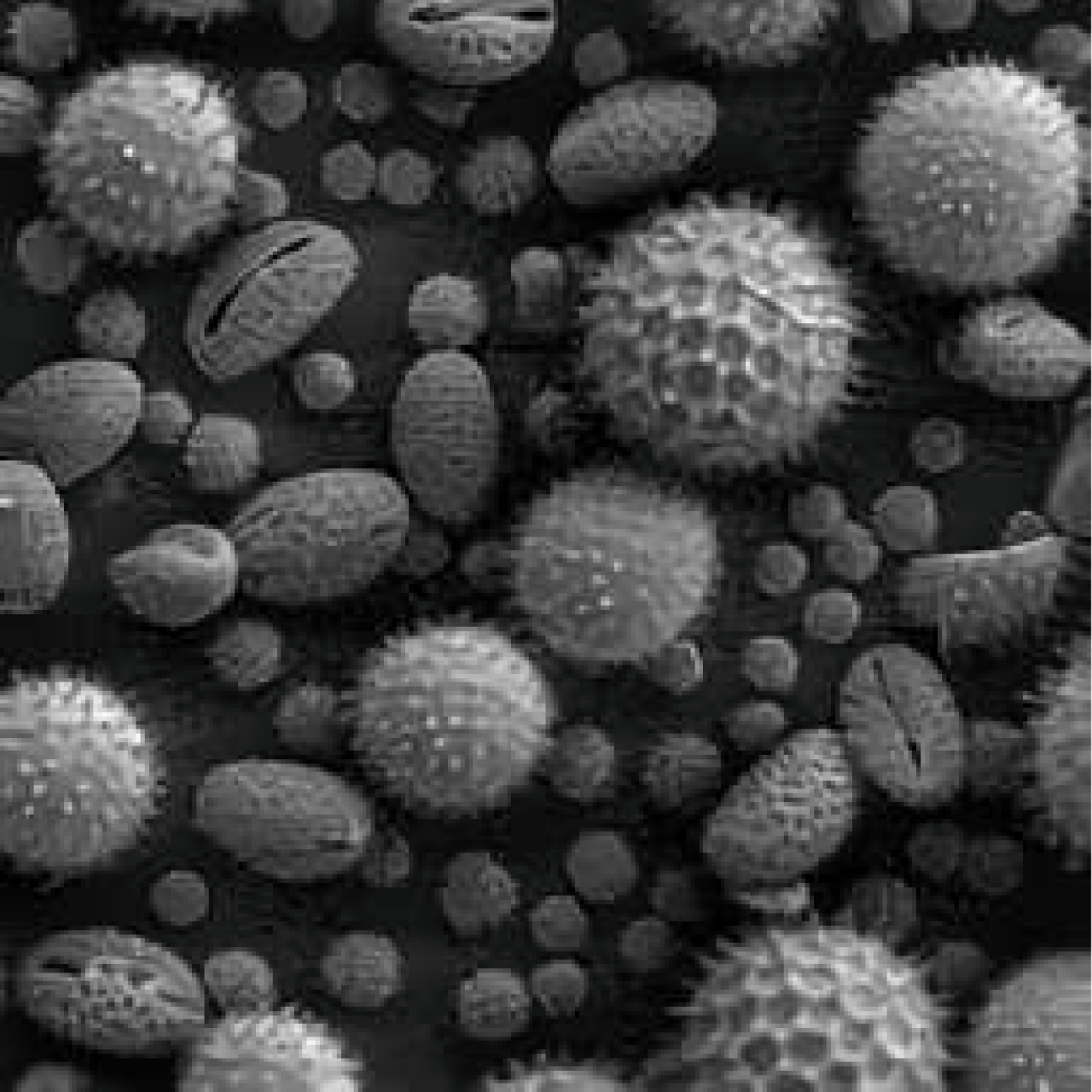}
            \vspace{2pt}
            \begin{adjustbox}{center,minipage=3.33\linewidth,trim={.35\width,.35\height,.25\width,.25\height},clip}
                \includegraphics[width=\linewidth]{figures/117/mbwddp.pdf}
            \end{adjustbox}
        \end{minipage}
    }
    \hfill
    \subfloat[][\parbox{0.31\columnwidth}{\centering (h) prNet-Small\\PSNR:28.67,SSIM:0.89}]{%
        \begin{minipage}[t]{0.31\columnwidth}
            \includegraphics[width=\linewidth]{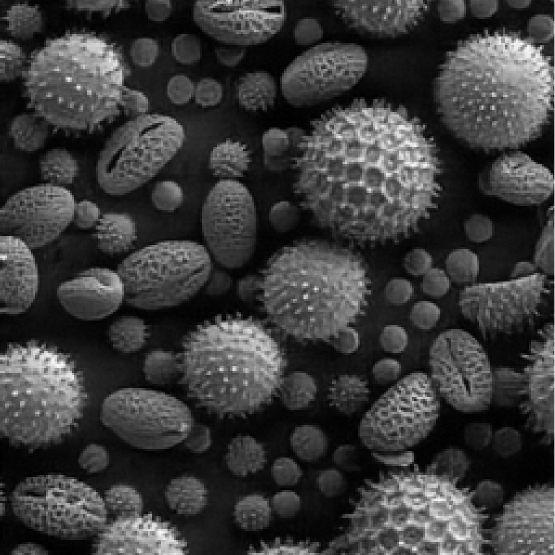}
            \vspace{2pt}
            \begin{adjustbox}{center,minipage=3.33\linewidth,trim={.35\width,.35\height,.25\width,.25\height},clip}
                \includegraphics[width=\linewidth]{figures/117/117_0_small_main_loop_28.671_0.888_28.102.pdf}
            \end{adjustbox}
        \end{minipage}
    }
    \hfill
    \subfloat[][\parbox{0.31\columnwidth}{\centering (i) prNet-Large-Adversarial (+TTA)\\PSNR:30.88,SSIM:0.92}]{%
        \begin{minipage}[t]{0.31\columnwidth}
            \includegraphics[width=\linewidth]{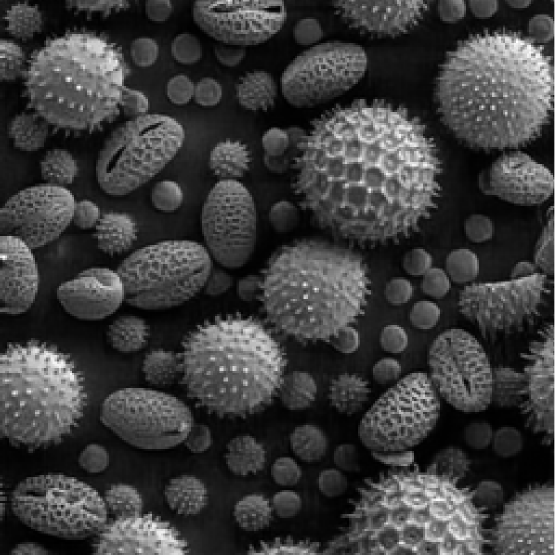}
            \vspace{2pt}
            \begin{adjustbox}{center,minipage=3.33\linewidth,trim={.35\width,.35\height,.25\width,.25\height},clip}
                \includegraphics[width=\linewidth]{figures/117/117_0_large_main_loop_adversarial_tta_30.882_0.928_28.102.pdf}
            \end{adjustbox}
        \end{minipage}
    }

    \caption{The outputs of various algorithms for the out-of-domain ``Pollen" test image subjected to $\alpha=3$ noise (SNR=28.10dB).}
    \label{fig:prnettttpollen}
\end{figure}

To evaluate the generalization capability of different algorithms, the results for both natural and unnatural test images are presented separately in Table \ref{table:prnettttresultsqua}. Although the pipelines are trained solely with natural images, the developed methods achieve superior reconstruction performance for both natural and unnatural images at the trained noise level $\alpha=3$, despite the distinct statistical properties of the latter. To evaluate the reconstruction performance visually, example reconstructions for an unnatural image from the test dataset are given in Fig. \ref{fig:prnettttpollen}. Notably, the performance of the compared methods such as prDeep declines significantly for unnatural image dataset, which is anticipated since the reconstruction depends on a regularization prior learned solely from natural images.

Table \ref{table:prnettttresultsqua} also reveals that the developed approaches outperform other methods across various noise levels ($\alpha = 2, 4$) in terms of reconstruction quality, despite being trained only for a specific noise level ($\alpha = 3$). This indicates the robustness of our methods to different noise conditions.

Phase retrieval algorithms are generally sensitive to initialization due to the inherent nonlinearity of the problem. To demonstrate the robustness of the developed approach to different initializations and image characteristics, PSNR and SSIM histograms are provided in Fig.~\ref{fig:prnetttthistograms} for the developed methods (with $\alpha=3$). These histograms include reconstructions obtained from $236$ distinct test images and $5$ Monte Carlo runs, implying that $5$ different initializations were used for each test image. The small spreads and high mean values clearly indicate the robustness of the developed approaches to varying initializations and image statistics.

\subsection{Limitations}

Since the measurement model assumes real-valued, non-negative objects, the global phase ambiguity does not arise in our setting. Additionally, to resolve conjugate inversion ambiguity during evaluation, we compare each reconstruction and its flipped version with the ground truth to ensure correct orientation. Nonetheless, spatial circular shift ambiguity remains a challenge. While natural images tend to exhibit balanced intensity distributions within the known support, which helps mitigate such ambiguities, this issue is largely underexplored in prior phase retrieval literature \cite{Goy2018LowPC, Uelwer2019PhaseRU}. Unlike some compared methods that resolve this ambiguity using ground-truth alignment, we deliberately refrain from such an approach to preserve a more realistic evaluation setting. Notably, the HIO-based initialization we employ is inherently robust to this ambiguity. However, certain unnatural test images do not fully occupy the known support, occasionally resulting in multiple plausible reconstructions from identical measurements. While refinement techniques such as the shrinkwrap method \cite{marcheshrinkwrap} could potentially resolve this, we chose not to incorporate them, as our focus remains on natural image reconstruction scenarios.

Moreover, we derive our data-consistency update under a Gaussian noise model, which is a standard approximation to Poisson photon-counting noise that is valid in the moderate-to-high count regimes typical of many phase retrieval setups. In photon-starved scenarios such as low-dose X-ray imaging or single-photon microscopy, this approximation breaks down, as the Gaussian model no longer accurately represents the underlying Poisson statistics~\cite{fannjiang2020numerics}. In principle, the framework can be extended to such regimes because the learned posterior-sampling structure of our pipeline is agnostic to the specific noise model, and adaptation requires only replacing the data-consistency operator with one matched to the exact Poisson likelihood, followed by retraining. Investigating this extension in low-photon imaging applications is an interesting direction for future work.

Additionally, perceptual quality metrics, which are commonly employed to assess image quality from a human perceptual perspective, are not reported in this work.  While such metrics are valuable for evaluating reconstructions intended for human interpretation, they often rely on deep learning models trained on natural RGB images. Since our focus is grayscale phase retrieval, for which the suitability of existing perceptual metrics remains unclear, we primarily rely on established distortion metrics to quantify reconstruction performance. 

Finally, our reported pipelines are trained at a single noise level ($\alpha = 3$) and rely on cross-noise-level generalization to handle mismatched test conditions. A straightforward extension is to sample $\alpha$ from a distribution at training time and condition the denoiser on its value, mirroring how diffusion models condition on the time index. We leave a systematic empirical study of this extension to future work.

%%%%%%%%%%%%%%%%%%%%%%%%%%%%%%%%%%%%%%%%%%%%%%%%%%%%%%%%%%%%%%%%%%%%%%%%%%%%%%%%%%%%%%%%%%%%%%%%%%
\section{Conclusion}
\label{sec:conclusionprnet}

%This paper introduces a novel approach to phase retrieval using Langevin dynamics for posterior sampling. We propose two architectures: prNet-Small for efficiency and prNet-Large for robustness via multiple initializations. Both models refine initial HIO estimates through a denoising-data consistency loop, trained using outputs from previous iterations, similar to algorithm unrolling \cite{Aggarwal2017MoDLMD}. prNet-Large-Adversarial incorporates a second denoiser with a Wasserstein loss to enhance perceptual quality. Extensive experiments demonstrate that our methods consistently outperform both classical and modern baselines, while maintaining low computational runtime. Our results suggest that the combination of denoisers with model-based methods in the Langevin dynamics framework shows promise for developing reliable stochastic solvers for nonlinear inverse problems.

This paper introduces a novel approach to phase retrieval using Langevin dynamics for posterior sampling. We propose three variants: prNet-Small for efficiency, prNet-Large for improved fidelity via parallel posterior samples, and prNet-Large-Adversarial, which improves prNet-Large with a learned aggregation denoiser. All three refine the initial HIO estimates through a denoising/data-consistency loop, trained on the outputs of previous iterations in a manner reminiscent of deep unrolling \cite{Monga2019AlgorithmUI}. Extensive experiments demonstrate that our methods substantially outperform both classical and modern baselines while maintaining low computational runtime. These results suggest that combining learned denoisers with model-based updates in a Langevin dynamics framework is a promising direction for building reliable stochastic solvers for nonlinear inverse problems.
%% SuppMat Citations:
\nocite{casella2024statistical, song2023pseudoinverse}

%%%%%%%%%%%%%%%%%%%%%%%%%%%%%%%%%%%%%%%%%%%%%%%%%%%%%%%%%%%%%%%%%%%%%%%%%%%%%%%%%%%%%%%%%%%%%%%%%%

\section*{Acknowledgments}
This work was supported in part by Scientific and Technological Research Council of Turkey (TUBITAK) under the Grant Number 120E505 and by the BAGEP Award of the Science Academy.
Figen S. Oktem thanks TUBITAK and the Science Academy for the support.
% \author da da \thanks var?
% This should be a simple paragraph before the References to thank those individuals and institutions who have supported your work on this article.

% {\appendix[Proof of the Zonklar Equations]
% Use $\backslash${\tt{appendix}} if you have a single appendix:
% Do not use $\backslash${\tt{section}} anymore after $\backslash${\tt{appendix}}, only $\backslash${\tt{section*}}.
% If you have multiple appendixes use $\backslash${\tt{appendices}} then use $\backslash${\tt{section}} to start each appendix.
% You must declare a $\backslash${\tt{section}} before using any $\backslash${\tt{subsection}} or using $\backslash${\tt{label}} ($\backslash${\tt{appendices}} by itself
%  starts a section numbered zero.)}


\bibliographystyle{IEEEtran}
\bibliography{ieeetip_bibliography}

@Article{kaya2025ddrm,
  author  = {Kaya, Mehmet Onurcan and Oktem, Figen S.},
  title   = {{DDRM-PR}: Fourier phase retrieval using denoising diffusion restoration models},
  number  = {5},
  pages   = {A95--A105},
  volume  = {64},
  journal = {Applied Optics},
  year    = {2025},
}

@Article{delbracio2023inversion,
  author  = {Delbracio, Mauricio and Milanfar, Peyman},
  title   = {Inversion by direct iteration: An alternative to denoising diffusion for image restoration},
  journal = {TMLR},
  year    = {2023},
}

@Article{Isil:19,
  author   = {{\c{C}}a{\u{g}}atay I{\c{s}}{\i}l and Figen S. Oktem and Aykut Ko{\c{c}}},
  title    = {Deep iterative reconstruction for phase retrieval},
  number   = {20},
  pages    = {5422--5431},
  volume   = {58},
  journal  = {Applied Optics},
  year     = {2019},
}

@Article{shechtman2015phase,
  author  = {Shechtman, Yoav and Eldar, Yonina C and Cohen, Oren and Chapman, Henry Nicholas and Miao, Jianwei and Segev, Mordechai},
  title   = {Phase retrieval with application to optical imaging: a contemporary overview},
  number  = {3},
  pages   = {87--109},
  volume  = {32},
  journal = {IEEE Signal Processing Magazine},
  year    = {2015},
}

@Article{chan2024tutorial,
  author  = {Chan, Stanley},
  title={Tutorial on diffusion models for imaging and vision},
  journal={Foundations and Trends in Computer Graphics and Vision},
  volume={16},
  number={4},
  pages={322--471},
  year={2024},
}

@article{netrapalli2013phase,
  title={Phase retrieval using alternating minimization},
  author={Netrapalli, Praneeth and Jain, Prateek and Sanghavi, Sujay},
  journal={Advances in Neural Information Processing Systems},
  volume={26},
  year={2013}
}

@inproceedings{chandra2019phasepack,
  title={Phasepack: A phase retrieval library},
  author={Chandra, Rohan and Goldstein, Tom and Studer, Christoph},
  booktitle={2019 13th International conference on Sampling Theory and Applications (SampTA)},
  pages={1--5},
  year={2019},
  organization={IEEE}
}

@Article{Wang_2024,
  author  = {Wang, Kaiqiang and Song, Li and Wang, Chutian and Ren, Zhenbo and Zhao, Guangyuan and Dou, Jiazhen and Di, Jianglei and Barbastathis, George and Zhou, Renjie and Zhao, Jianlin and Lam, Edmund Y.},
  title   = {On the use of deep learning for phase recovery},
  number  = {1},
  volume  = {13},
  journal = {Light: Science \& Applications},
  year    = {2024},
}

@Article{Nishizaki2020AnalysisON,
  author  = {Yohei Nishizaki and Ryoichi Horisaki and Katsuhisa Kitaguchi and Mamoru Saito and Jun Tanida},
  title   = {Analysis of non-iterative phase retrieval based on machine learning},
  pages   = {136 - 141},
  volume  = {27},
  journal = {Optical Review},
  year    = {2020},
}

@article{li2023diffusion,
author = {Li, Xin and Ren, Yulin and Jin, Xin and Lan, Cuiling and Wang, Xingrui and Zeng, Wenjun and Wang, Xinchao and Chen, Zhibo},
title = {Diffusion Models for Image Restoration and Enhancement: A Comprehensive Survey},
year = {2025},
volume = {133},
number = {11},
journal = {Int. J. Comput. Vision},
pages = {8078--8108},
}

@InProceedings{whang2022deblurring,
  author    = {Whang, Jay and Delbracio, Mauricio and Talebi, Hossein and Saharia, Chitwan and Dimakis, Alexandros G. and Milanfar, Peyman},
  booktitle = {CVPR},
  title     = {Deblurring via stochastic refinement},
  year      = {2022},
}

@inproceedings{
song2024solving,
title={Solving Inverse Problems with Latent Diffusion Models via Hard Data Consistency},
author={Bowen Song and Soo Min Kwon and Zecheng Zhang and Xinyu Hu and Qing Qu and Liyue Shen},
booktitle={ICLR},
year={2024},
}

@Article{Maiden:17,
  author    = {Andrew Maiden and Daniel Johnson and Peng Li},
  title     = {Further improvements to the ptychographical iterative engine},
  number    = {7},
  pages     = {736--745},
  volume    = {4},
  journal   = {Optica},
  publisher = {Optica Publishing Group},
  year      = {2017},
}

@InProceedings{Loshchilov2017DecoupledWD,
  author    = {Ilya Loshchilov and Frank Hutter},
  booktitle = {ICLR},
  title     = {Decoupled Weight Decay Regularization},
  year      = {2017},
}

@Article{10004774,
  author   = {Zhao, Zhizhen and Ye, Jong Chul and Bresler, Yoram},
  title    = {Generative Models for Inverse Imaging Problems: From mathematical foundations to physics-driven applications},
  number   = {1},
  pages    = {148-163},
  volume   = {40},
  journal  = {IEEE Signal Processing Magazine},
  year     = {2023},
}

@InProceedings{Blau2017ThePT,
  author    = {Yochai Blau and Tomer Michaeli},
  booktitle = {CVPR},
  title     = {The Perception-Distortion Tradeoff},
  year      = {2018},
}

@Article{Monga2019AlgorithmUI,
  author  = {Vishal Monga and Yuelong Li and Yonina C. Eldar},
  title   = {Algorithm Unrolling: Interpretable, Efficient Deep Learning for Signal and Image Processing},
  pages   = {18-44},
  volume  = {38},
  journal = {IEEE Signal Processing Magazine},
  year    = {2019},
}

@Article{Gladrow2019DigitalPH,
  author  = {Jannes Gladrow},
  title   = {Digital phase-only holography using deep conditional generative models},
  journal = {ArXiv},
  year    = {2019},
}

@InProceedings{Uelwer2019PhaseRU,
  author    = {Tobias Uelwer and Alexander Oberstrass and Stefan Harmeling},
  booktitle = {ICPR},
  title     = {Phase Retrieval Using Conditional Generative Adversarial Networks},
  year      = {2020},
}

@InProceedings{Deng2019PhysicsED,
  author  = {Mo Deng and Alexandre Goy and Kwabena Arthur and George Barbastathis},
  title   = {Physics Embedded Deep Neural Network for Phase Retrieval under Low Photon Conditions},
  booktitle = {COSI},
  year    = {2019},
}

@InProceedings{Gulrajani2017ImprovedTO,
  author    = {Ishaan Gulrajani and Faruk Ahmed and Mart{\'i}n Arjovsky and Vincent Dumoulin and Aaron C. Courville},
  booktitle = {NeurIPS},
  title     = {Improved Training of Wasserstein GANs},
  year      = {2017},
}

@Article{Luo2022UnderstandingDM,
  author  = {Calvin Luo},
  title   = {Understanding Diffusion Models: A Unified Perspective},
  journal = {ArXiv},
  year    = {2022},
}

@article{kaya2026i2i,
  title={{I2I-PR}: Deep iterative refinement for phase retrieval using image-to-image diffusion models},
  author={Kaya, Mehmet Onurcan and Oktem, Figen S},
  journal={Digital Signal Processing},
  pages={106184},
  year={2026},
  publisher={Elsevier}
}

@Article{isil2024deep,
  author  = {I{\c{s}}{\i}l, {\c{C}}a{\u{g}}atay and Oktem, Figen S.},
  title   = {Deep Plug-and-Play {HIO} Approach for Phase Retrieval},
  number  = {5},
  pages   = {A84--A94},
  volume  = {64},
  journal = {Applied Optics},
  year    = {2025},
}

@InProceedings{Bansal2022ColdDI,
  author    = {Bansal, Arpit and Borgnia, Eitan and Chu, Hong-Min and Li, Jie and Kazemi, Hamid and Huang, Furong and Goldblum, Micah and Geiping, Jonas and Goldstein, Tom},
  booktitle = {NeurIPS},
  title     = {Cold diffusion: Inverting arbitrary image transforms without noise},
  year      = {2023},
}

@Article{Goy2018LowPC,
  author  = {Alexandre Goy and Kwabena Arthur and Shuai Li and George Barbastathis},
  title   = {Low Photon Count Phase Retrieval Using Deep Learning.},
  pages   = {243902},
  volume  = {121 24},
  journal = {Physical review letters},
  year    = {2018},
}

@InProceedings{jaiswal2023physics,
  author    = {Jaiswal, Ajay and Zhang, Xingguang and Chan, Stanley H. and Wang, Zhangyang},
  booktitle = {ICCV},
  title     = {Physics-driven turbulence image restoration with stochastic refinement},
  year      = {2023},
}

@InProceedings{Loshchilov2016SGDRSG,
  author    = {Ilya Loshchilov and Frank Hutter},
  booktitle = {ICLR},
  title     = {SGDR: Stochastic Gradient Descent with Warm Restarts},
  year      = {2017},
}

@InProceedings{Isil:20,
  author    = {{\c{C}}a{\u{g}}atay I{\c{s}}{\i}l and Figen S. Oktem},
  booktitle = {COSI},
  title     = {Model-based Phase Retrieval with Deep Denoiser Prior},
  year      = {2020},
}

@InProceedings{Wang2020WhenDD,
  author  = {Yaotian Wang and Xiaohang Sun and Jason W. Fleischer},
  title   = {When deep denoising meets iterative phase retrieval},
  booktitle = {ICML},
  year    = {2020},
}

@Article{marcheshrinkwrap,
  author  = {Marchesini, Stefano and He, H. and Chapman, Henry and Hau-Riege, S.P. and Noy, Aleksandr and Howells, Malcolm and Weierstall, Uwe and Spence, J},
  title   = {X-ray image reconstruction from a diffraction pattern alone},
  volume  = {68},
  journal = {Physical Review B},
  year    = {2003},
}

@Article{Naimipour2020UPRAM,
  author  = {Naveed Naimipour and Shahin Khobahi and Mojtaba Soltanalian},
  title   = {UPR: A Model-Driven Architecture for Deep Phase Retrieval},
  journal = {ACSSC},
  year    = {2020},
}

@Article{Wang2020PhaseRW,
  author  = {Chang-Jen Wang and Chao-Kai Wen and Shang-Ho Lawrence Tsai and Shi Jin},
  title   = {Phase Retrieval With Learning Unfolded Expectation Consistent Signal Recovery Algorithm},
  pages   = {780-784},
  volume  = {27},
  journal = {IEEE Signal Processing Letters},
  year    = {2020},
}

@inproceedings{Tayal2020InversePD,
  title={Unlocking inverse problems using deep learning: Breaking symmetries in phase retrieval},
  author={Tayal, Kshitij and Lai, Chieh-Hsin and Manekar, Raunak and Zhuang, Zhong and Kumar, Vipin and Sun, Ju},
  booktitle={NeurIPS 2020 Workshop on Deep Learning and Inverse Problems},
  year={2020}
}

@Article{hayes1982,
  author   = {Monson Hayes},
  title    = {The reconstruction of a multidimensional sequence from the phase or magnitude of its Fourier transform},
  number   = {2},
  pages    = {140-154},
  volume   = {30},
  journal  = {IEEE Transactions on Acoustics, Speech, and Signal Processing},
  year     = {1982},
}

@Article{waldspurger2015phase,
  author  = {Waldspurger, Irene and d'Aspremont, Alexandre and Mallat, St{\'e}phane},
  title   = {Phase recovery, maxcut and complex semidefinite programming},
  pages   = {47--81},
  volume  = {149},
  journal = {Mathematical Programming},
  year    = {2015},
}

@InProceedings{Arjovsky2017WassersteinG,
  author  = {Mart{\'i}n Arjovsky and Soumith Chintala and L{\'e}on Bottou},
  title   = {Wasserstein GAN},
  booktitle = {ICML},
  year    = {2017},
}

@Article{El_Helou_2020,
  author    = {El Helou, Majed and Susstrunk, Sabine},
  title     = {Blind Universal Bayesian Image Denoising With Gaussian Noise Level Learning},
  pages     = {4885--4897},
  volume    = {29},
  journal   = {IEEE Transactions on Image Processing},
  publisher = {Institute of Electrical and Electronics Engineers (IEEE)},
  year      = {2020},
}

@InProceedings{deng2009imagenet,
  author    = {Deng, Jia and Dong, Wei and Socher, Richard and Li, Li-Jia and Li, Kai and Fei-Fei, Li},
  booktitle = {CVPR},
  title     = {Imagenet: A large-scale hierarchical image database},
  year      = {2009},
}

@Article{lopeztaipa,
  author  = {L{\'o}pez-Tapia, Santiago and Molina, Rafael and Katsaggelos, Aggelos},
  title   = {Deep learning approaches to inverse problems in imaging: Past, present and future},
  pages   = {103285},
  volume  = {119},
  journal = {Digital Signal Processing},
  year    = {2021},
}

@Article{fienup1978reconstruction,
  author  = {Fienup, James R.},
  title   = {Reconstruction of an object from the modulus of its Fourier transform},
  number  = {1},
  pages   = {27--29},
  volume  = {3},
  journal = {Optics Letters},
  year    = {1978},
}

@Article{Romano2016TheLE,
  author  = {Romano, Yaniv and Elad, Michael and Milanfar, Peyman},
  title   = {The little engine that could: Regularization by denoising (RED)},
  number  = {4},
  pages   = {1804--1844},
  volume  = {10},
  journal = {SIAM Journal on Imaging Sciences},
  year    = {2017},
}

@InProceedings{ronneberger2015u,
  author       = {Ronneberger, Olaf and Fischer, Philipp and Brox, Thomas},
  booktitle    = {MICCAI},
  title        = {U-net: Convolutional networks for biomedical image segmentation},
  year         = {2015},
}

@Article{marchesini2007invited,
  author  = {Marchesini, Stefano},
  title   = {Invited article: A unified evaluation of iterative projection algorithms for phase retrieval},
  number  = {1},
  volume  = {78},
  journal = {Review of scientific instruments},
  year    = {2007},
}

@Article{fienup1982comparison,
  author  = {James R. Fienup},
  title   = {Phase retrieval algorithms: a comparison},
  number  = {15},
  pages   = {2758--2769},
  volume  = {21},
  journal = {Applied Optics},
  year    = {1982},
}

@InProceedings{Kawar2021SNIPSSN,
  author    = {Bahjat Kawar and Gregory Vaksman and Michael Elad},
  booktitle = {NeurIPS},
  title     = {SNIPS: Solving Noisy Inverse Problems Stochastically},
  year      = {2021},
}

@InProceedings{zhang2024wrong,
  author    = {Wenjie Zhang and Yuxiang Wan and Zhong Zhuang and Ju Sun},
  booktitle = {IS\&T Electronic Imaging Symposium},
  title     = {What{\textquoteright}s Wrong with End-to-End Learning for Phase Retrieval?},
  year      = {2024},
}

@Article{Naimipour2020UnfoldedAF,
    AUTHOR = {Naimipour, Naveed and Khobahi, Shahin and Soltanalian, Mojtaba and Safavi, Haleh and Shaw, Harry C.},
    TITLE = {Unfolded Algorithms for Deep Phase Retrieval},
    JOURNAL = {Algorithms},
    VOLUME = {17},
    YEAR = {2024},
    NUMBER = {12},
    ARTICLE-NUMBER = {587}
}

@InProceedings{Hand2018PhaseRU,
  author    = {Paul Hand and Oscar Leong and Vladislav Voroninski},
  booktitle = {NeurIPS},
  title     = {Phase Retrieval Under a Generative Prior},
  year      = {2018},
}

@InProceedings{pmlr-v80-metzler18a,
  author    = {Metzler, Christopher and Schniter, Phillip and Veeraraghavan, Ashok and Baraniuk, Richard},
  booktitle = {ICML},
  title     = {pr{D}eep: Robust Phase Retrieval with a Flexible Deep Network},
  year      = {2018},
}

@Article{gs1978,
  author  = {Gerchberg, R. W. and Saxton, W. O.},
  title   = {A practical algorithm for the determination of phase from image and diffraction plane pictures},
  pages   = {237-250},
  volume  = {35},
  journal = {Optik},
  year    = {1972},
}

@Article{stefanoqianpty,
  author  = {Qian, Jianliang and Yang, C. and Schirotzek, A. and Maia, Filipe and Marchesini, Stefano},
  title   = {Efficient algorithms for ptychographic phase retrieval, in Inverse Problems and Applications},
  pages   = {261-280},
  volume  = {615},
  journal = {Contemp. Math},
  year    = {2014},
}

@InProceedings{jaganathan2013sparse,
  author    = {Jaganathan, Kishore and Oymak, Samet and Hassibi, Babak},
  booktitle = {ISIT},
  title     = {Sparse phase retrieval: Convex algorithms and limitations},
  year      = {2013},
}

@InProceedings{8099502,
  author    = {Ledig, Christian and Theis, Lucas and Husz{\'a}r, Ferenc and Caballero, Jose and Cunningham, Andrew and Acosta, Alejandro and Aitken, Andrew and Tejani, Alykhan and Totz, Johannes and Wang, Zehan and Shi, Wenzhe},
  booktitle = {CVPR},
  title     = {Photo-Realistic Single Image Super-Resolution Using a Generative Adversarial Network},
  year      = {2017},
}

@InProceedings{Shoushtari2022DOLPHDM,
  author    = {Shirin Shoushtari and Jiaming Liu and Kamilov, {Ulugbek S.}},
  booktitle = {ACSSC},
  title     = {Diffusion Models for Phase Retrieval in Computational Imaging},
  year      = {2023},
}

@InProceedings{zhang2017learning,
  author    = {Zhang, Kai and Zuo, Wangmeng and Gu, Shuhang and Zhang, Lei},
  booktitle = {CVPR},
  title     = {Learning Deep {CNN} Denoiser Prior for Image Restoration},
  year      = {2017},
}

@Article{Dong_2023,
  author  = {Dong, Jonathan and Valzania, Lorenzo and Maillard, Antoine and Pham, Thanh-an and Gigan, Sylvain and Unser, Michael},
  title   = {Phase Retrieval: From Computational Imaging to Machine Learning},
  number  = {1},
  pages   = {45--57},
  volume  = {40},
  journal = {IEEE Signal Processing Magazine},
  year    = {2023},
}

@Article{Jin2016DeepCN,
  author  = {Kyong Hwan Jin and Michael T. McCann and Emmanuel Froustey and Michael A. Unser},
  title   = {Deep Convolutional Neural Network for Inverse Problems in Imaging},
  pages   = {4509-4522},
  volume  = {26},
  journal = {IEEE Transactions on Image Processing},
  year    = {2016},
}

@InProceedings{Shevkunov2021DeepCN,
  author    = {Igor Shevkunov and Jarkko Kilpel{\"a}inen and Karen Eguiazarian},
  booktitle = {BiOS},
  title     = {Deep convolutional neural network-based lensless quantitative phase retrieval},
  year      = {2021},
}

@InBook{Dimakis_2022,
  author    = {Dimakis, Alexandros G.},
  booktitle = {Mathematical Aspects of Deep Learning},
  title     = {Deep Generative Models and Inverse Problems},
  editor    = {Grohs, Philipp and Kutyniok, GittaEditors},
  pages     = {400--421},
  publisher = {Cambridge University Press},
  place     = {Cambridge},
  year      = {2022},
}

@Article{miyasawa1961empirical,
  author  = {Miyasawa, Koichi},
  title   = {An empirical Bayes estimator of the mean of a normal population},
  number  = {181-188},
  pages   = {1--2},
  volume  = {38},
  journal = {Bull. Inst. Internat. Statist},
  year    = {1961},
}

@InProceedings{He2015DeepRL,
  author    = {Kaiming He and X. Zhang and Shaoqing Ren and Jian Sun},
  booktitle = {CVPR},
  title     = {Deep Residual Learning for Image Recognition},
  year      = {2016},
}

@Article{Liu2023PRISTANetDI,
  author  = {Aoxu Liu and Xiaohong Fan and Yin Yang and Jianping Zhang},
  title   = {PRISTA-Net: Deep Iterative Shrinkage Thresholding Network for Coded Diffraction Patterns Phase Retrieval},
  journal = {ArXiv},
  year    = {2023},
}

@Article{elad2023image,
  author    = {Elad, Michael and Kawar, Bahjat and Vaksman, Gregory},
  title     = {Image denoising: The deep learning revolution and beyond---a survey paper},
  number    = {3},
  pages     = {1594--1654},
  volume    = {16},
  journal   = {SIAM Journal on Imaging Sciences},
  publisher = {SIAM},
  year      = {2023},
}

@InProceedings{6737048,
  author    = {Venkatakrishnan, Singanallur V. and Bouman, Charles A. and Wohlberg, Brendt},
  booktitle = {GlobalSIP},
  title     = {Plug-and-Play priors for model based reconstruction},
  year      = {2013},
}

@article{shorten2019survey,
  title={A survey on image data augmentation for deep learning},
  author={Shorten, Connor and Khoshgoftaar, Taghi M},
  journal={Journal of big data},
  volume={6},
  number={1},
  pages={1--48},
  year={2019}
}

@inproceedings{Kimura2024UnderstandingTA,
  title={Understanding test-time augmentation},
  author={Kimura, Masanari},
  booktitle={ICONIP},
  year={2021}
}

@inproceedings{shanmugam2021better,
  title={Better aggregation in test-time augmentation},
  author={Shanmugam, Divya and Blalock, Davis and Balakrishnan, Guha and Guttag, John},
  booktitle={ICCV},
  year={2021}
}

@incollection{casado2020ensemble,
  title={Ensemble methods for object detection},
  author={Casado-Garc{\'\i}a, {\'A}ngela and Heras, J{\'o}nathan},
  booktitle={ECAI 2020},
  year={2020}
}

@article{wang2019aleatoric,
  title={Aleatoric uncertainty estimation with test-time augmentation for medical image segmentation with convolutional neural networks},
  author={Wang, Guotai and Li, Wenqi and Aertsen, Michael and Deprest, Jan and Ourselin, S{\'e}bastien and Vercauteren, Tom},
  journal={Neurocomputing},
  volume={338},
  pages={34--45},
  year={2019}
}

@article{saharia2022image,
  title={Image super-resolution via iterative refinement},
  author={Saharia, Chitwan and Ho, Jonathan and Chan, William and Salimans, Tim and Fleet, David J and Norouzi, Mohammad},
  journal={IEEE Transactions on Pattern Analysis and Machine Intelligence},
  volume={45},
  number={4},
  pages={4713--4726},
  year={2022}
}

@inproceedings{saharia2022palette,
  title={Palette: Image-to-image diffusion models},
  author={Saharia, Chitwan and Chan, William and Chang, Huiwen and Lee, Chris and Ho, Jonathan and Salimans, Tim and Fleet, David and Norouzi, Mohammad},
  booktitle={ACM SIGGRAPH},
  year={2022}
}

@article{chang2021LPR,
  title={Large-scale phase retrieval},
  author={Chang, Xuyang and Bian, Liheng and Zhang, Jun},
  journal={elight},
  volume={1},
  number={1},
  pages={4},
  year={2021},
  publisher={Springer}
}

@inproceedings{
      meng2022sdedit,
      title={{SDE}dit: Guided Image Synthesis and Editing with Stochastic Differential Equations},
      author={Chenlin Meng and Yutong He and Yang Song and Jiaming Song and Jiajun Wu and Jun-Yan Zhu and Stefano Ermon},
      booktitle={ICLR},
      year={2022},
}

@INPROCEEDINGS{kaya2025_mlsp,
  author={Kaya, Mehmet Onurcan and Oktem, Figen S.},
  booktitle={2025 IEEE 35th International Workshop on Machine Learning for Signal Processing (MLSP)}, 
  title={{prNet}: Efficient and Robust Phase Retrieval via Stochastic Refinement}, 
  year={2025},
  volume={},
  number={},
  pages={01-06}
}

@inproceedings{li2024difffpr,
  title={{DiffFPR}: Diffusion prior for oversampled Fourier phase retrieval},
  author={Li, Ji and Wang, Chao},
  booktitle={ICML},
  year={2024}
}

@inproceedings{zhang2025improving,
  title={Improving diffusion inverse problem solving with decoupled noise annealing},
  author={Zhang, Bingliang and Chu, Wenda and Berner, Julius and Meng, Chenlin and Anandkumar, Anima and Song, Yang},
  booktitle={CVPR},
  year={2025}
}

@inproceedings{timofte2016seven,
  title={Seven ways to improve example-based single image super resolution},
  author={Timofte, Radu and Rothe, Rasmus and Van Gool, Luc},
  booktitle={CVPR},
  year={2016}
}

@article{zhang2021plug,
  title={Plug-and-play image restoration with deep denoiser prior},
  author={Zhang, Kai and Li, Yawei and Zuo, Wangmeng and Zhang, Lei and Van Gool, Luc and Timofte, Radu},
  journal={IEEE Transactions on Pattern Analysis and Machine Intelligence},
  volume={44},
  number={10},
  pages={6360--6376},
  year={2021}
}

@inproceedings{terris2024equivariant,
  title={Equivariant plug-and-play image reconstruction},
  author={Terris, Matthieu and Moreau, Thomas and Pustelnik, Nelly and Tachella, Julian},
  booktitle={CVPR},
  year={2024}
}

@inproceedings{NIPS2011_86e8f7ab,
 author = {Bergstra, James and Bardenet, R\'{e}mi and Bengio, Yoshua and K\'{e}gl, Bal\'{a}zs},
 booktitle = {NeurIPS},
 title = {Algorithms for Hyper-Parameter Optimization},
 year = {2011}
}

@book{casella2024statistical,
  title={Statistical inference},
  author={Casella, George and Berger, Roger},
  year={2024},
  publisher={Chapman and Hall/CRC}
}

@inproceedings{
chung2023dps,
title={Diffusion Posterior Sampling for General Noisy Inverse Problems},
author={Hyungjin Chung and Jeongsol Kim and Michael Thompson Mccann and Marc Louis Klasky and Jong Chul Ye},
booktitle={ICLR},
year={2023},
}

@inproceedings{
song2023pseudoinverse,
title={Pseudoinverse-Guided Diffusion Models for Inverse Problems},
author={Jiaming Song and Arash Vahdat and Morteza Mardani and Jan Kautz},
booktitle={ICLR},
year={2023},
}

@article{fannjiang2020numerics,
  title={The numerics of phase retrieval},
  author={Fannjiang, Albert and Strohmer, Thomas},
  journal={Acta Numerica},
  volume={29},
  pages={125--228},
  year={2020},
}

@article{laumont2022bayesian,
  title={Bayesian imaging using plug \& play priors: when langevin meets tweedie},
  author={Laumont, R{\'e}mi and Bortoli, Valentin De and Almansa, Andr{\'e}s and Delon, Julie and Durmus, Alain and Pereyra, Marcelo},
  journal={SIAM Journal on Imaging Sciences},
  volume={15},
  number={2},
  pages={701--737},
  year={2022},
}

@article{shastri2025fast,
  title={Fast and robust phase retrieval via deep expectation-consistent approximation},
  author={Shastri, Saurav K and Schniter, Philip},
  journal={IEEE Transactions on Computational Imaging},
  volume={11},
  pages={116--128},
  year={2025},
}

@inproceedings{peer2023diffphase,
  title={DiffPhase: Generative diffusion-based STFT phase retrieval},
  author={Peer, Tal and Welker, Simon and Gerkmann, Timo},
  booktitle={ICASSP},
  year={2023},
}

@inproceedings{cha2024red,
  title={{RED-PhaseCut}: a variational sampler using diffusion model for fourier phase retrieval},
  author={Cha, Eunju},
  booktitle={EUSIPCO},
  year={2024},
}

@article{wu2024principled,
  title={Principled probabilistic imaging using diffusion models as plug-and-play priors},
  author={Wu, Zihui and Sun, Yu and Chen, Yifan and Zhang, Bingliang and Yue, Yisong and Bouman, Katherine},
  journal={NeurIPS},
  year={2024}
}

\appendices

\onecolumn

\setcounter{page}{1}

% \title{prNet: Data-Driven Phase Retrieval via\\ Stochastic Refinement: Supplementary Material}
% \author{Mehmet Onurcan Kaya~\orcidlink{0009-0006-2606-3992} and Figen S. Oktem~\orcidlink{0000-0002-7882-5120},~\IEEEmembership{Member,~IEEE}}
% \maketitle
\begin{center}
    {\LARGE\bfseries
    prNet: Data-Driven Phase Retrieval\\via
    Stochastic Refinement\\[0.5em]
    Supplementary Material}

    \vspace{1em}

    {\large
    Mehmet Onurcan Kaya and Figen S. Oktem}
\end{center}

% \vspace{1cm}
% \section*{Implementation Details}

\section*{Derivation Details}

This section provides detailed derivations for the equations in Section~\ref{sec:methodprnet}. Equation references refer to the main paper unless stated otherwise.

\subsection{Derivation of Eq.~\eqref{eq:eq10}: Simplified Forward Model}

Starting from the original measurement model,
\begin{equation}
\mathbf{y}^2 = |\mathbf{A}\mathbf{x}|^2 + \mathbf{w}, \quad \quad \mathbf{w} \sim \mathcal{N}\left(\mathbf{0}, \alpha^2 \, \mathrm{diag}(|\mathbf{A}\mathbf{x}|^2)\right),
\label{eq:measurementmodel_supp}
\end{equation}
the conditional density of the squared measurements given $\mathbf{x}$ is
% which implies the conditional distribution of the squared measurements \( \mathbf{y}^2 \) given the signal \( \mathbf{x} \) is:
\begin{equation}    
\begin{aligned}
p(\mathbf{y}^2 \mid \mathbf{x}) &= \mathcal{N}\left( |\mathbf{A}\mathbf{x}|^2 , \alpha^2 \, \mathrm{diag}(|\mathbf{A}\mathbf{x}|^2)\right)
\\&= \prod_{i} \frac{1}{\sqrt{2\pi \alpha^2 |[\mathbf{A}\mathbf{x}]_i|^2}} \exp\left( -\frac{(y_i^2 - |[\mathbf{A}\mathbf{x}]_i|^2)^2}{2 \alpha^2 |[\mathbf{A}\mathbf{x}]_i|^2} \right).
\end{aligned}
\end{equation}

Since the covariance matrix is diagonal, the measurements are independent conditioned on \( \mathbf{x} \), and we can restrict our analysis to the univariate case without loss of generality. Therefore, let us consider the scalar case:
% Since the covariance is diagonal, the entries are conditionally independent given $\mathbf{x}$, and it suffices to analyze the scalar case
% 
%
\begin{equation}
y_i^2 = |[\mathbf{A}\mathbf{x}]_i|^2 + w_i, \quad \quad w_i \sim \mathcal{N}\left(0, \alpha^2 |[\mathbf{A}\mathbf{x}]_i|^2 \right).
\end{equation}

This implies that conditioned on \(\mathbf{x}\), the distribution of \(y_i^2\) is Gaussian
\begin{equation}
y_i^2 \mid \mathbf{x} \sim \mathcal{N}\left( \mu_i, \sigma_i^2 \right), \quad \text{where} \quad \mu_i = |[\mathbf{A}\mathbf{x}]_i|^2, \quad \sigma_i^2 = \alpha^2 |[\mathbf{A}\mathbf{x}]_i|^2.
\end{equation}

% We wish to approximate the distribution of \(y_i = \sqrt{y_i^2}\) using the delta method, which allows us to approximate the distribution of a nonlinear function of a random variable.
To obtain a tractable model for $y_i = \sqrt{y_i^2}$, we apply the delta method~\cite{casella2024statistical}, which states that for $Z \sim \mathcal{N}(\mu, \sigma^2)$ and a smooth function $g$, the distribution of $g(Z)$ is approximately $\mathcal{N}(g(\mu),\, \sigma^2 [g'(\mu)]^2)$. Taking $Z = y_i^2$ and $g(Z) = \sqrt{Z}$ yields
\[
g'(\mu_i) = \frac{1}{2\sqrt{\mu_i}} = \frac{1}{2|[\mathbf{A}\mathbf{x}]_i|} \quad
\Rightarrow \quad \text{Var}(y_i) \approx \alpha^2 |[\mathbf{A}\mathbf{x}]_i|^2 \cdot \left( \frac{1}{2|[\mathbf{A}\mathbf{x}]_i|} \right)^2 = \left( \frac{\alpha}{2} \right)^2
\]
% \noindent
% \textbf{Delta method.} Given a random variable \(Z \sim \mathcal{N}(\mu, \sigma^2)\), and a smooth function \(g(Z)\), the delta method approximates the distribution of \(g(Z)\) as $
% \mathcal{N}(g(\mu), \sigma^2 [g'(\mu)]^2)
% $.
% \href{https://en.wikipedia.org/wiki/Delta_method}{https://en.wikipedia.org/wiki/Delta\_method}
% In our case, we take \(Z = y_i^2\), and \(g(Z) = \sqrt{Z}\). Applying the delta method:
% Therefore, the delta method implies the approximate distribution:
and hence
\begin{equation}
y_i \mid \mathbf{x} \sim \mathcal{N}\left( |[\mathbf{A}\mathbf{x}]_i|, \left( \frac{\alpha}{2} \right)^2 \right).
\end{equation}

% This scalar derivation justifies the simplified measurement model in Eq.~\eqref{eq:simplifiedmeasurementmodel_recap}, where the measurement noise has a fixed variance \( (\alpha/2)^2 \) across all entries. Thus, returning to the vector-valued setting, we adopt the simplified forward model:
The variance is independent of $i$, thus collecting the entries gives the vector-valued simplified forward model
\begin{equation}
\mathbf{y} = |\mathbf{A}\mathbf{x}| + \mathbf{w}, \quad \quad \mathbf{w} \sim \mathcal{N}\left( \mathbf{0}, \left(\frac{\alpha}{2}\right)^2 \mathbf{I} \right),
\label{eq:simplifiedmeasurementmodel_recap}
\end{equation}
% which approximates Eq.~\eqref{eq:measurementmodel} in the moderate-to-high SNR regime, where the relative variance $\sigma_i / \mu_i$ is small and the Gaussian approximation under $\sqrt{\cdot}$ is accurate.
which provides a tractable approximation of the original model in Eq.~\eqref{eq:measurementmodel} under moderate-to-high SNR conditions when the variance is small (so that its values are concentrated in a small neighborhood), thus, the distribution of the transformed random variable is approximately Gaussian.
%This simplification enables a closed-form expression for the log-likelihood gradient \( \nabla_{\mathbf{x}} \log p(\mathbf{y} \mid \mathbf{x}_t) \), facilitating its use in iterative inference or gradient-based optimization.

\subsection{Derivation of Eq.~\eqref{eq:eq11}: Gradient of the Log-Likelihood}

% We have two models.
We combine two models: the simplified forward model from Eq.~\eqref{eq:simplifiedmeasurementmodel_recap},
% 1. Simplified forward model in Eq.~\eqref{eq:simplifiedmeasurementmodel}:
\begin{equation}
    \mathbf{y} = \left| \mathbf{A} \mathbf{x} \right| + \mathbf{w}, \quad \mathbf{w} \sim \mathcal{N} \left( \mathbf{0}, \left( \frac{\alpha}{2} \right)^2 \mathbf{I} \right),
\end{equation}
and the diffusion process
% 2. Diffusion process definition: 
\begin{equation}
\mathbf{x}_t = \mathbf{x} + \mathbf{v}, \quad \mathbf{v} \sim \mathcal{N}(\mathbf{0}, \sigma_t^2 \mathbf{I}) .
\end{equation}
% BU HALA APPROXIMATION, OBTAINED DEMEYELİM.
% Bu forma varmak için
% 1. Independence btw. x_t and y gerekiyor
% 2. \sigma_t^2 küçük olmalı, large SNR gerekiyor
% Burada çok ara stepi atladım. Çoğu diffusion for inverse problems paperı da atlıyor.
% İki modelimiz var:
% y = |Ax| + w where w ~ N(0, (alpha/2)^2 I)
% x_t = x + v where v ~ N(0, sigma_t^2 I)
% Ve biz p(y|x_t) bulmak istiyoruz.
% x GT görüntü. x_t mevcut adımdaki tahminimiz, additive Gaussian ile hatasını modelledik çünkü düzgün eğitirsek önceki adımlardan perfect tahmin geliyor ve bu adıma gelmeden üstüne biz gürültü v ekliyoruz gibi düşünüyoruz.
% p(y | x_t)
% = p(y - |Ax_t| | x_t)
% = p(|Ax| + w - |A(x+v)| | x_t)
% = p(|Ax| + w - |Ax| - J(x)*v | x_t) (first order Taylor expansion around x)
% = p(w - J(x)*v | x_t)
% Burada w - J(x)*v ikisinin de zero-mean Gaussian olduğunu ve varyanslarını biliyoruz. Ama x_t nasıl etkiliyor bilmiyoruz. Onu nasıl kullanacağız?
% İşte, independence assumptionı bunun için gerekli x_t independent olmalı ki 
% p(w - J(x)*v | x_t) = p(w - Jv)
% diyebilelim
% x_t aslında bunlardan etkileniyor independent değil. Ama bu assumption common in the literature, OK.
% Sonra buradan (alpha/2)^2  I + sigma_t^2 * J(x) * J(x)^H olarak varyans geliyor.
% J(x) yani |Ax| fonksiyonunun gt. olan x'teki gradyanını
%%%%%%%%%%%%%%%%%%%%%%%%%%%%%%%%%%%%%%%%%%%%%%%%%
% Therefore,
Our goal is to characterize $p(\mathbf{y} \mid \mathbf{x}_t)$. Substituting both models and applying a first-order Taylor expansion of $|\mathbf{A}(\mathbf{x}+\mathbf{v})|$ around $\mathbf{v} = \mathbf{0}$ gives
\begin{equation}
\begin{aligned}
&p(\mathbf{y} | \mathbf{x}_t)\\
&=p(\mathbf{y} - \left| \mathbf{A} \mathbf{x}_t \right| | \mathbf{x}_t)\\
&=p(\left| \mathbf{A} \mathbf{x} \right| + \mathbf{w} - \left| \mathbf{A} \mathbf{x}_t \right| | \mathbf{x}_t)\\
&=p(\left| \mathbf{A} \mathbf{x} \right| + \mathbf{w} - \left| \mathbf{A} (\mathbf{x} + \mathbf{v}) \right| | \mathbf{x}_t)\\
&=p(\left| \mathbf{A} \mathbf{x} \right| + \mathbf{w} - \left| \mathbf{A} \mathbf{x} \right| - \mathbf{J}(\mathbf{x})  \mathbf{v} | \mathbf{x}_t), \quad &\text{first-order Taylor expansion around } \mathbf{x}\\
&=p( \mathbf{w} - \mathbf{J}(\mathbf{x})  \mathbf{v} | \mathbf{x}_t), &\text{here, } \mathbf{J}(\mathbf{x}) \text{ is the Jacobian of } \left| \mathbf{A} \mathbf{x} \right| \text{ at } \mathbf{x}\\
&=p( \mathbf{w} - \mathbf{J}(\mathbf{x})  \mathbf{v}) &\text{assuming independence btw. } \mathbf{w} - \mathbf{J}(\mathbf{x})  \mathbf{v} \text{ and } \mathbf{x}_t\\
&=\mathcal{N}(\mathbf{0}, 
\left( \frac{\alpha}{2} \right)^2 \mathbf{I} + \sigma_t^2 \mathbf{J}(\mathbf{x}) \mathbf{J}(\mathbf{x})^H
)
\end{aligned}
\end{equation}
where $\mathbf{J}(\mathbf{x})$ denotes the Jacobian of $|\mathbf{A}\mathbf{x}|$ at $\mathbf{x}$. The independence approximation is standard in diffusion-based posterior sampling~\cite{chung2023dps,song2023pseudoinverse} and is most accurate in the small-$\sigma_t$ regime, where $\mathbf{x}_t \approx \mathbf{x}$ and the residual $\mathbf{w} - \mathbf{J}\mathbf{v}$ is approximately decoupled from $\mathbf{x}_t$. This is consistent with our warm-start initialization, which already provides a reasonable estimate of $\mathbf{x}$.

% Independence assumption is valid in the small-$\sigma_t$ (i.e., late-time) regime where $\mathbf{x}_t \approx \mathbf{x}$, making the backward influence of $\mathbf{w} - \mathbf{Jv}$ on $\mathbf{x}_t$ negligible. We already start from a fair estimate of $\mathbf{x}$.

Therefore, the conditional likelihood $p(\mathbf{y} | \mathbf{x}_t)$ corresponds to a multivariate Gaussian distribution with mean $\left| \mathbf{A} \mathbf{x}_t \right|$ and covariance matrix $\left( \frac{\alpha}{2} \right)^2 \mathbf{I} + \sigma_t^2 \mathbf{J}(\mathbf{x}) \mathbf{J}(\mathbf{x})^H$.
The Jacobian admits the closed form
\[
\mathbf{J}(\mathbf{x}) = \mathrm{diag}\!\left(\frac{\mathbf{A}\mathbf{x}}{|\mathbf{A}\mathbf{x}|}\right) \mathbf{A},
\]
and for unitary $\mathbf{A}$ the diagonal entries have unit modulus and $\mathbf{A}\mathbf{A}^H = \mathbf{I}$, yielding $\mathbf{J}(\mathbf{x})\mathbf{J}(\mathbf{x})^H = \mathbf{I}$. The covariance therefore reduces to $\bigl(\bigl(\tfrac{\alpha}{2}\bigr)^2 + \sigma_t^2\bigr)\mathbf{I}$, and the log-likelihood becomes
\begin{equation}
\log p(\mathbf{y} \mid \mathbf{x}_t) = -\frac{\nicefrac{1}{2}}{\left(\tfrac{\alpha}{2}\right)^2 + \sigma_t^2}\, \bigl\| \mathbf{y} - |\mathbf{A}\mathbf{x}_t| \bigr\|^2 + \mathrm{const}.
\end{equation}
Differentiating with respect to $\mathbf{x}_t$ yields
\begin{equation}
\nabla_{\mathbf{x}} \log p(\mathbf{y} \mid \mathbf{x}_t) = -\frac{\nicefrac{1}{2}}{\left(\tfrac{\alpha}{2}\right)^2 + \sigma_t^2}\, \nabla_{\mathbf{x}} \bigl\| \mathbf{y} - |\mathbf{A}\mathbf{x}_t| \bigr\|^2,
\end{equation}
which is the expression in Eq.~\eqref{eq:eq11}.

% The Jacobian of $f(\mathbf{x}) = |\mathbf{A} \mathbf{x}|$ is:

% $$
% \mathbf{J}(\mathbf{x}) = \operatorname{diag}\left( \frac{\mathbf{A} \mathbf{x}}{|\mathbf{A} \mathbf{x}|} \right) \cdot \mathbf{A}
% $$

% And, when $\mathbf{A}$ is unitary, $\mathbf{J}(\mathbf{x})\mathbf{J}(\mathbf{x})^H = \mathbf{I}$.

% With this simplification, the log-likelihood takes the form:
% \begin{equation}
%     \log p(\mathbf{y} | \mathbf{x}_t) = -\frac{\nicefrac{1}{2}}{ \left( \frac{\alpha}{2} \right)^2 + \sigma_t^2} \left\| \mathbf{y} - \left| \mathbf{A} \mathbf{x}_t \right| \right\|^2 + \text{const}
% \end{equation}
% Differentiating both sides with respect to $\mathbf{x}$ yields the gradient of the log-likelihood:
% \begin{equation}
%     \nabla_{\mathbf{x}} \log p(\mathbf{y} | \mathbf{x}_t) = -\frac{\nicefrac{1}{2}}{ \left( \frac{\alpha}{2} \right)^2 + \sigma_t^2} \nabla_{\mathbf{x}} \left\| \mathbf{y} - \left| \mathbf{A} \mathbf{x}_t \right| \right\|^2,
% \end{equation}
% which results in the expression shown in Eq.~\eqref{eq:loggradientleastsquareslike}.

\subsection{Derivation of Eq.~\eqref{eq:eq12}: Sub-gradient Expression}

Let $f(\mathbf{x}) = \bigl\|\mathbf{y} - |\mathbf{A}\mathbf{x}|\bigr\|^2 = \sum_i \bigl(y_i - |(\mathbf{A}\mathbf{x})_i|\bigr)^2$. Using the convention $\partial |u| / \partial u = u / |u|$ for $u \neq 0$ (a sub-gradient at the origin), we have for each component
\[
\frac{\partial}{\partial (\mathbf{A}\mathbf{x})_i}\, \bigl(y_i - |(\mathbf{A}\mathbf{x})_i|\bigr)^2
= 2\,\bigl(|(\mathbf{A}\mathbf{x})_i| - y_i\bigr)\, \frac{(\mathbf{A}\mathbf{x})_i}{|(\mathbf{A}\mathbf{x})_i|}.
\]
Applying the chain rule and stacking the components into matrix form gives
\[
\nabla_{\mathbf{x}} f(\mathbf{x})
= 2\, \mathbf{A}^H \!\left[\, \bigl(|\mathbf{A}\mathbf{x}| - \mathbf{y}\bigr) \odot \frac{\mathbf{A}\mathbf{x}}{|\mathbf{A}\mathbf{x}|} \right].
\]
Using the identity $|\mathbf{A}\mathbf{x}| \odot \tfrac{\mathbf{A}\mathbf{x}}{|\mathbf{A}\mathbf{x}|} = \mathbf{A}\mathbf{x}$ and splitting the terms,
\[
\nabla_{\mathbf{x}} f(\mathbf{x})
= 2\,\mathbf{A}^H \mathbf{A}\, \mathbf{x} \;-\; 2\, \mathbf{A}^H \!\left[\frac{\mathbf{A}\mathbf{x}}{|\mathbf{A}\mathbf{x}|} \odot \mathbf{y} \right].
\]
For unitary $\mathbf{A}$, $\mathbf{A}^\dagger = \mathbf{A}^H$ and $\mathbf{A}^H \mathbf{A}\mathbf{x} = \mathbf{x}$, so
\[
\nabla_{\mathbf{x}} \bigl\|\mathbf{y} - |\mathbf{A}\mathbf{x}|\bigr\|^2 = 2\!\left(\mathbf{x} - \mathbf{A}^\dagger\!\left[\tfrac{\mathbf{A}\mathbf{x}}{|\mathbf{A}\mathbf{x}|} \odot \mathbf{y}\right]\right),
\]
which matches the sub-gradient expression in Eq.~\eqref{eq:eq12}.

\clearpage
\section*{Uncertainty Quantification via Posterior Sampling}

A fundamental limitation of point estimators, such as MAP or MMSE estimators,
is that they summarize the posterior distribution $p(\mathbf{x}|\mathbf{y})$ with a single estimate.
While this produces a reconstruction that is optimal in a distortion sense, it discards the uncertainty and multimodality of the posterior distribution. Phase retrieval is a severely ill-posed inverse problem with intrinsic ambiguities, such as sign, translation, and rotation. Because the phase information is lost during measurement process, the mapping from the underlying image to its Fourier magnitude measurements is many-to-one. Although uniqueness may be established up to these trivial ambiguities, multiple distinct images remain consistent with the same observations. Consequently, a single point estimate cannot capture the full range of plausible solutions.

Our framework addresses this directly. By drawing multiple samples through stochastic Langevin dynamics, prNet-Large generates $K$ diverse 
reconstructions for each input measurement. These samples reflect different plausible solutions supported by the learned posterior model, rather than simply introducing random perturbations to a single reconstruction. Regions that are strongly constrained by the measurements tend to exhibit high agreement across samples, whereas regions that are weakly constrained due to noise or information loss typically show greater variability.

This naturally enables pixel-wise uncertainty estimation without any additional machinery. Given $K$
posterior samples $\{\hat{\mathbf{x}}^{(k)}\}_{k=1}^{K}$, the predictive uncertainty at each pixel is
estimated as the sample standard deviation:
\begin{equation}
    \boldsymbol{\sigma}[n] = \sqrt{\frac{1}{K-1}\sum_{k=1}^{K}\left(\hat{\mathbf{x}}^{(k)}[n] - \bar{\mathbf{x}}[n]\right)^2},
\end{equation}
where $\bar{\mathbf{x}} = \frac{1}{K}\sum_{k=1}^{K}\hat{\mathbf{x}}^{(k)}$ is the posterior mean estimate.
A well-calibrated uncertainty map should spatially correlate with the actual reconstruction error
$|\mathbf{x} - \bar{\mathbf{x}}|$: pixels where the model is uncertain should be the same pixels where
the mean prediction deviates from the ground truth.

Figures show this comparison for a representative subset of test images. Each column displays
the ground truth, the mean prediction (with its PSNR and SSIM), the predicted uncertainty map $\boldsymbol{\sigma}$,
and the actual error map $|\mathbf{x} - \bar{\mathbf{x}}|$. Both maps are visualized with a power-scaling
transform to enhance contrast. The spatial agreement between the two rightmost panels confirms that our
stochastic sampler produces meaningfully calibrated uncertainty estimates. This is a unique advantage of
the Langevin-based posterior sampling framework: uncertainty is not a post-hoc addition, but an intrinsic
byproduct of the sampling process, obtained at no additional cost beyond the multiple reconstructions
already used for ensemble averaging.

This capability has direct practical value. In applications where a downstream decision depends on a
specific image feature, such as identifying a structural boundary or detecting a localized signal,
the uncertainty map indicates which features are reliably reconstructed and which should be treated with
caution. Point estimators provide no such guidance.

\begin{figure}[h]
    \centering
    \begin{tabular*}{\textwidth}{@{\extracolsep{\fill}}
    >{\centering\arraybackslash}p{0.22\textwidth}
    >{\centering\arraybackslash}p{0.225\textwidth}
    >{\centering\arraybackslash}p{0.215\textwidth}
    >{\centering\arraybackslash}p{0.23\textwidth}}
    Ground Truth & Mean Prediction, $\bar{\mathbf{x}}$ & Predicted Uncertainty, $\boldsymbol{\sigma}$ & Actual Error, $|\mathbf{x} - \bar{\mathbf{x}}|$ \\
    \end{tabular*}
    \\[0.5em]
    \includegraphics[width=\linewidth,trim={0 0 0 18mm},clip]{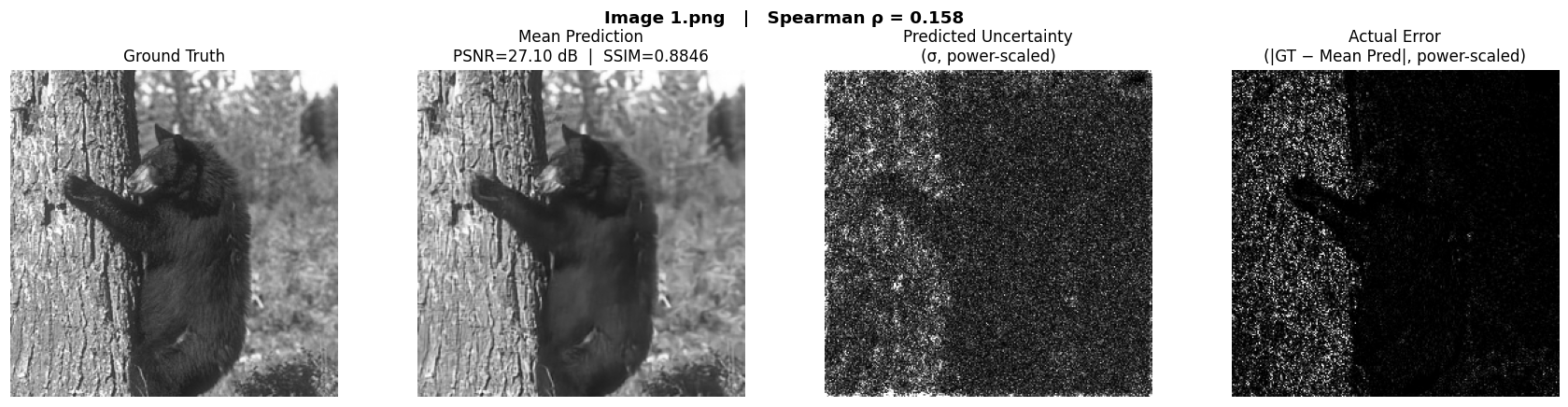}
    \includegraphics[width=\linewidth,trim={0 0 0 18mm},clip]{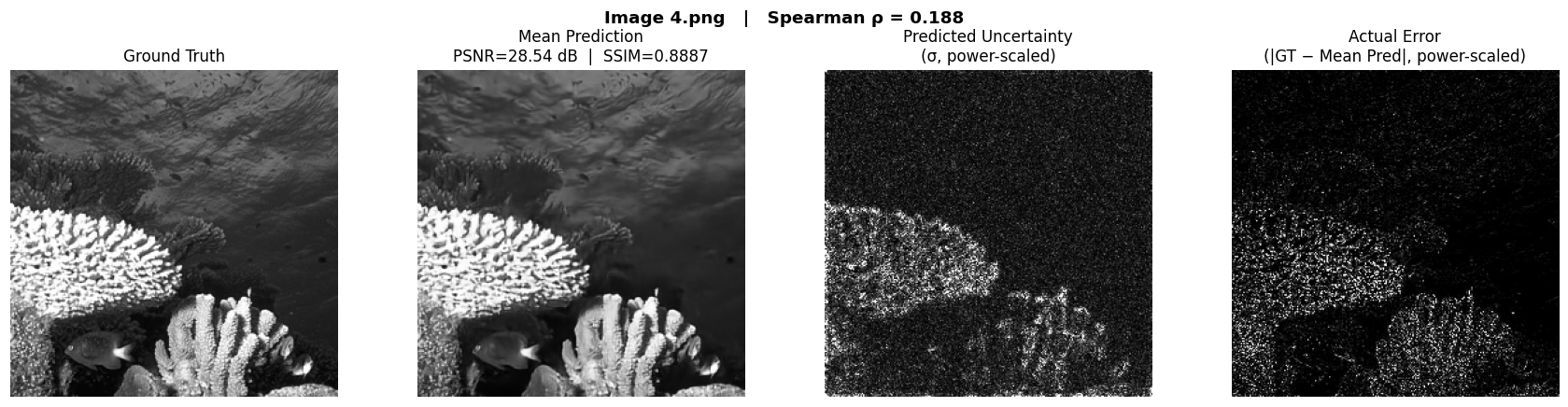}
    \includegraphics[width=\linewidth,trim={0 0 0 18mm},clip]{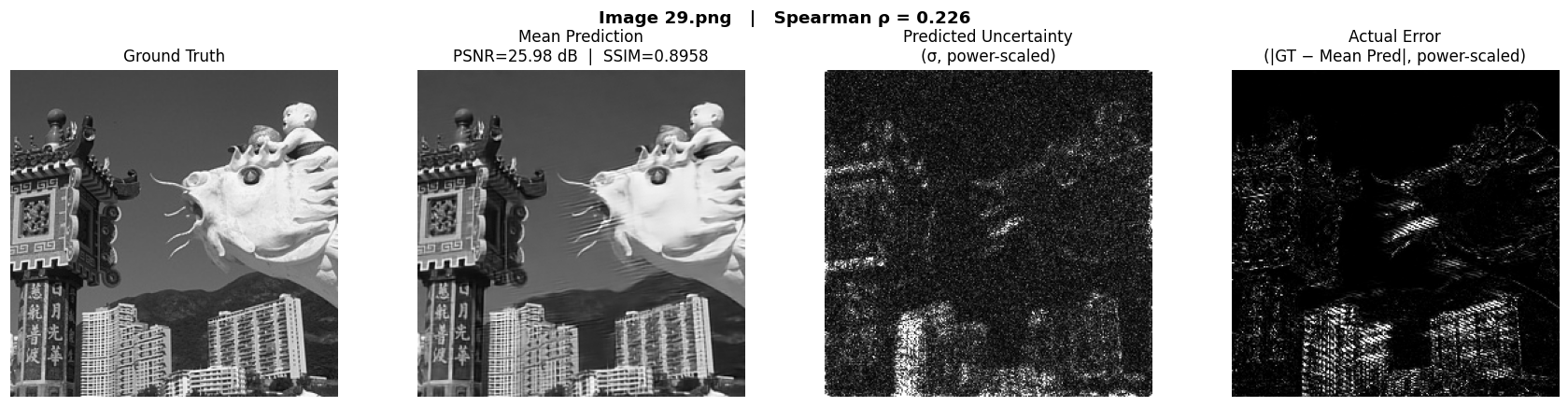}
    \includegraphics[width=\linewidth,trim={0 0 0 18mm},clip]{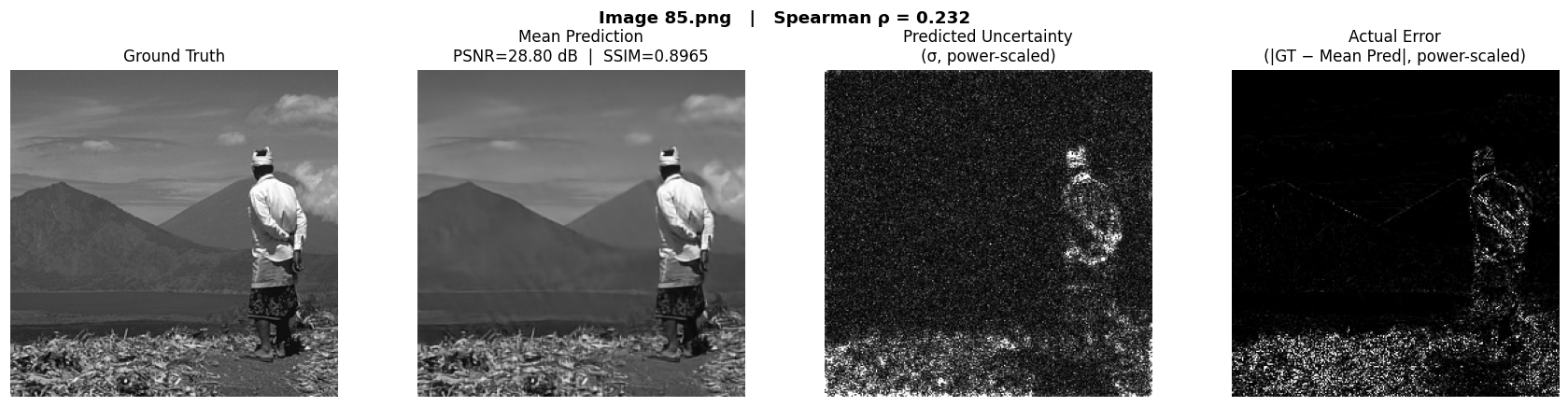}
    \includegraphics[width=\linewidth,trim={0 0 0 18mm},clip]{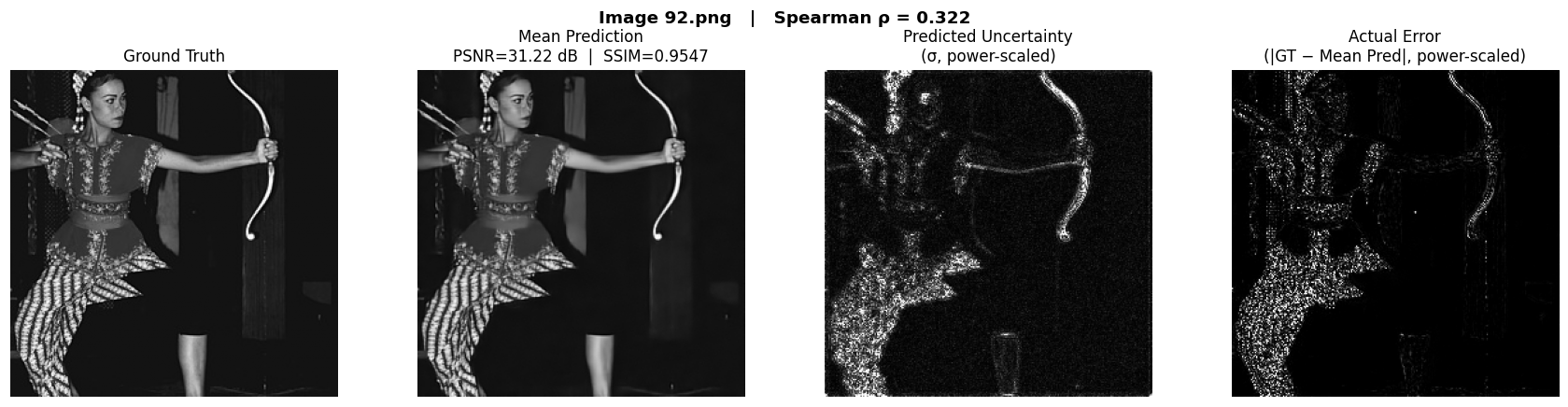}
    \includegraphics[width=\linewidth,trim={0 0 0 18mm},clip]{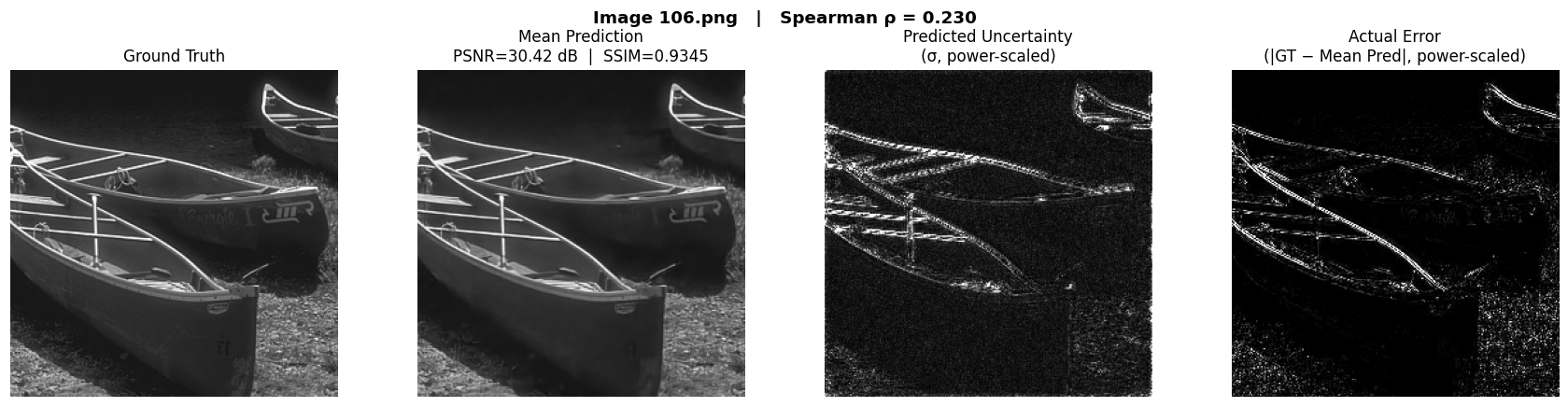}

    \caption{Uncertainty estimation with prNet-Large. The predicted pixel-wise uncertainty from posterior samples matches the actual reconstruction error.}
    \label{fig:placeholder}
\end{figure}

\begin{figure}[h]
    \centering
    \begin{tabular*}{\textwidth}{@{\extracolsep{\fill}}
    >{\centering\arraybackslash}p{0.22\textwidth}
    >{\centering\arraybackslash}p{0.225\textwidth}
    >{\centering\arraybackslash}p{0.215\textwidth}
    >{\centering\arraybackslash}p{0.23\textwidth}}
    Ground Truth & Mean Prediction, $\bar{\mathbf{x}}$ & Predicted Uncertainty, $\boldsymbol{\sigma}$ & Actual Error, $|\mathbf{x} - \bar{\mathbf{x}}|$ \\
    \end{tabular*}
    \\[0.5em]
    \includegraphics[width=\linewidth,trim={0 0 0 18mm},clip]{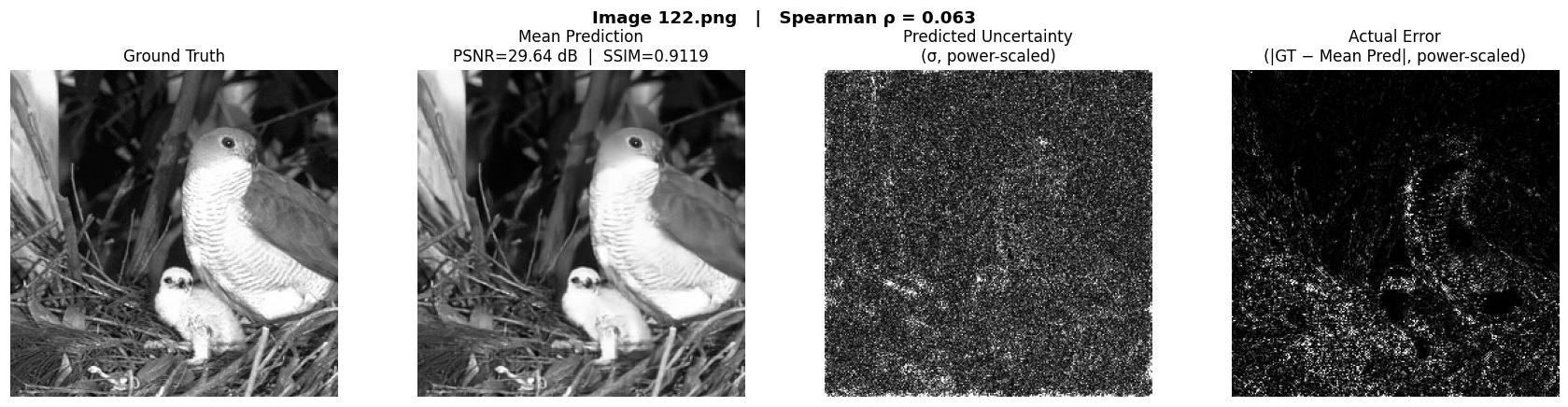}
    \includegraphics[width=\linewidth,trim={0 0 0 18mm},clip]{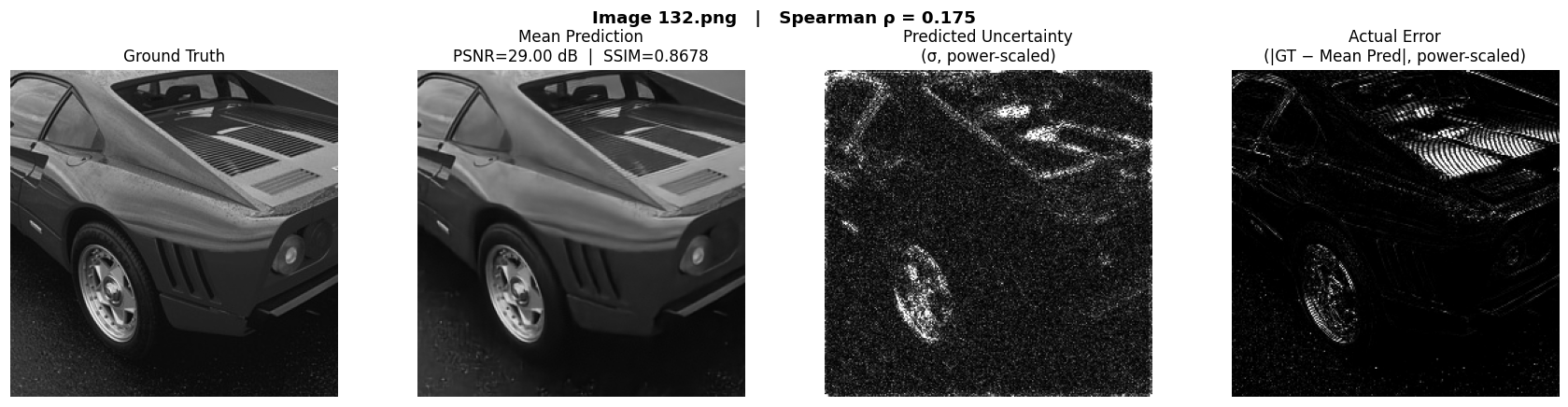}
    \includegraphics[width=\linewidth,trim={0 0 0 18mm},clip]{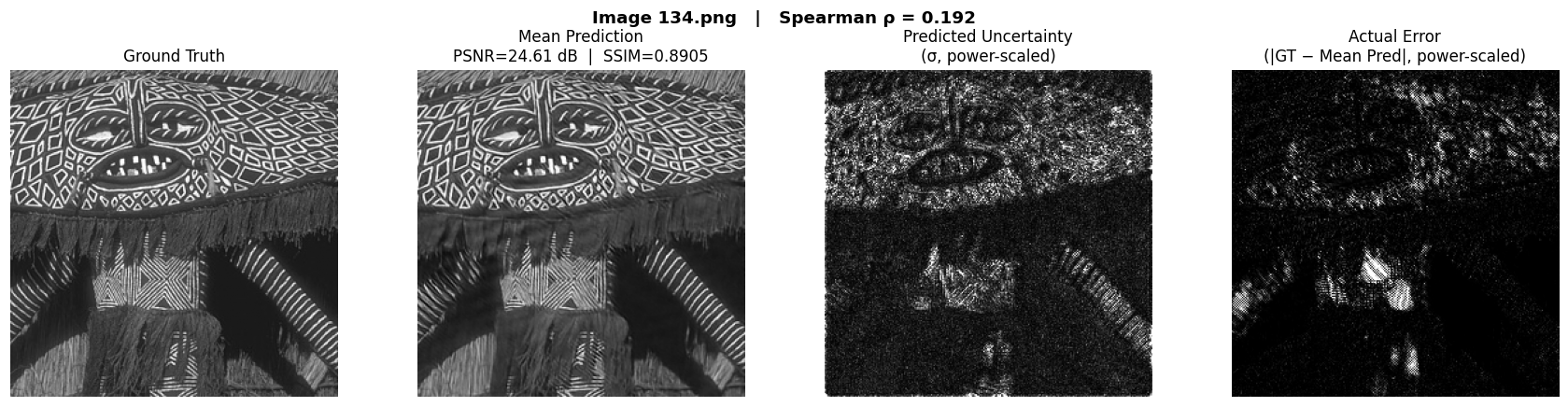}
    \includegraphics[width=\linewidth,trim={0 0 0 18mm},clip]{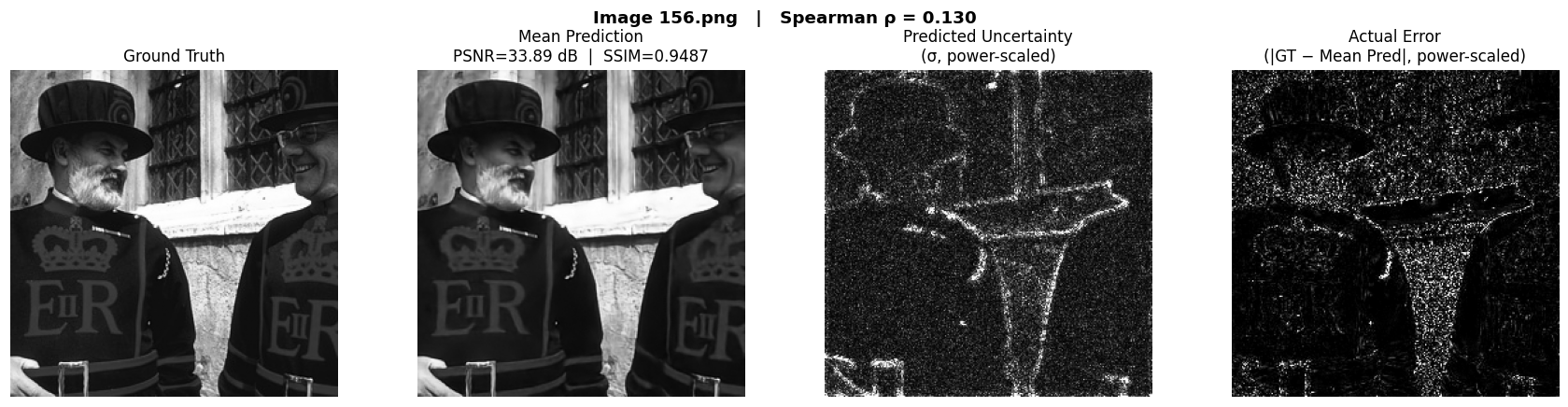}
    \includegraphics[width=\linewidth,trim={0 0 0 18mm},clip]{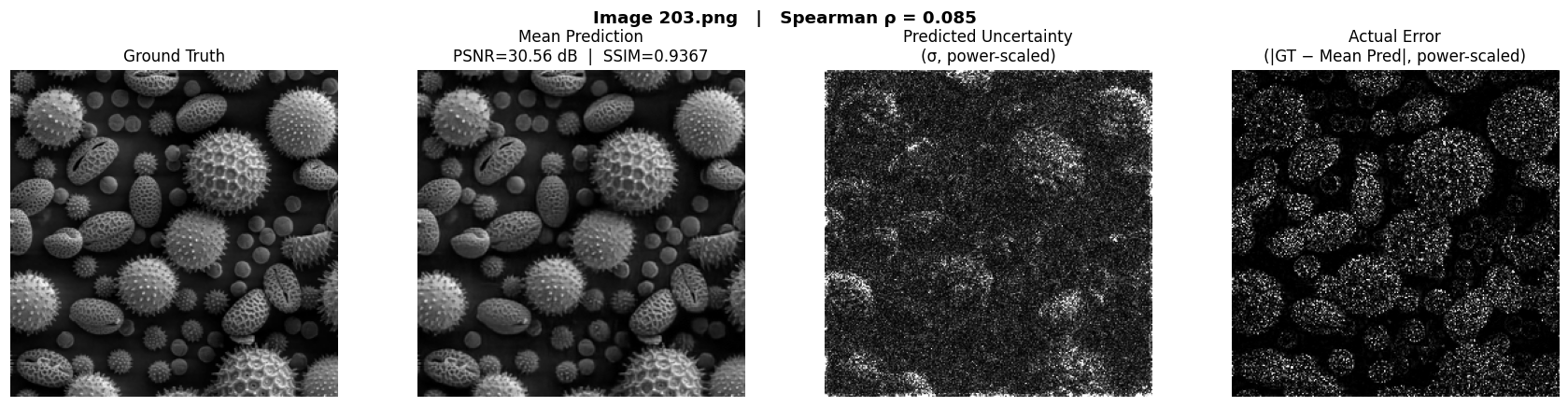}
    \includegraphics[width=\linewidth,trim={0 0 0 18mm},clip]{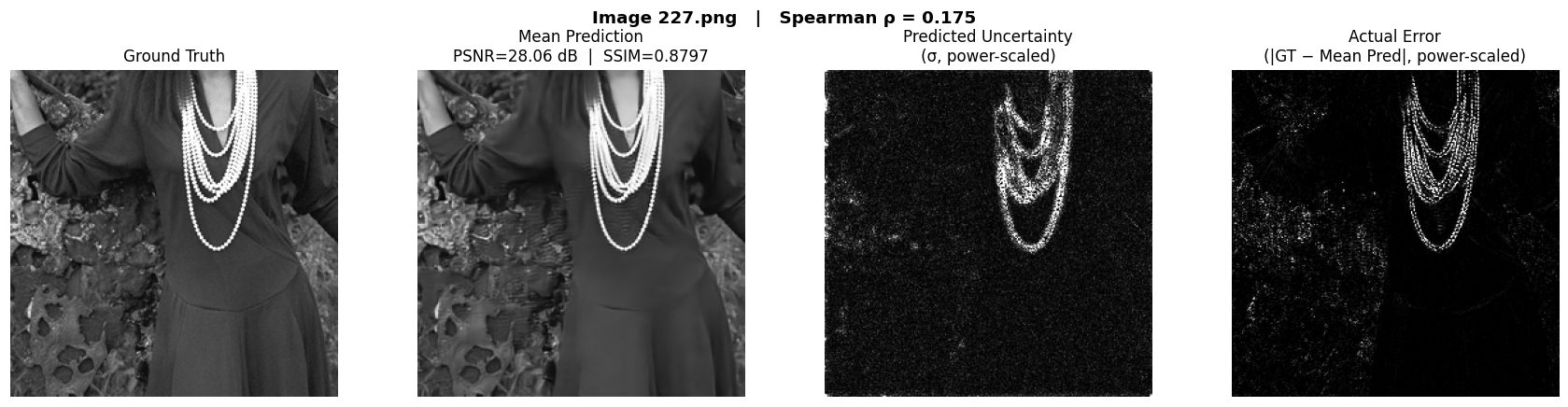}

    \caption{Uncertainty estimation with prNet-Large. The predicted pixel-wise uncertainty from posterior samples matches the actual reconstruction error.}
    \label{fig:placeholder}
\end{figure}

\section*{Additional Qualitative Results}

% ============================================================
%  Figure: 83
% ============================================================
\begin{figure}[h]
    \centering
    \begin{minipage}{0.66\textwidth}
    \centering
    \subfloat[][\parbox{0.31\columnwidth}{\centering (a) Ground truth}]{%
        \includegraphics[width=0.31\columnwidth]{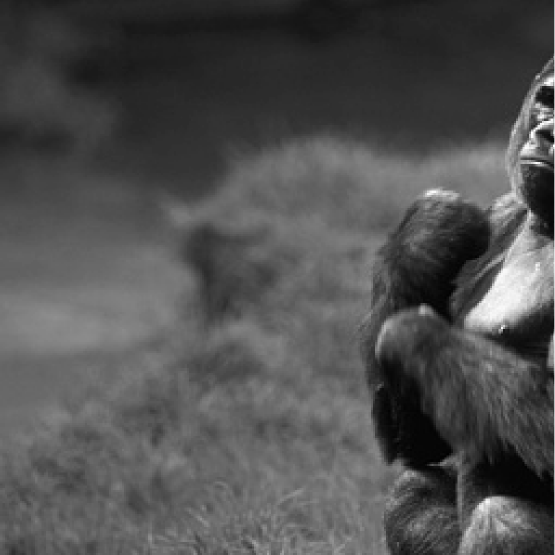}
    }
    \hfill
    \subfloat[][\parbox{0.31\columnwidth}{\centering (b) Initialization stage\\PSNR:14.99,SSIM:0.33}]{%
        \includegraphics[width=0.31\columnwidth]{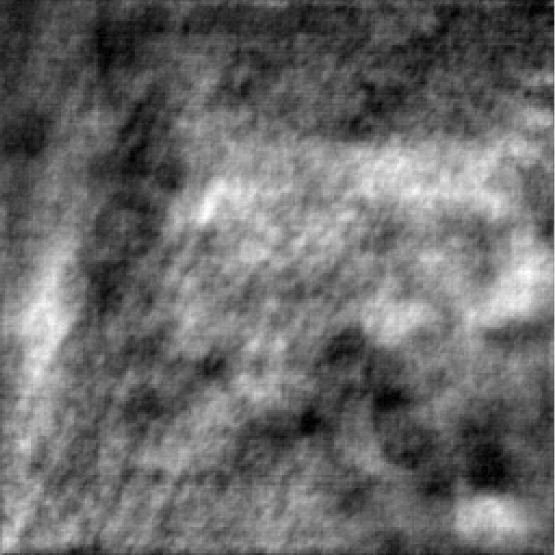}
    }
    \hfill
    \subfloat[][\parbox{0.31\columnwidth}{\centering (c) prNet-Small\\PSNR:20.26,SSIM:0.75}]{%
        \includegraphics[width=0.31\columnwidth]{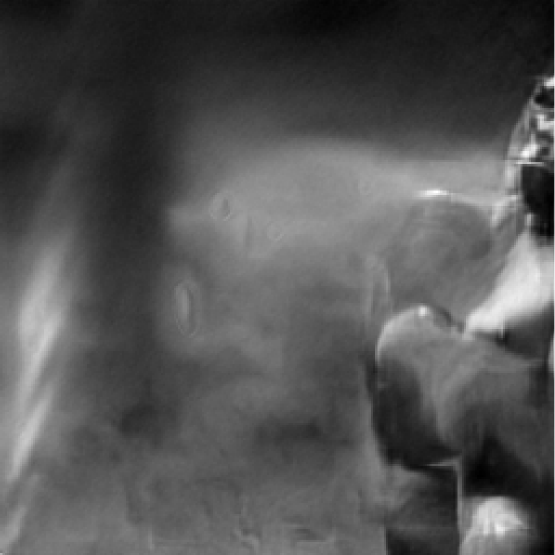}
    }

    \subfloat[][\parbox{0.31\columnwidth}{\centering (d) Main loop-Large\\PSNR:34.52,SSIM:0.94}]{%
        \includegraphics[width=0.31\columnwidth]{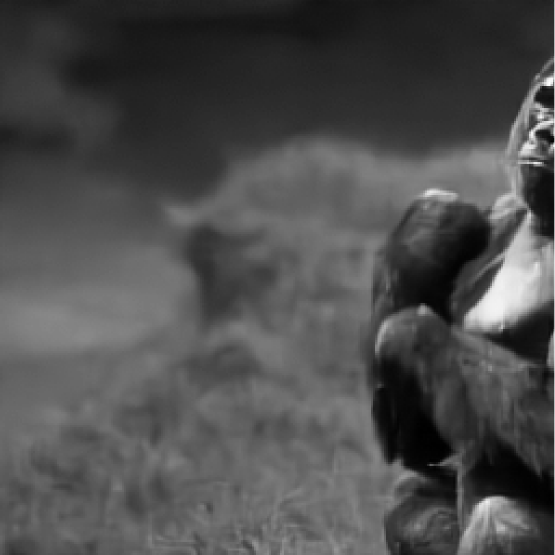}
    }
    \hfill
    \subfloat[][\parbox{0.31\columnwidth}{\centering (e) prNet-Large\\PSNR:34.64,SSIM:0.94}]{%
        \includegraphics[width=0.31\columnwidth]{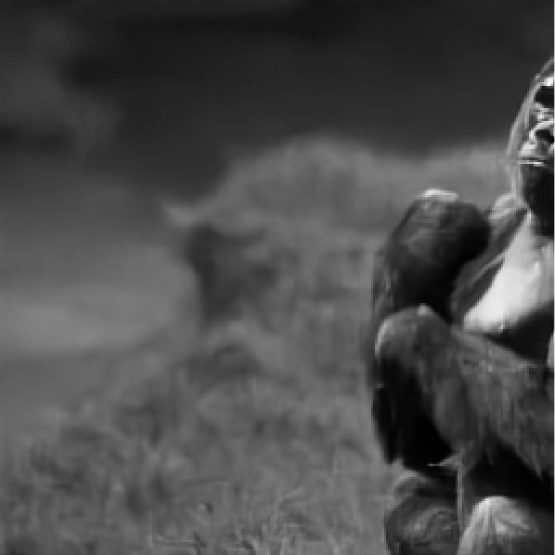}
    }
    \hfill
    \subfloat[][\parbox{0.31\columnwidth}{\centering (f) prNet-Large-Adversarial (+TTA)\\PSNR:34.94,SSIM:0.94}]{%
        \includegraphics[width=0.31\columnwidth]{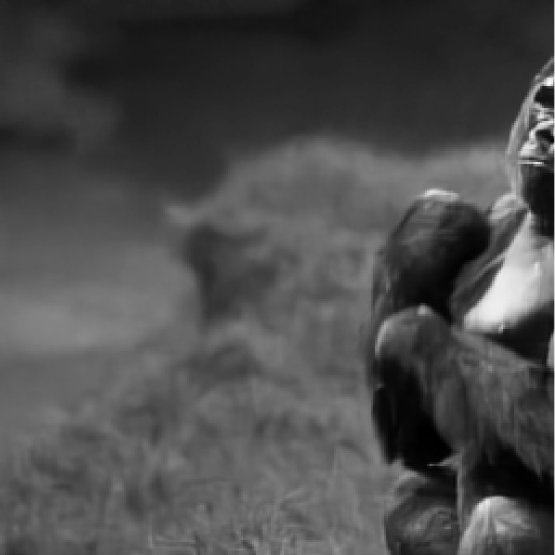}
    }
    \end{minipage}
    \caption{Intermediate reconstruction results for the ``Primate" test image at a noise level of $\alpha=3$ (SNR=30.00dB).}
    \label{fig:intermediate_83}
\end{figure}

% ============================================================
%  Figure: 182
% ============================================================
\begin{figure}[h]
    \centering
    \begin{minipage}{0.66\textwidth}
    \centering
    \subfloat[][\parbox{0.31\columnwidth}{\centering (a) Ground truth}]{%
        \includegraphics[width=0.31\columnwidth]{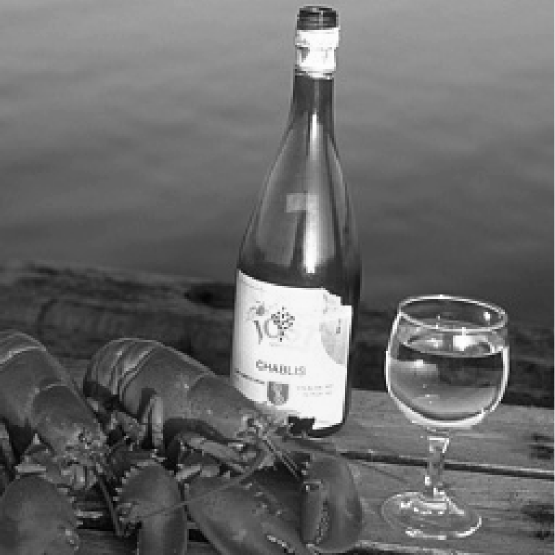}
    }
    \hfill
    \subfloat[][\parbox{0.31\columnwidth}{\centering (b) Initialization stage\\PSNR:18.39,SSIM:0.41}]{%
        \includegraphics[width=0.31\columnwidth]{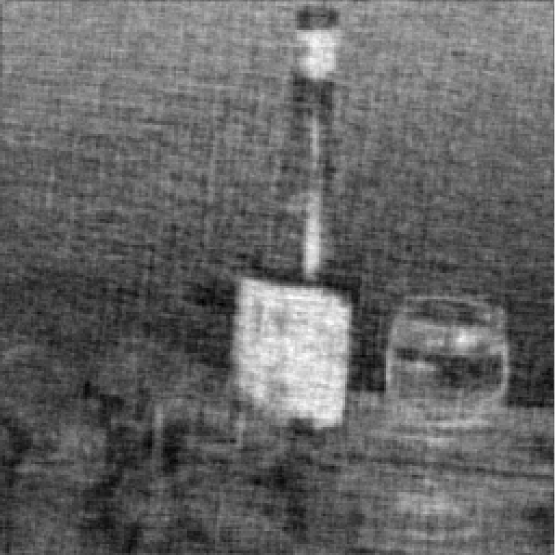}
    }
    \hfill
    \subfloat[][\parbox{0.31\columnwidth}{\centering (c) prNet-Small\\PSNR:33.59,SSIM:0.93}]{%
        \includegraphics[width=0.31\columnwidth]{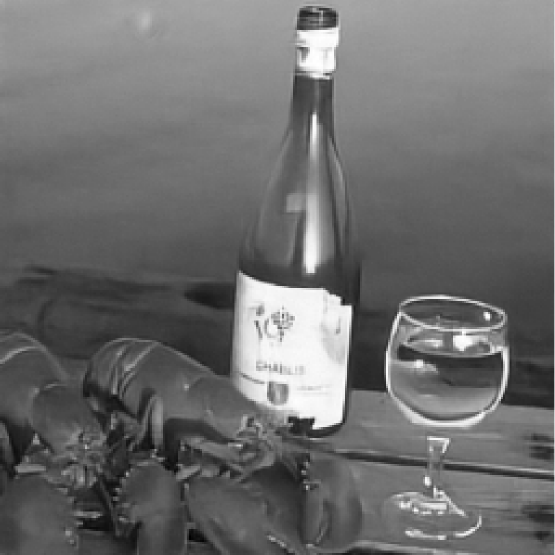}
    }

    \subfloat[][\parbox{0.31\columnwidth}{\centering (d) Main loop-Large\\PSNR:35.28,SSIM:0.95}]{%
        \includegraphics[width=0.31\columnwidth]{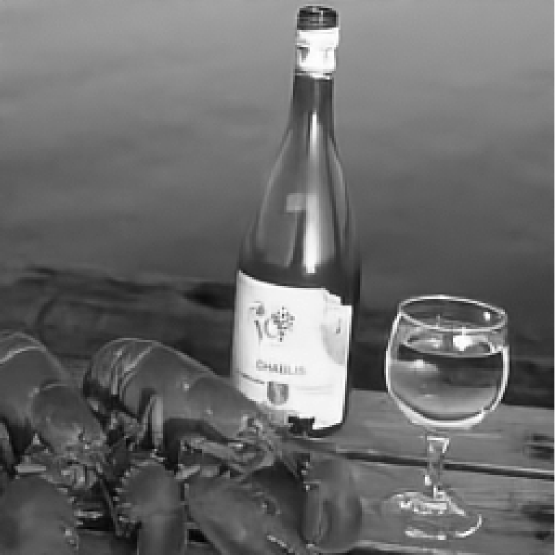}
    }
    \hfill
    \subfloat[][\parbox{0.31\columnwidth}{\centering (e) prNet-Large\\PSNR:35.34,SSIM:0.95}]{%
        \includegraphics[width=0.31\columnwidth]{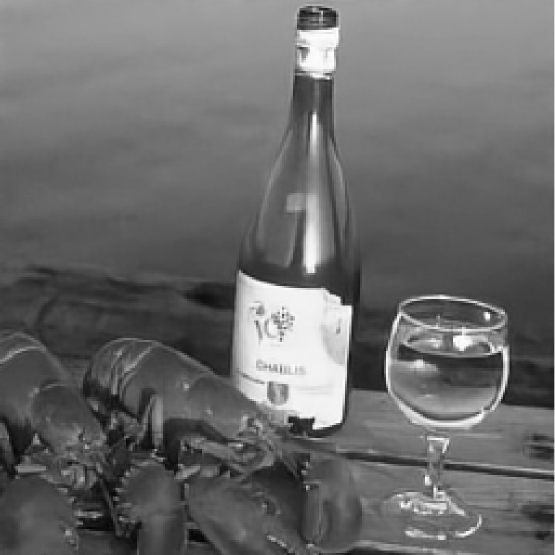}
    }
    \hfill
    \subfloat[][\parbox{0.31\columnwidth}{\centering (f) prNet-Large-Adversarial (+TTA)\\PSNR:35.51,SSIM:0.95}]{%
        \includegraphics[width=0.31\columnwidth]{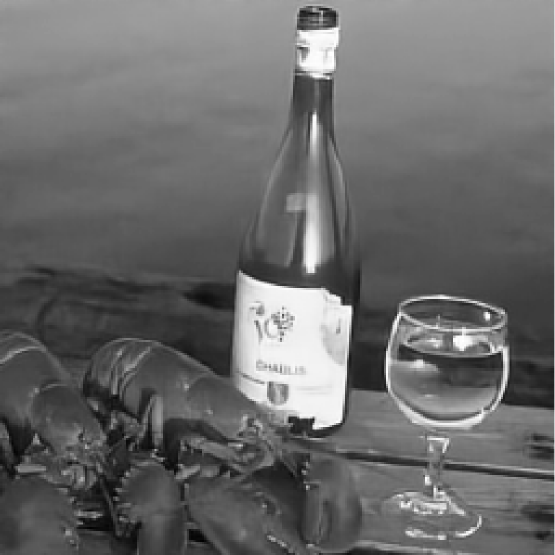}
    }
    \end{minipage}
    \caption{Intermediate reconstruction results for the ``Wine" test image at a noise level of $\alpha=3$ (SNR=31.81dB).}
    \label{fig:intermediate_182}
\end{figure}

% ============================================================
%  Figure: 157
% ============================================================
\begin{figure}[h]
    \centering
    \begin{minipage}{0.66\textwidth}
    \centering
    \subfloat[][\parbox{0.31\columnwidth}{\centering (a) Ground truth}]{%
        \includegraphics[width=0.31\columnwidth]{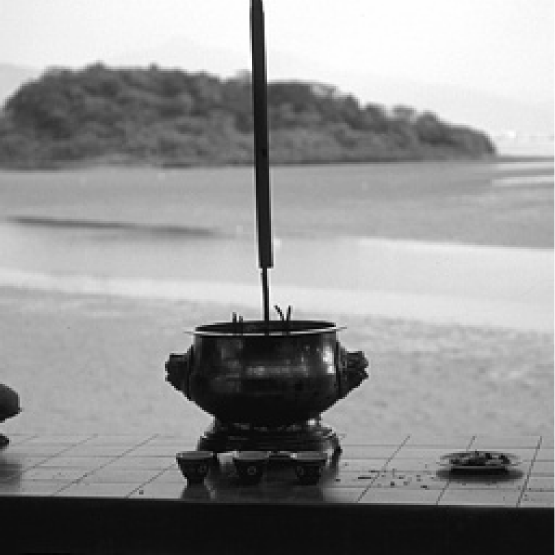}
    }
    \hfill
    \subfloat[][\parbox{0.31\columnwidth}{\centering (b) Initialization stage\\PSNR:17.11,SSIM:0.57}]{%
        \includegraphics[width=0.31\columnwidth]{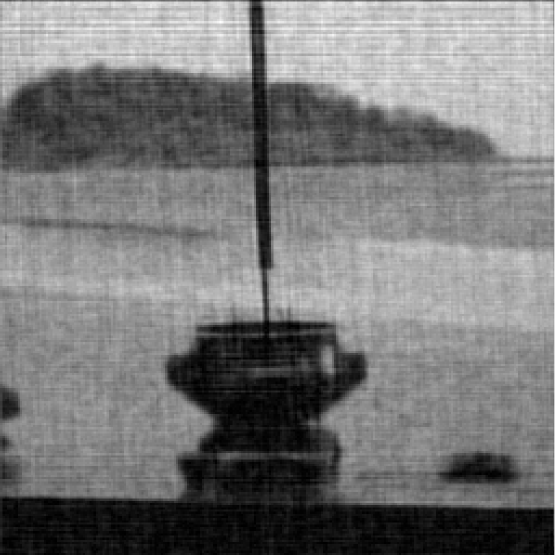}
    }
    \hfill
    \subfloat[][\parbox{0.31\columnwidth}{\centering (c) prNet-Small\\PSNR:36.02,SSIM:0.95}]{%
        \includegraphics[width=0.31\columnwidth]{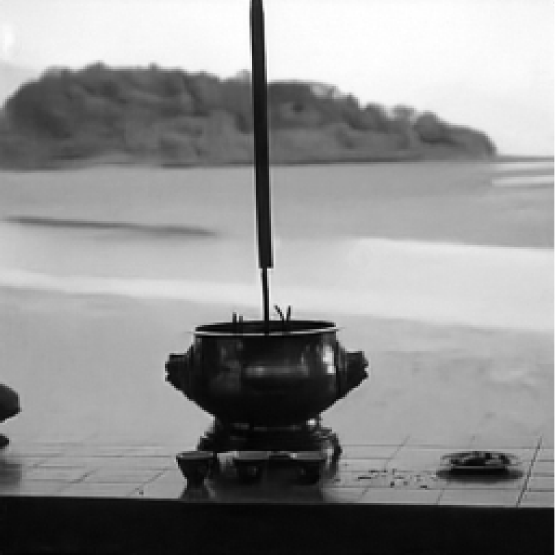}
    }

    \subfloat[][\parbox{0.31\columnwidth}{\centering (d) Main loop-Large\\PSNR:37.92,SSIM:0.96}]{%
        \includegraphics[width=0.31\columnwidth]{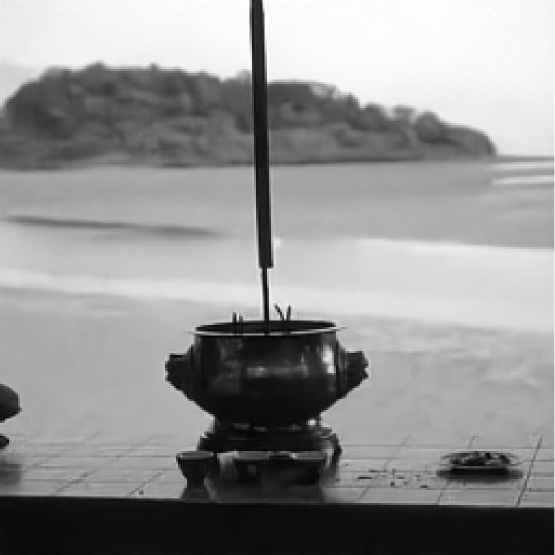}
    }
    \hfill
    \subfloat[][\parbox{0.31\columnwidth}{\centering (e) prNet-Large\\PSNR:38.02,SSIM:0.96}]{%
        \includegraphics[width=0.31\columnwidth]{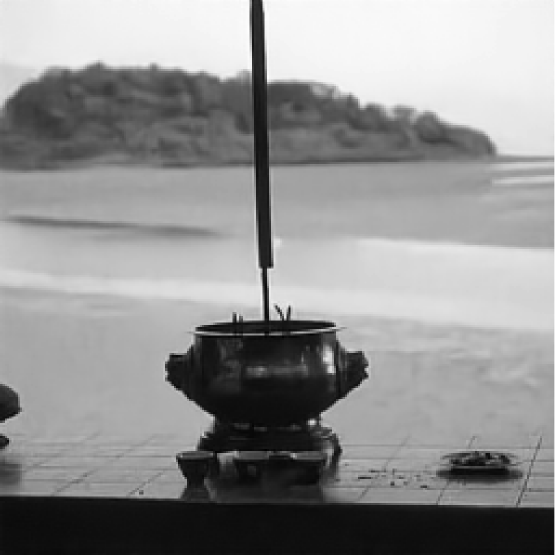}
    }
    \hfill
    \subfloat[][\parbox{0.31\columnwidth}{\centering (f) prNet-Large-Adversarial (+TTA)\\PSNR:38.36,SSIM:0.96}]{%
        \includegraphics[width=0.31\columnwidth]{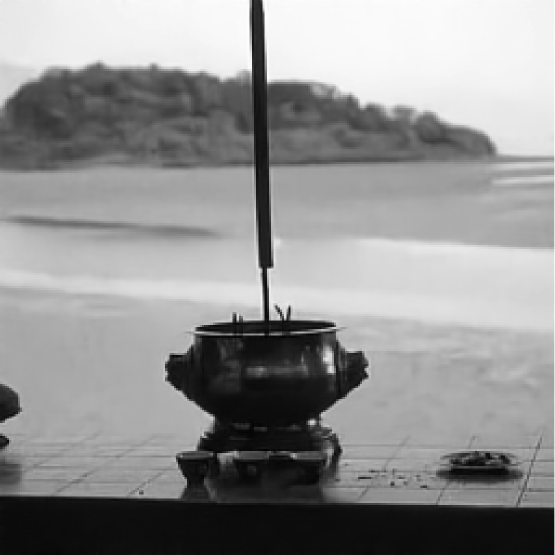}
    }
    \end{minipage}
    \caption{Intermediate reconstruction results for the ``Incensory" test image at a noise level of $\alpha=3$ (SNR=31.77dB).}
    \label{fig:intermediate_157}
\end{figure}

% ============================================================
%  Figure: 53
% ============================================================
\begin{figure}[h]
    \centering
    \begin{minipage}{0.66\textwidth}
    \centering
    \subfloat[][\parbox{0.31\columnwidth}{\centering (a) Ground truth}]{%
        \includegraphics[width=0.31\columnwidth]{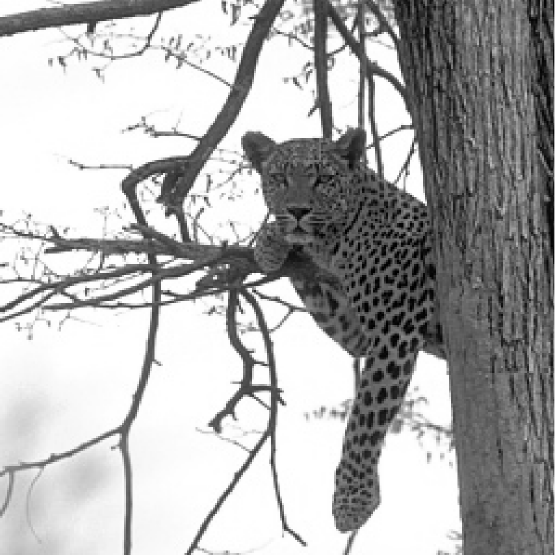}
    }
    \hfill
    \subfloat[][\parbox{0.31\columnwidth}{\centering (b) Initialization stage\\PSNR:12.52,SSIM:0.44}]{%
        \includegraphics[width=0.31\columnwidth]{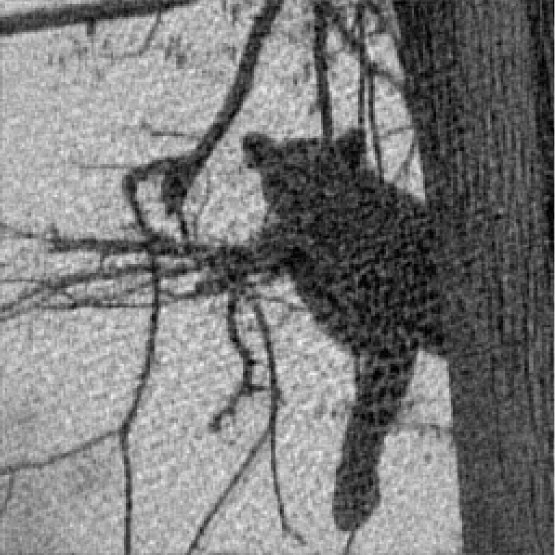}
    }
    \hfill
    \subfloat[][\parbox{0.31\columnwidth}{\centering (c) prNet-Small\\PSNR:30.76,SSIM:0.95}]{%
        \includegraphics[width=0.31\columnwidth]{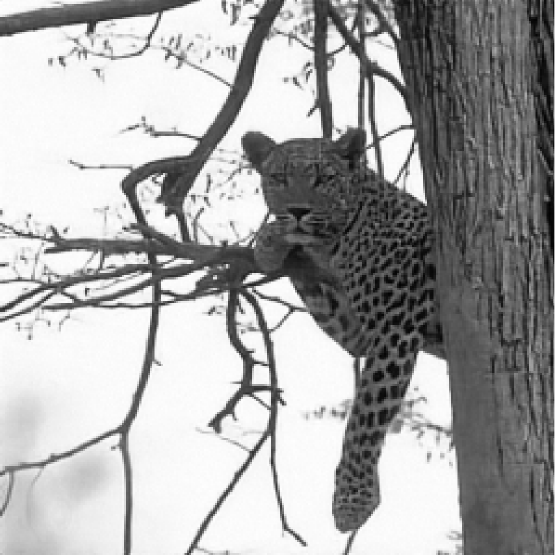}
    }

    \subfloat[][\parbox{0.31\columnwidth}{\centering (d) Main loop-Large\\PSNR:32.19,SSIM:0.96}]{%
        \includegraphics[width=0.31\columnwidth]{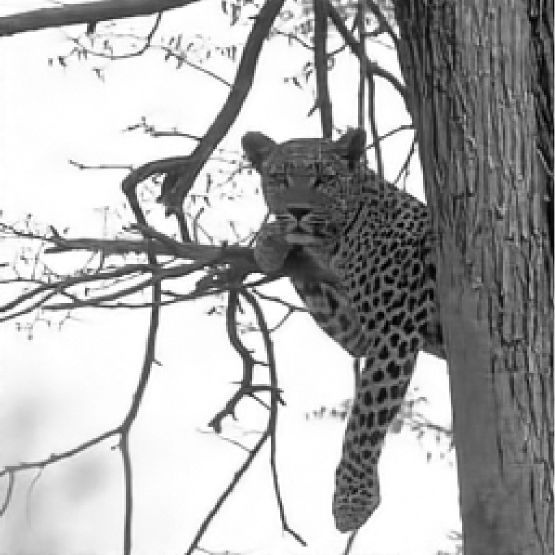}
    }
    \hfill
    \subfloat[][\parbox{0.31\columnwidth}{\centering (e) prNet-Large\\PSNR:32.37,SSIM:0.96}]{%
        \includegraphics[width=0.31\columnwidth]{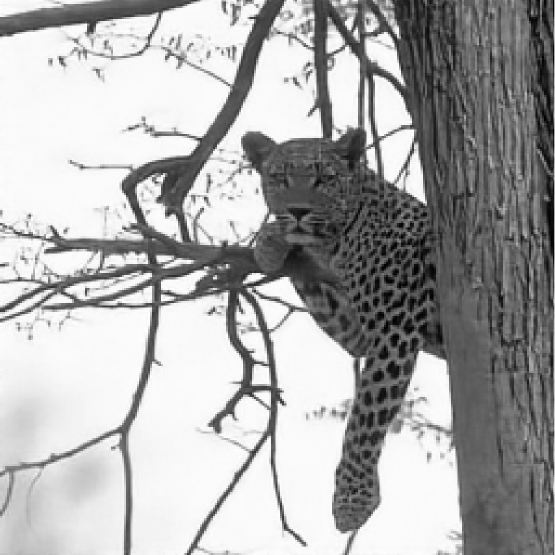}
    }
    \hfill
    \subfloat[][\parbox{0.31\columnwidth}{\centering (f) prNet-Large-Adversarial (+TTA)\\PSNR:32.60,SSIM:0.96}]{%
        \includegraphics[width=0.31\columnwidth]{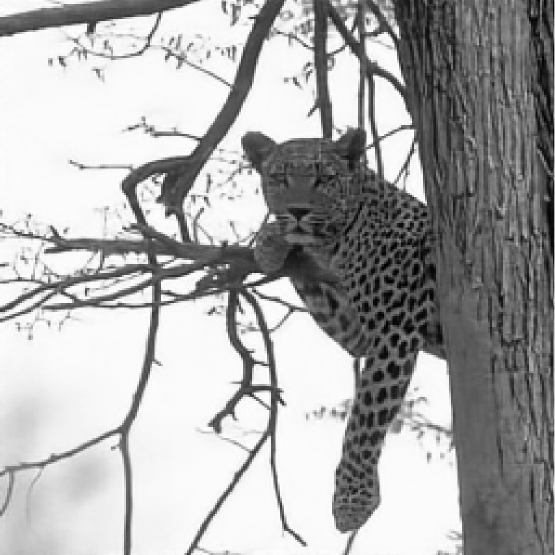}
    }
    \end{minipage}
    \caption{Intermediate reconstruction results for the ``Leopard" test image at a noise level of $\alpha=3$ (SNR=34.24dB).}
    \label{fig:intermediate_53}
\end{figure}

% ============================================================
%  Figure: 177
% ============================================================
\begin{figure}[h]
    \centering
    \begin{minipage}{0.66\textwidth}
    \centering
    \subfloat[][\parbox{0.31\columnwidth}{\centering (a) Ground truth}]{%
        \includegraphics[width=0.31\columnwidth]{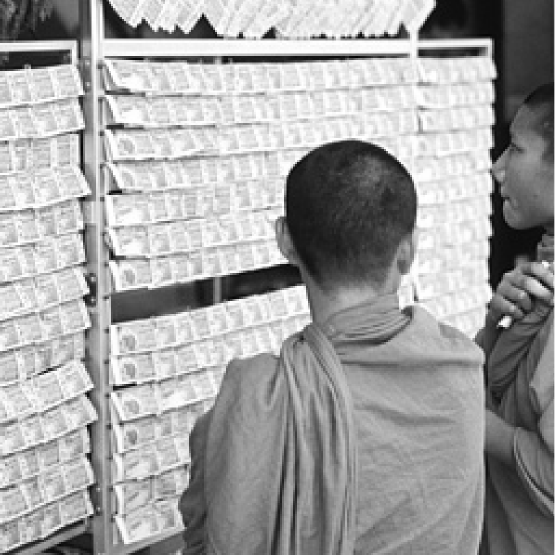}
    }
    \hfill
    \subfloat[][\parbox{0.31\columnwidth}{\centering (b) Initialization stage\\PSNR:13.32,SSIM:0.44}]{%
        \includegraphics[width=0.31\columnwidth]{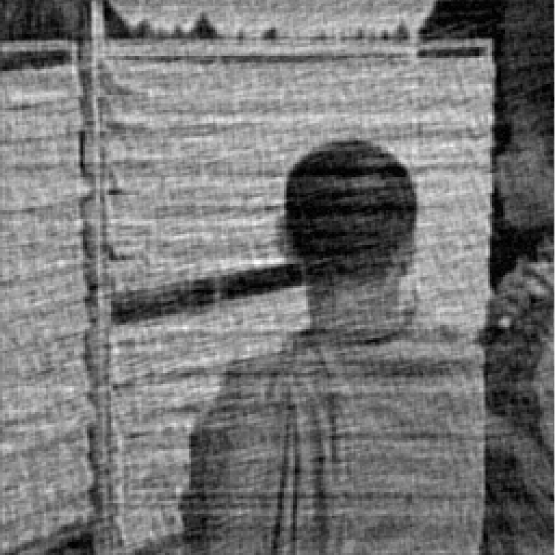}
    }
    \hfill
    \subfloat[][\parbox{0.31\columnwidth}{\centering (c) prNet-Small\\PSNR:29.26,SSIM:0.89}]{%
        \includegraphics[width=0.31\columnwidth]{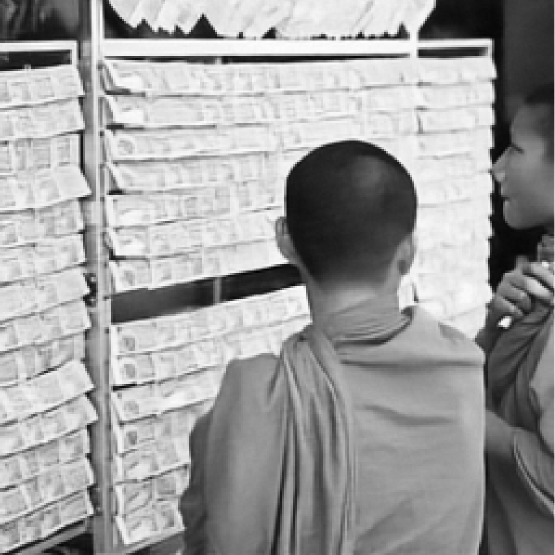}
    }

    \subfloat[][\parbox{0.31\columnwidth}{\centering (d) Main loop-Large\\PSNR:31.61,SSIM:0.93}]{%
        \includegraphics[width=0.31\columnwidth]{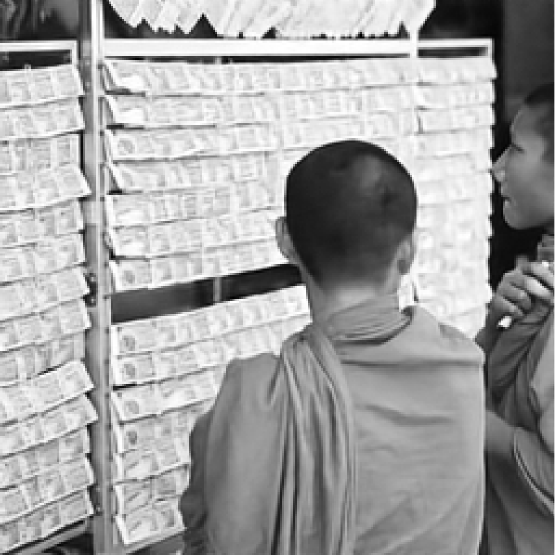}
    }
    \hfill
    \subfloat[][\parbox{0.31\columnwidth}{\centering (e) prNet-Large\\PSNR:31.67,SSIM:0.93}]{%
        \includegraphics[width=0.31\columnwidth]{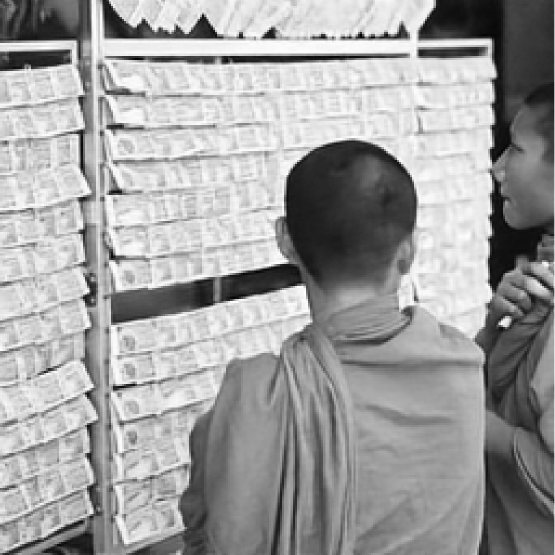}
    }
    \hfill
    \subfloat[][\parbox{0.31\columnwidth}{\centering (f) prNet-Large-Adversarial (+TTA)\\PSNR:31.77,SSIM:0.93}]{%
        \includegraphics[width=0.31\columnwidth]{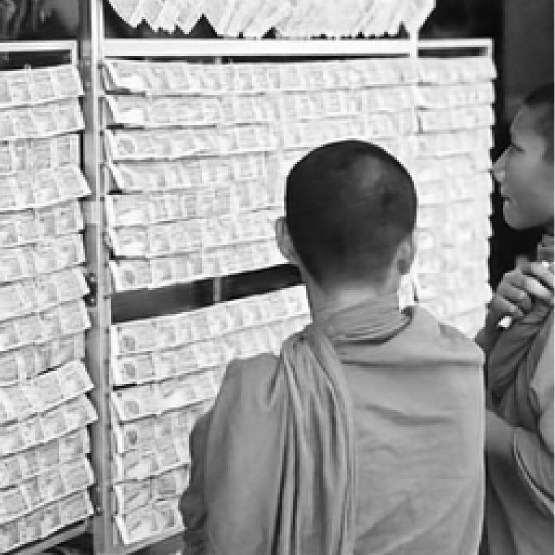}
    }
    \end{minipage}
    \caption{Intermediate reconstruction results for the ``Monks" test image at a noise level of $\alpha=3$ (SNR=34.04dB).}
    \label{fig:intermediate_177}
\end{figure}

% ============================================================
%  Figure: 44
% ============================================================
\begin{figure}[h]
    \centering
    \begin{minipage}{0.66\textwidth}
    \centering
    \subfloat[][\parbox{0.31\columnwidth}{\centering (a) Ground truth}]{%
        \includegraphics[width=0.31\columnwidth]{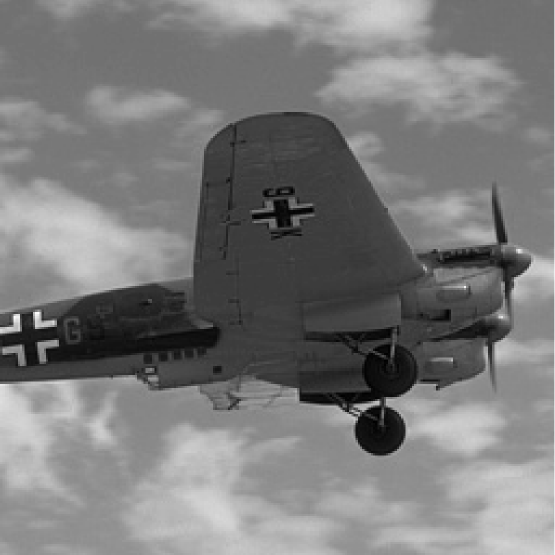}
    }
    \hfill
    \subfloat[][\parbox{0.31\columnwidth}{\centering (b) Initialization stage\\PSNR:24.04,SSIM:0.54}]{%
        \includegraphics[width=0.31\columnwidth]{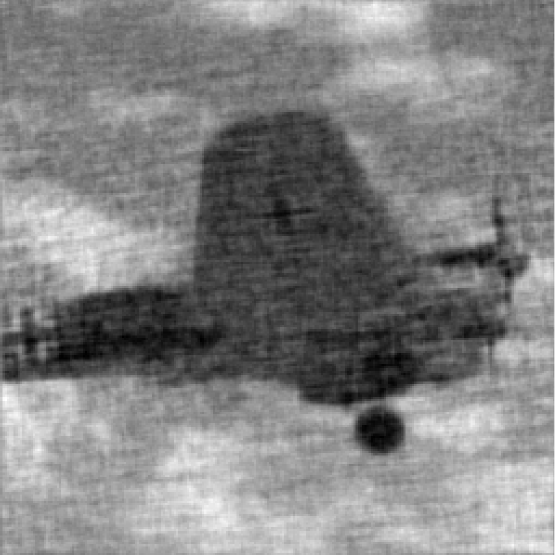}
    }
    \hfill
    \subfloat[][\parbox{0.31\columnwidth}{\centering (c) prNet-Small\\PSNR:35.71,SSIM:0.95}]{%
        \includegraphics[width=0.31\columnwidth]{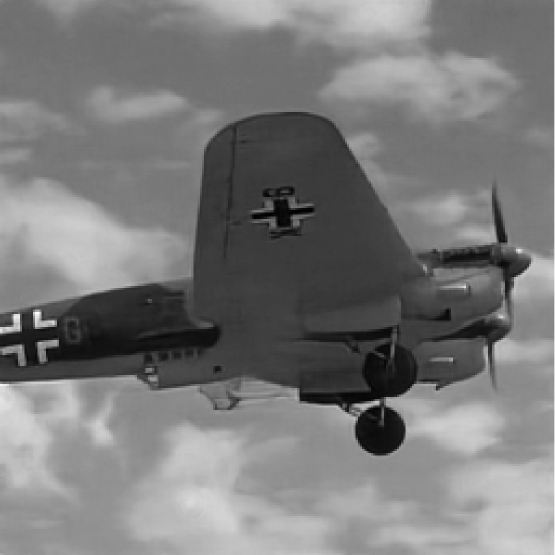}
    }

    \subfloat[][\parbox{0.31\columnwidth}{\centering (d) Main loop-Large\\PSNR:38.14,SSIM:0.96}]{%
        \includegraphics[width=0.31\columnwidth]{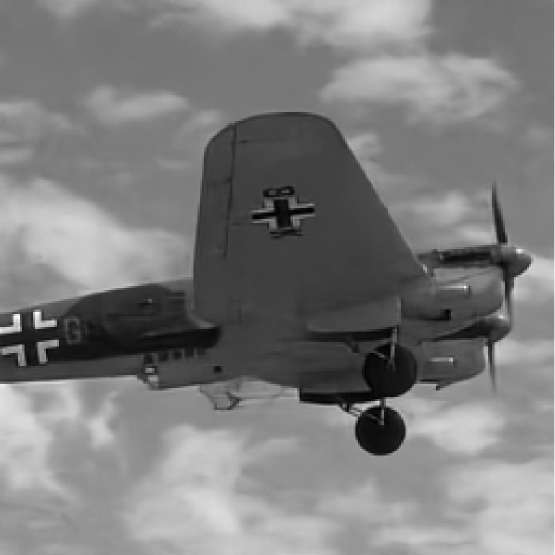}
    }
    \hfill
    \subfloat[][\parbox{0.31\columnwidth}{\centering (e) prNet-Large\\PSNR:38.15,SSIM:0.96}]{%
        \includegraphics[width=0.31\columnwidth]{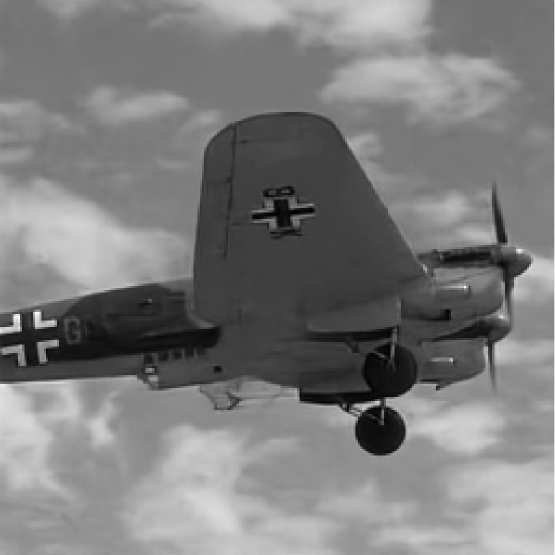}
    }
    \hfill
    \subfloat[][\parbox{0.31\columnwidth}{\centering (f) prNet-Large-Adversarial (+TTA)\\PSNR:38.29,SSIM:0.96}]{%
        \includegraphics[width=0.31\columnwidth]{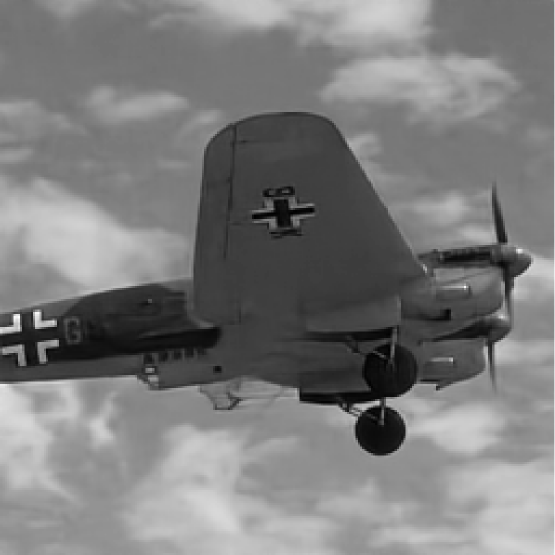}
    }
    \end{minipage}
    \caption{Intermediate reconstruction results for the ``Aircraft" test image at a noise level of $\alpha=3$ (SNR=32.16dB).}
    \label{fig:intermediate_44}
\end{figure}

% ============================================================
%  Figure: 19
% ============================================================
\begin{figure}[h]
    \centering
    \begin{minipage}{0.66\textwidth}
    \centering
    \subfloat[][\parbox{0.31\columnwidth}{\centering (a) Ground truth}]{%
        \includegraphics[width=0.31\columnwidth]{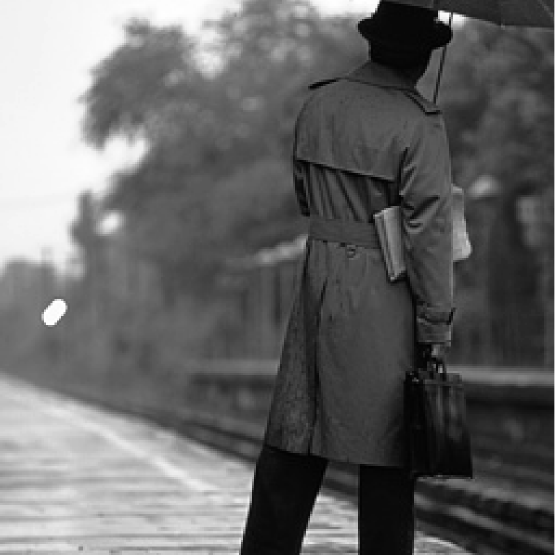}
    }
    \hfill
    \subfloat[][\parbox{0.31\columnwidth}{\centering (b) Initialization stage\\PSNR:22.88,SSIM:0.62}]{%
        \includegraphics[width=0.31\columnwidth]{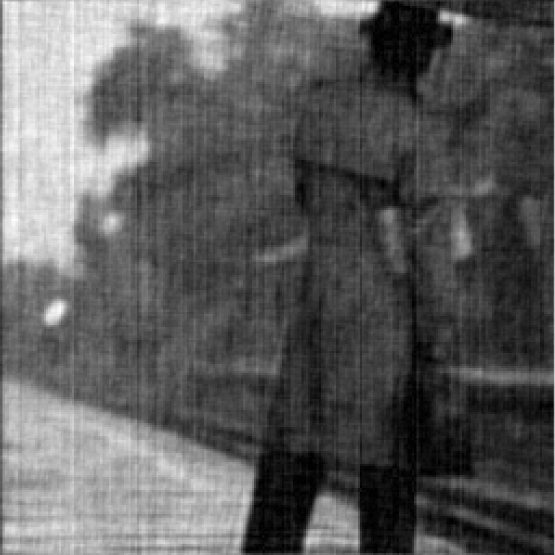}
    }
    \hfill
    \subfloat[][\parbox{0.31\columnwidth}{\centering (c) prNet-Small\\PSNR:35.95,SSIM:0.95}]{%
        \includegraphics[width=0.31\columnwidth]{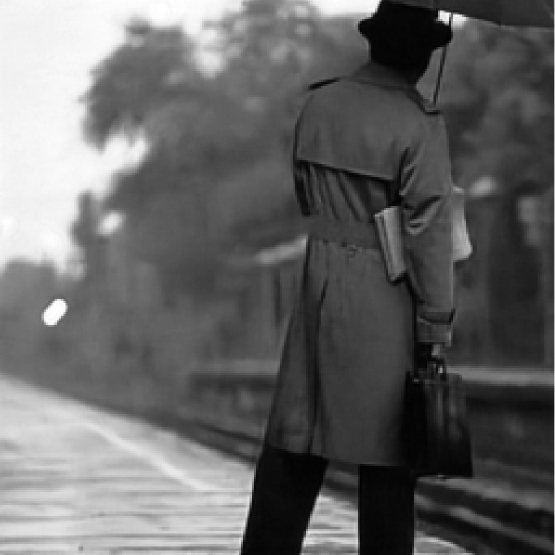}
    }

    \subfloat[][\parbox{0.31\columnwidth}{\centering (d) Main loop-Large\\PSNR:38.08,SSIM:0.96}]{%
        \includegraphics[width=0.31\columnwidth]{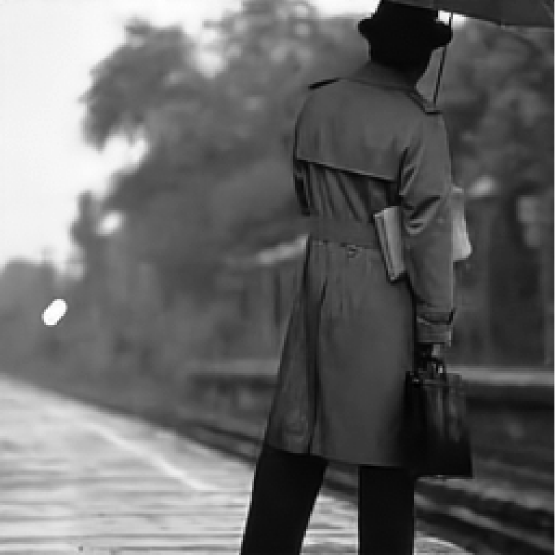}
    }
    \hfill
    \subfloat[][\parbox{0.31\columnwidth}{\centering (e) prNet-Large\\PSNR:38.16,SSIM:0.96}]{%
        \includegraphics[width=0.31\columnwidth]{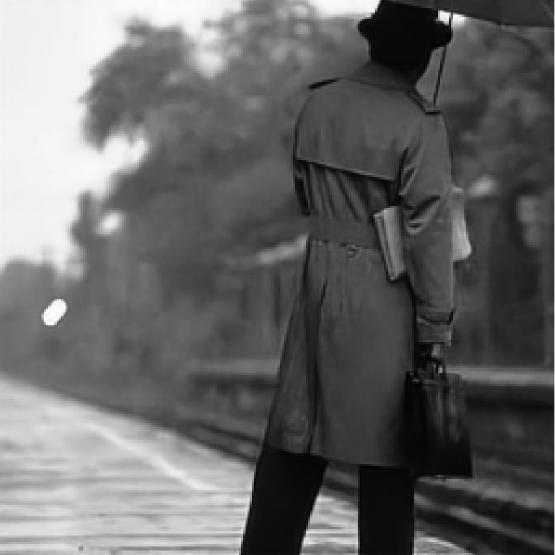}
    }
    \hfill
    \subfloat[][\parbox{0.31\columnwidth}{\centering (f) prNet-Large-Adversarial (+TTA)\\PSNR:38.43,SSIM:0.96}]{%
        \includegraphics[width=0.31\columnwidth]{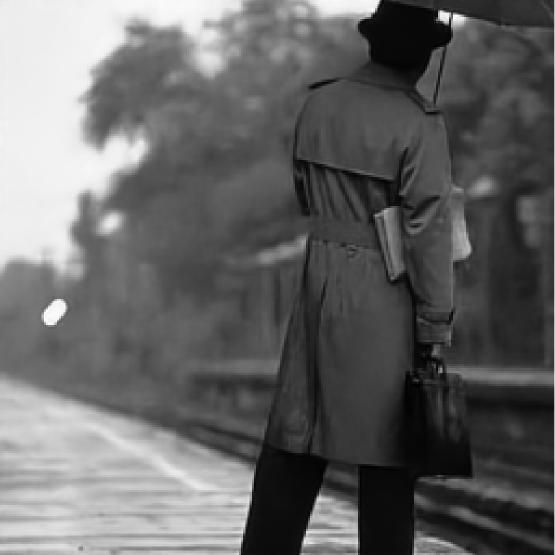}
    }
    \end{minipage}
    \caption{Intermediate reconstruction results for the ``Man" test image at a noise level of $\alpha=3$ (SNR=31.19dB).}
    \label{fig:intermediate_19}
\end{figure}

% ============================================================
%  Figure: 12
% ============================================================
\begin{figure}[h]
    \centering
    \begin{minipage}{0.66\textwidth}
    \centering
    \subfloat[][\parbox{0.31\columnwidth}{\centering (a) Ground truth}]{%
        \includegraphics[width=0.31\columnwidth]{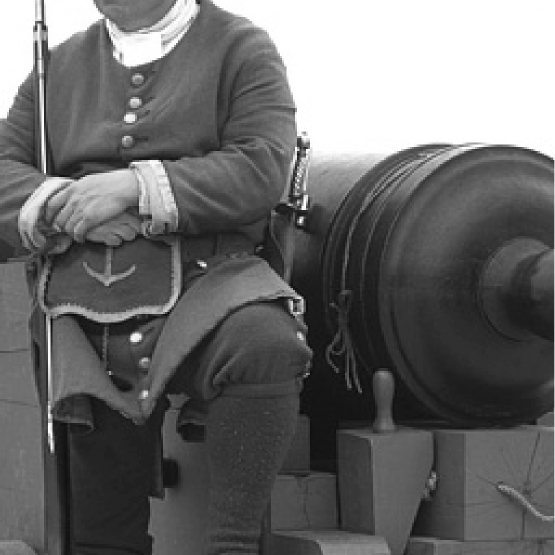}
    }
    \hfill
    \subfloat[][\parbox{0.31\columnwidth}{\centering (b) Initialization stage\\PSNR:19.20,SSIM:0.50}]{%
        \includegraphics[width=0.31\columnwidth]{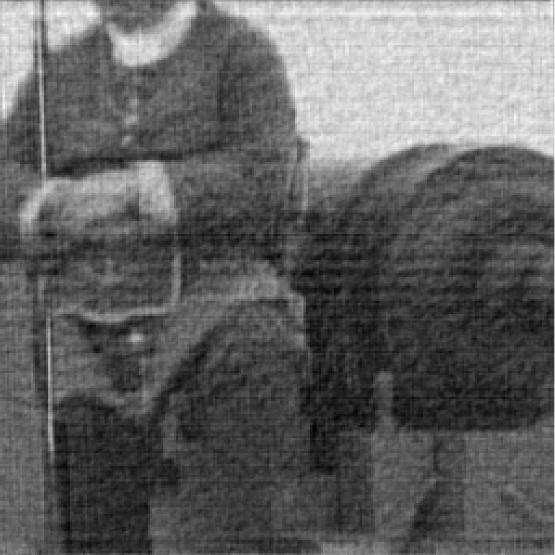}
    }
    \hfill
    \subfloat[][\parbox{0.31\columnwidth}{\centering (c) prNet-Small\\PSNR:35.42,SSIM:0.94}]{%
        \includegraphics[width=0.31\columnwidth]{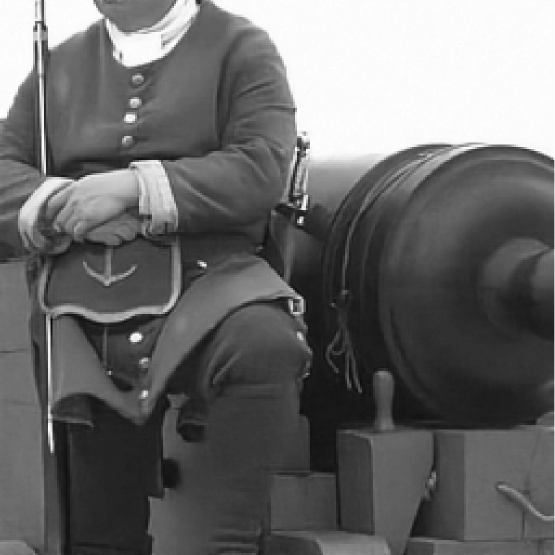}
    }

    \subfloat[][\parbox{0.31\columnwidth}{\centering (d) Main loop-Large\\PSNR:36.70,SSIM:0.95}]{%
        \includegraphics[width=0.31\columnwidth]{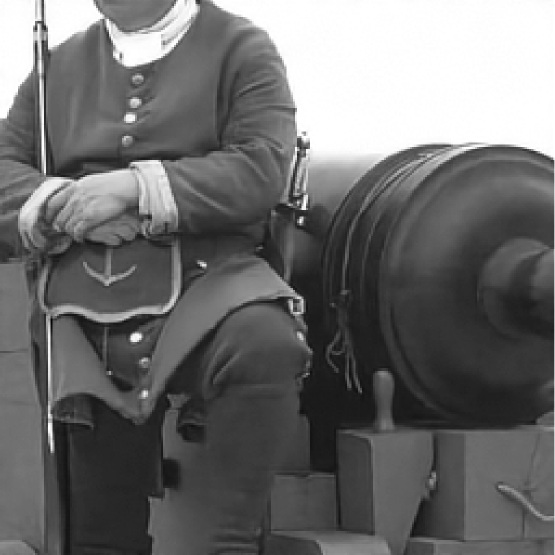}
    }
    \hfill
    \subfloat[][\parbox{0.31\columnwidth}{\centering (e) prNet-Large\\PSNR:36.71,SSIM:0.95}]{%
        \includegraphics[width=0.31\columnwidth]{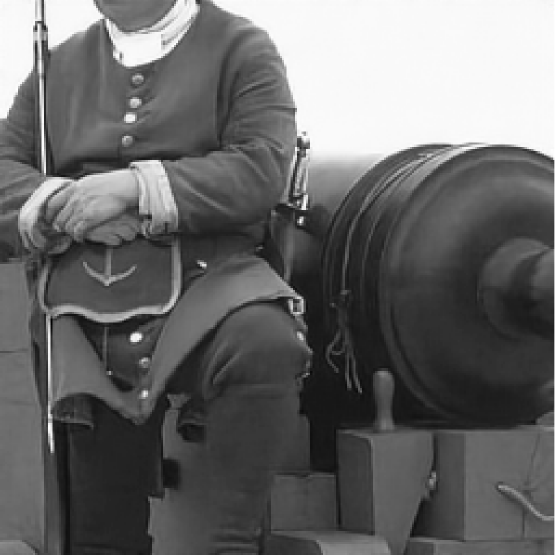}
    }
    \hfill
    \subfloat[][\parbox{0.31\columnwidth}{\centering (f) prNet-Large-Adversarial (+TTA)\\PSNR:36.96,SSIM:0.96}]{%
        \includegraphics[width=0.31\columnwidth]{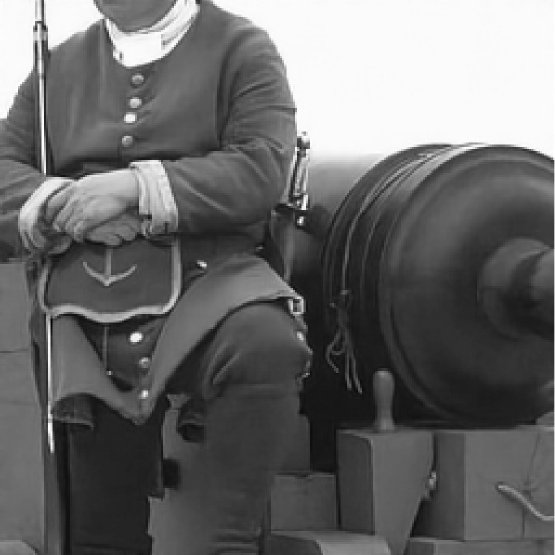}
    }
    \end{minipage}
    \caption{Intermediate reconstruction results for the ``Gunner" test image at a noise level of $\alpha=3$ (SNR=31.45dB).}
    \label{fig:intermediate_12}
\end{figure}

\makeatletter
\addtolength{\textheight}{1cm}%   new height first...
\clearpage                        %   ...then break the page
\setlength{\@fptop}{0pt}                 % pack float columns flush to the top
\setlength{\@fpsep}{10pt plus 2pt}       % small fixed gap between stacked floats
\setlength{\@fpbot}{0pt plus 1fil}       % dump all leftover space at the bottom
\makeatother
\twocolumn[{
\section*{Pixel-Wise Error Maps}
}]

\begin{figure}[!t]
    \centering

    % First row
    \subfloat[][\parbox{0.31\columnwidth}{\centering (a) Ground truth}\label{fig:error_map_gt}]{%
        \includegraphics[width=0.31\columnwidth]{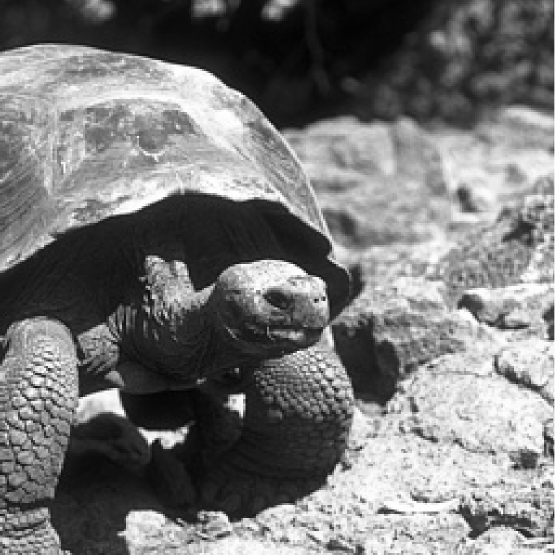}
    }
    \hfill
    \subfloat[][\parbox{0.31\columnwidth}{\centering (b) prDeep~\cite{pmlr-v80-metzler18a}\\PSNR:25.35,SSIM:0.71}\label{fig:error_map_prdeep}]{%
        \includegraphics[width=0.31\columnwidth]{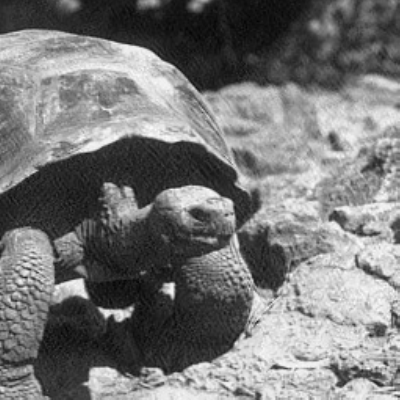}
    }
    \hfill
    \subfloat[][\parbox{0.31\columnwidth}{\centering (c) DIR~\cite{Isil:19}\\PSNR:26.49,SSIM:0.73}\label{fig:error_map_dir}]{%
        \includegraphics[width=0.31\columnwidth]{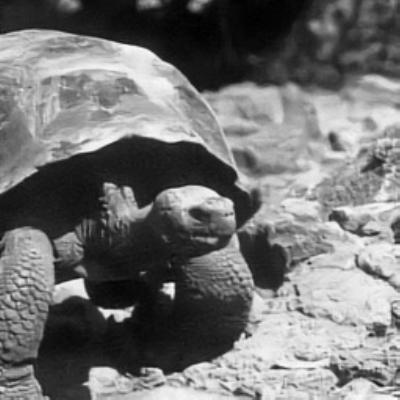}
    }

    % First row
    \subfloat[][\parbox{0.31\columnwidth}{\centering (d) DiffFPR$^{\ddagger}$~\cite{li2024difffpr}\\PSNR:25.06,SSIM:0.71}]{%
        \includegraphics[width=0.31\columnwidth]{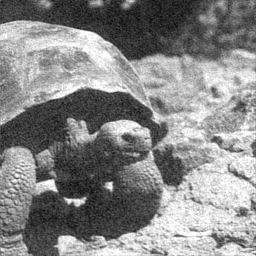}
    }
    \hfill
    \subfloat[][\parbox{0.31\columnwidth}{\centering (e) DAPS$^{\ddagger}$~\cite{zhang2025improving}\\PSNR:14.10,SSIM:0.70}\label{fig:error_map_prdeep}]{%
        \includegraphics[width=0.31\columnwidth]{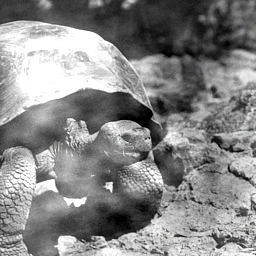}
    }
    \hfill
    \subfloat[][\parbox{0.31\columnwidth}{\centering (f) LatentDAPS$^{\ddagger}$~\cite{zhang2025improving}\\PSNR:16.73,SSIM:0.76}\label{fig:error_map_dir}]{%
        \includegraphics[width=0.31\columnwidth]{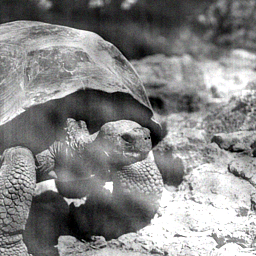}
    }
    
    % Second row
    \subfloat[][\parbox{0.31\columnwidth}{\centering (g) PnP-HIO~\cite{Isil:20}\\PSNR:27.87,SSIM:0.77}\label{fig:error_map_mbwddp}]{%
        \includegraphics[width=0.31\columnwidth]{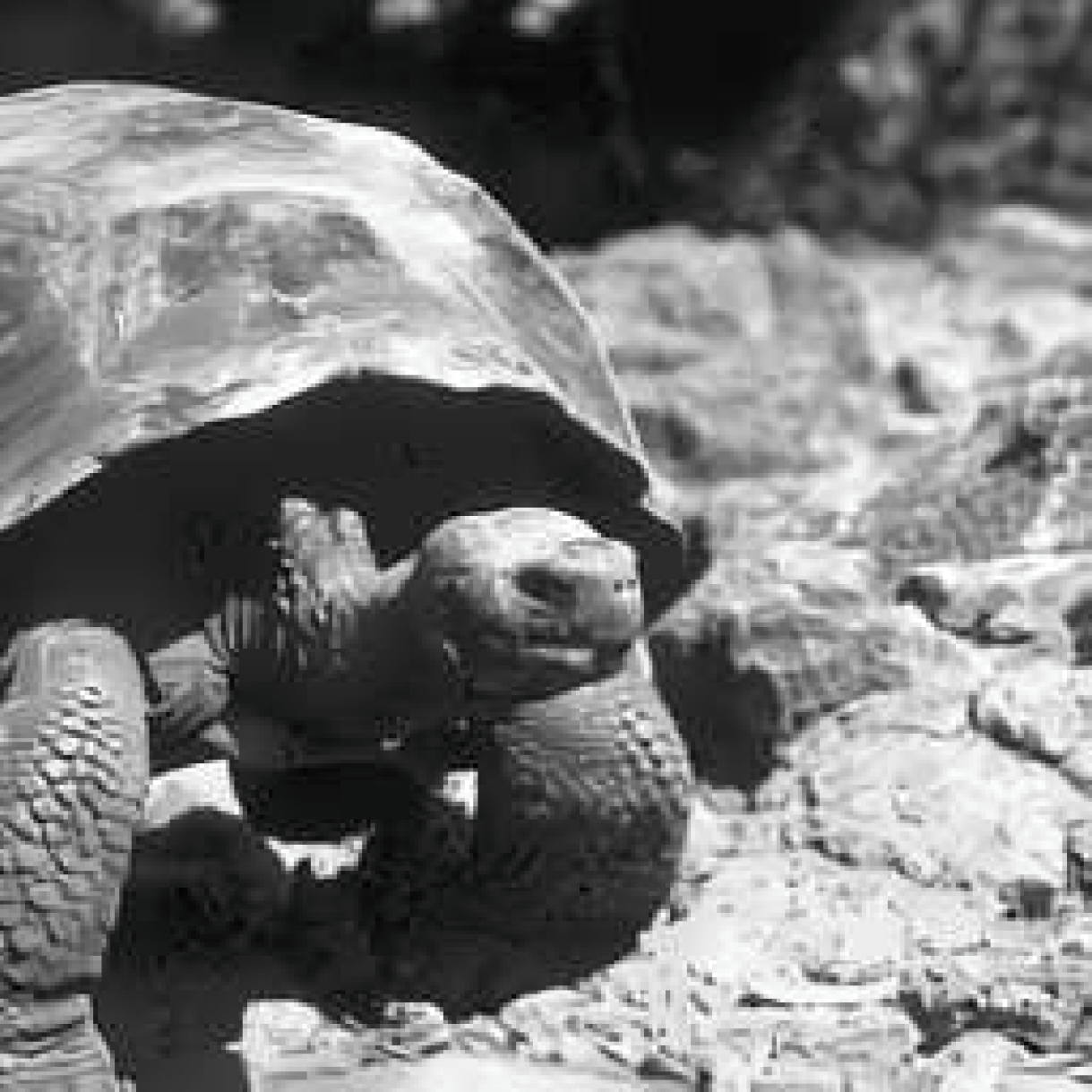}
    }
    \hfill
    \subfloat[][\parbox{0.31\columnwidth}{\centering (h) prNet-Small\\PSNR:29.74,SSIM:0.91}\label{fig:error_map_small}]{%
        \includegraphics[width=0.31\columnwidth]{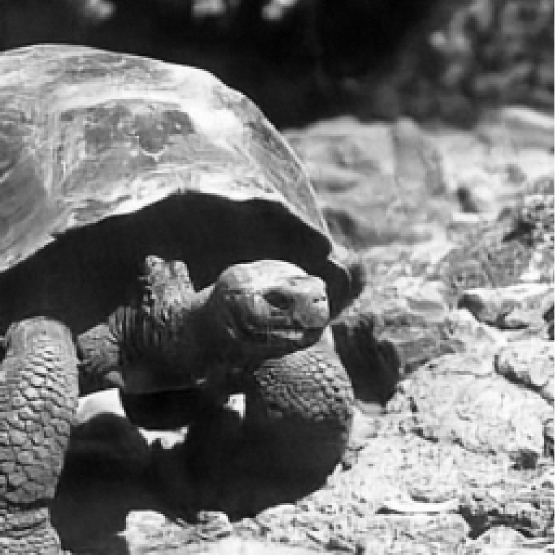}
    }
    \hfill
    \subfloat[][\parbox{0.31\columnwidth}{\centering (i) prNet-Large-Adversarial (+TTA)\\PSNR:31.31,SSIM:0.93}\label{fig:error_map_large}]{%
        \includegraphics[width=0.31\columnwidth]{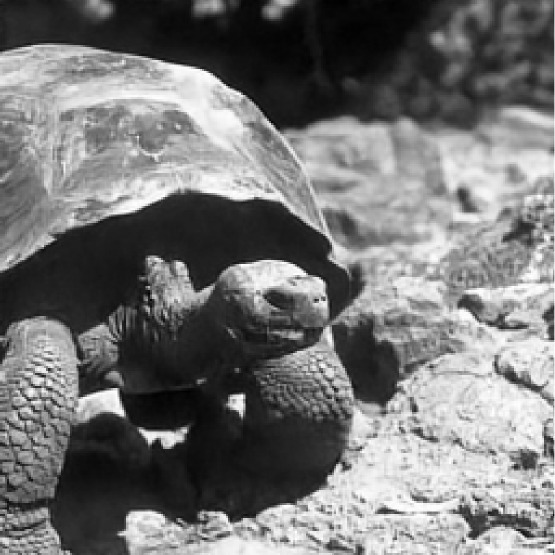}
    }

    \caption{The outputs of various algorithms for the ``Turtle" test image subjected to $\alpha=3$ noise (SNR=31.89dB).}
    \label{fig:error_map_prnettttturtle}
\end{figure}

\begin{figure}[!t]
    \centering
    \subfloat[][\parbox{0.31\columnwidth}{\centering (a) Ground truth}]{%
        \includegraphics[width=0.31\columnwidth]{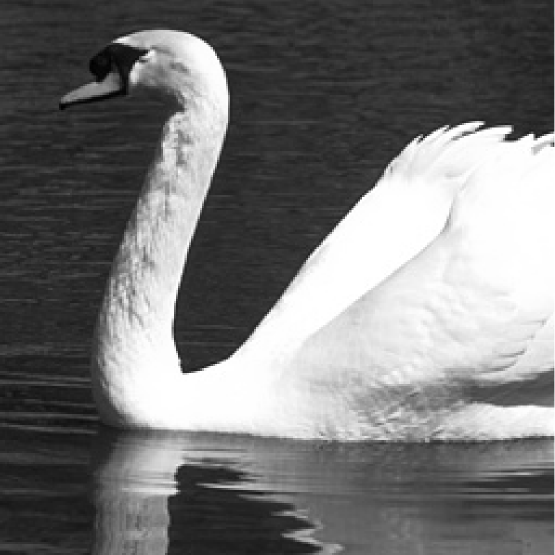}
    }
    \hfill
    \subfloat[][\parbox{0.31\columnwidth}{\centering (b) Initialization stage\\PSNR:14.28,SSIM:0.39}]{%
        \includegraphics[width=0.31\columnwidth]{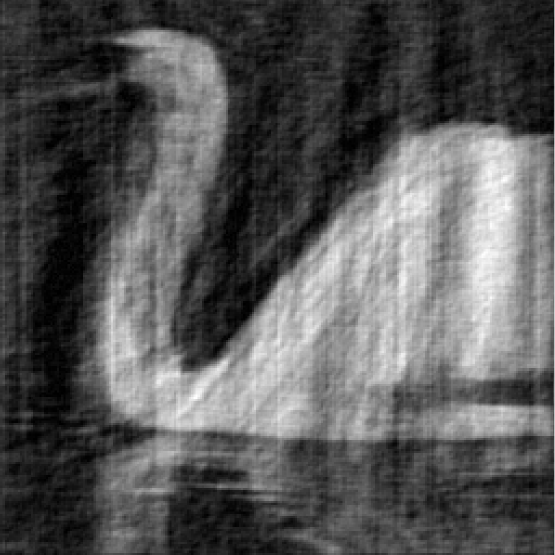}
    }
    \hfill
    \subfloat[][\parbox{0.31\columnwidth}{\centering (c) prNet-Small\\PSNR:24.97,SSIM:0.84}]{%
        \includegraphics[width=0.31\columnwidth]{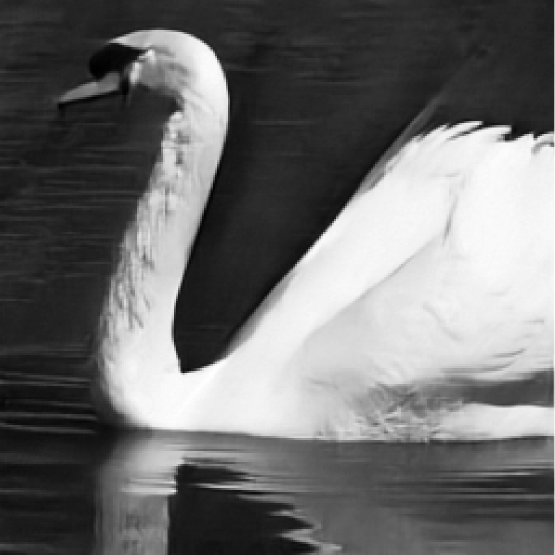}
    }

    \subfloat[][\parbox{0.31\columnwidth}{\centering (d) Main loop-Large\\PSNR:34.48,SSIM:0.93}]{%
        \includegraphics[width=0.31\columnwidth]{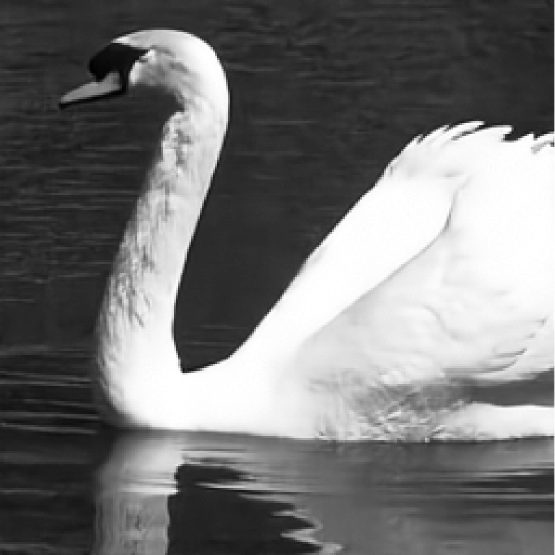}
    }
    \hfill
    \subfloat[][\parbox{0.31\columnwidth}{\centering (e) prNet-Large\\PSNR:35.02,SSIM:0.93}]{%
        \includegraphics[width=0.31\columnwidth]{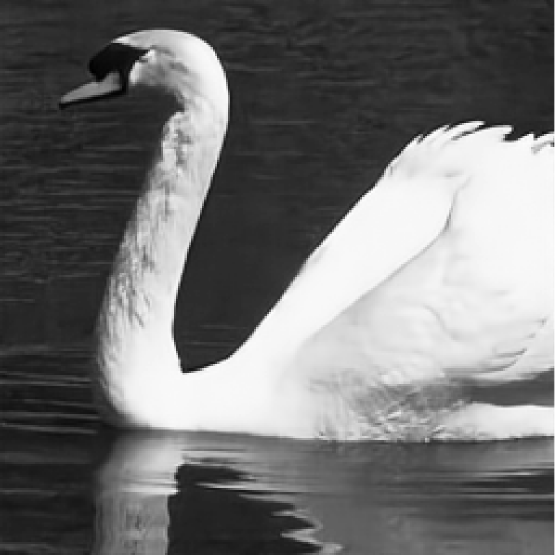}
    }
    \hfill
    \subfloat[][\parbox{0.31\columnwidth}{\centering (f) prNet-Large-Adversarial (+TTA)\\PSNR:35.21,SSIM:0.93}]{%
        \includegraphics[width=0.31\columnwidth]{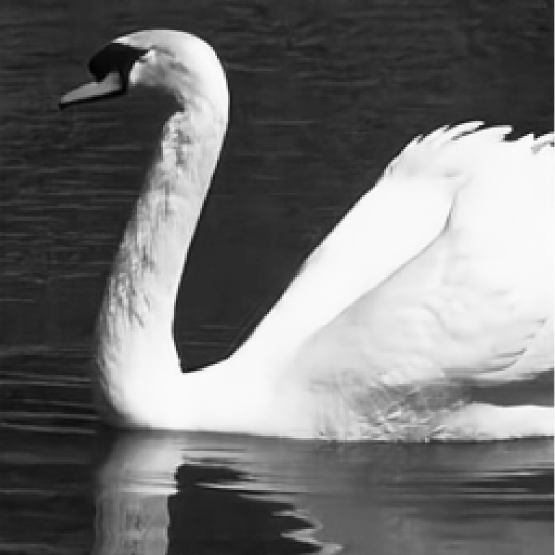}
    }
    \caption{Intermediate reconstruction results for the ``Swan" test image at a noise level of $\alpha=3$ (SNR=31.69dB).}

    \label{fig:error_map_prnettttintermediate}
\end{figure}

\begin{figure}[!t]
    \centering

    % First row
    \subfloat[][\parbox{0.31\columnwidth}{\centering (a) Ground truth}\label{fig:error_map_gt}]{%
        \includegraphics[width=0.31\columnwidth]{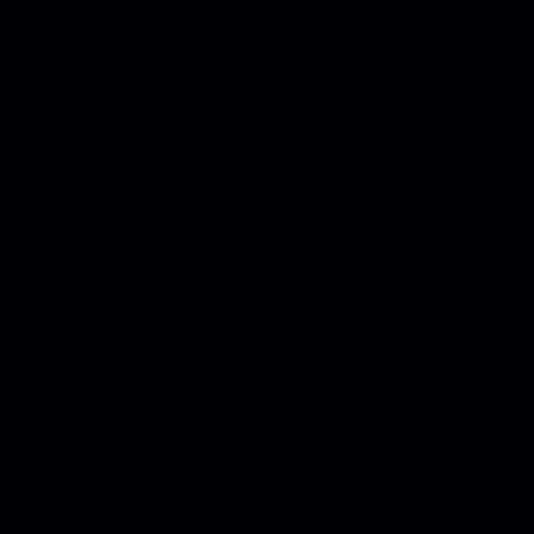}
    }
    \hfill
    \subfloat[][\parbox{0.31\columnwidth}{\centering (b) prDeep~\cite{pmlr-v80-metzler18a}\\PSNR:25.35,SSIM:0.71}\label{fig:error_map_prdeep}]{%
        \includegraphics[width=0.31\columnwidth]{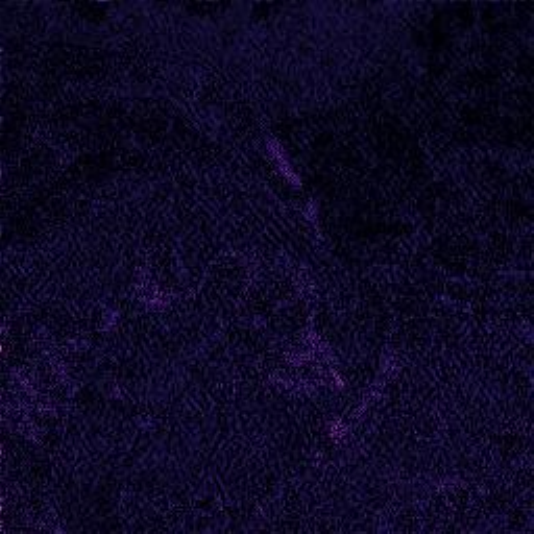}
    }
    \hfill
    \subfloat[][\parbox{0.31\columnwidth}{\centering (c) DIR~\cite{Isil:19}\\PSNR:26.49,SSIM:0.73}\label{fig:error_map_dir}]{%
        \includegraphics[width=0.31\columnwidth]{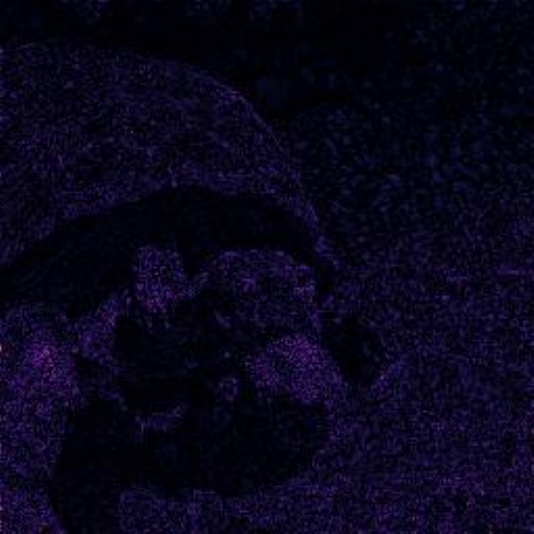}
    }

    % First row
    \subfloat[][\parbox{0.31\columnwidth}{\centering (d) DiffFPR$^{\ddagger}$~\cite{li2024difffpr}\\PSNR:25.06,SSIM:0.71}]{%
        \includegraphics[width=0.31\columnwidth]{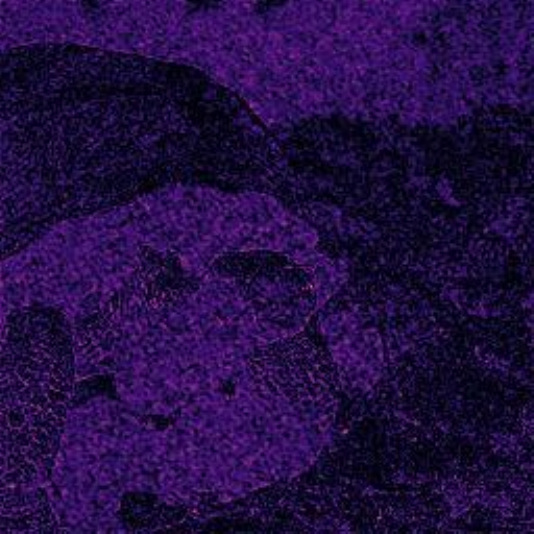}
    }
    \hfill
    \subfloat[][\parbox{0.31\columnwidth}{\centering (e) DAPS$^{\ddagger}$~\cite{zhang2025improving}\\PSNR:14.10,SSIM:0.70}\label{fig:error_map_prdeep}]{%
        \includegraphics[width=0.31\columnwidth]{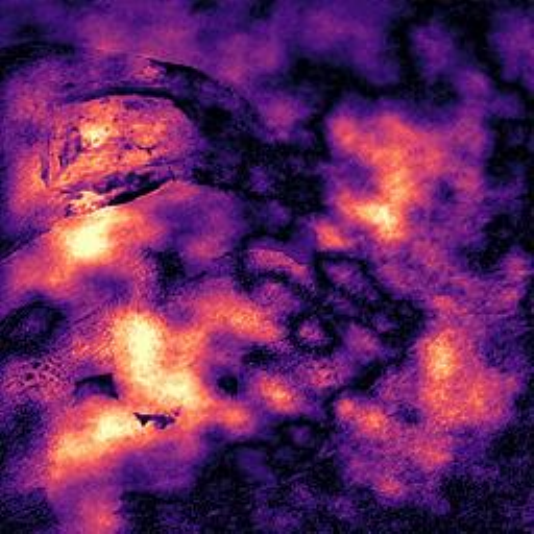}
    }
    \hfill
    \subfloat[][\parbox{0.31\columnwidth}{\centering (f) LatentDAPS$^{\ddagger}$~\cite{zhang2025improving}\\PSNR:16.73,SSIM:0.76}\label{fig:error_map_dir}]{%
        \includegraphics[width=0.31\columnwidth]{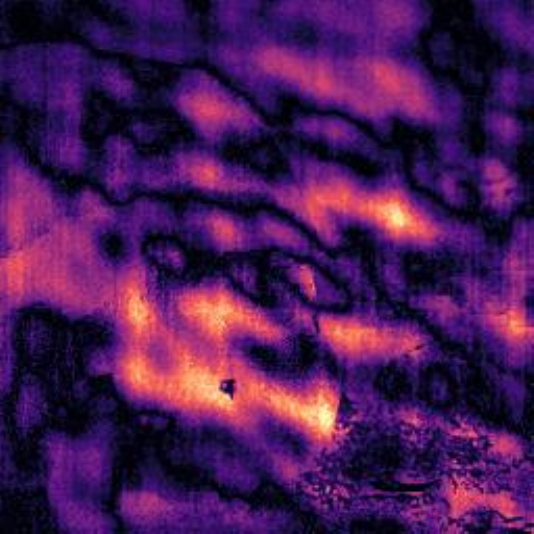}
    }
    
    % Second row
    \subfloat[][\parbox{0.31\columnwidth}{\centering (g) PnP-HIO~\cite{Isil:20}\\PSNR:27.87,SSIM:0.77}\label{fig:error_map_mbwddp}]{%
        \includegraphics[width=0.31\columnwidth]{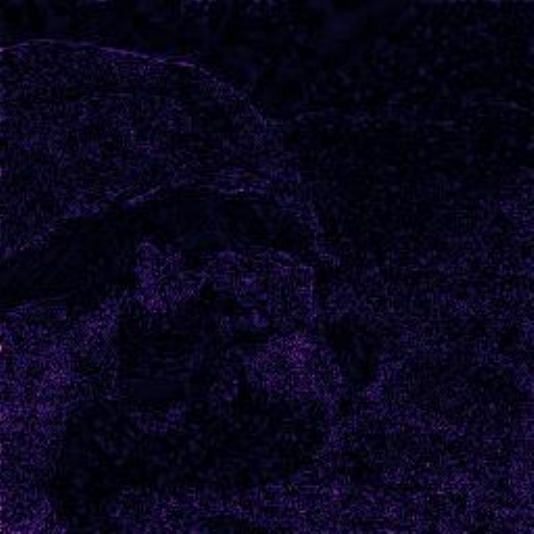}
    }
    \hfill
    \subfloat[][\parbox{0.31\columnwidth}{\centering (h) prNet-Small\\PSNR:29.74,SSIM:0.91}\label{fig:error_map_small}]{%
        \includegraphics[width=0.31\columnwidth]{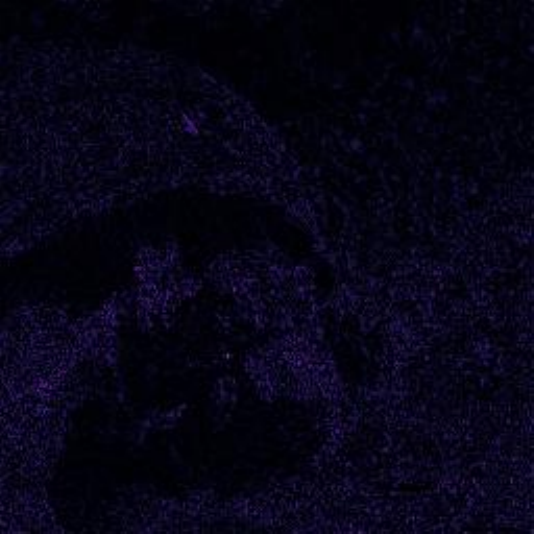}
    }
    \hfill
    \subfloat[][\parbox{0.31\columnwidth}{\centering (i) prNet-Large-Adversarial (+TTA)\\PSNR:31.31,SSIM:0.93}\label{fig:error_map_large}]{%
        \includegraphics[width=0.31\columnwidth]{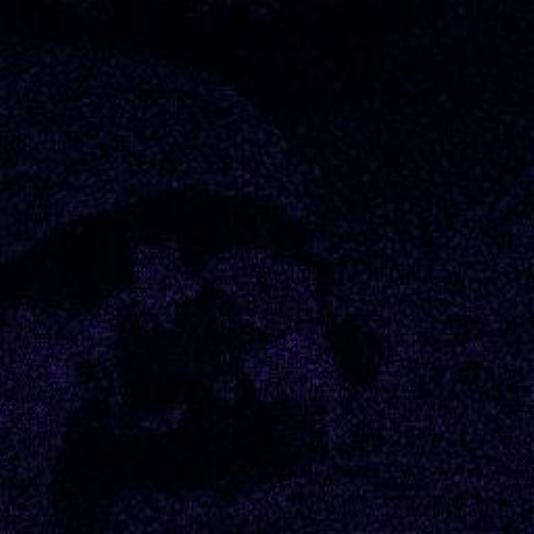}
    }

    \caption{Error maps of the reconstructions shown in Fig.~\ref{fig:error_map_prnettttturtle}. Brighter colors indicate larger absolute error.}
\end{figure}

%%%%%%%%%%%%%%%%%%%%%%%%%%%%%%%%%%%%%%%%%

\begin{figure}[!t]
    \centering
    \subfloat[][\parbox{0.31\columnwidth}{\centering (a) Ground truth}]{%
        \includegraphics[width=0.31\columnwidth]{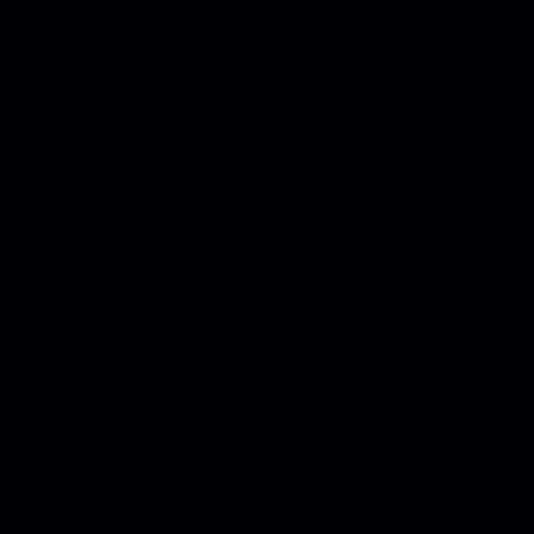}
    }
    % \subfloat[][]{%
    %     \phantom{\includegraphics[width=0.31\columnwidth]{figures/figures_errormaps_aligned/fig_3/gt.pdf}}
    % }
    \hfill
    \subfloat[][\parbox{0.31\columnwidth}{\centering (b) Initialization stage\\PSNR:14.28,SSIM:0.39}]{%
        \includegraphics[width=0.31\columnwidth]{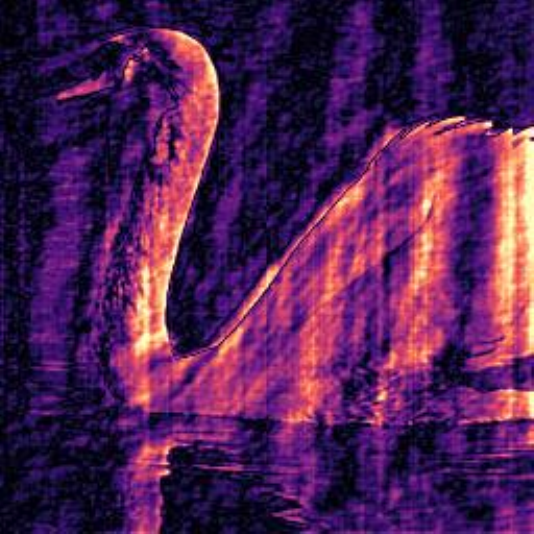}
    }
    \hfill
    \subfloat[][\parbox{0.31\columnwidth}{\centering (c) prNet-Small\\PSNR:24.97,SSIM:0.84}]{%
        \includegraphics[width=0.31\columnwidth]{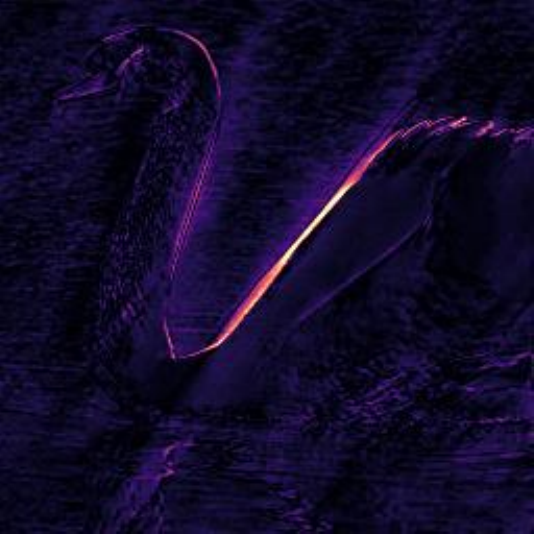}
    }

    \subfloat[][\parbox{0.31\columnwidth}{\centering (d) Main loop-Large\\PSNR:34.48,SSIM:0.93}]{%
        \includegraphics[width=0.31\columnwidth]{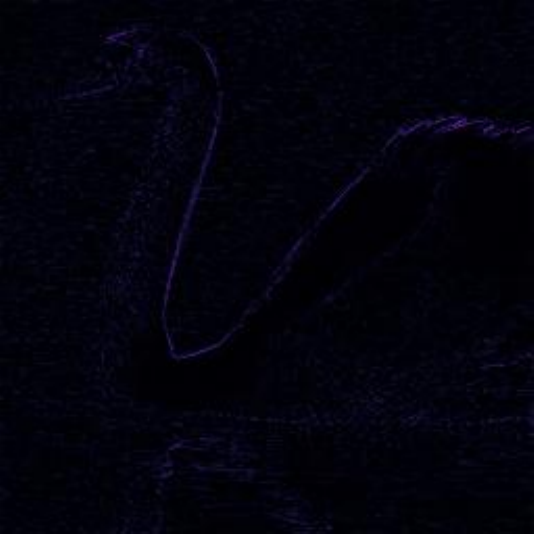}
    }
    \hfill
    \subfloat[][\parbox{0.31\columnwidth}{\centering (e) prNet-Large\\PSNR:35.02,SSIM:0.93}]{%
        \includegraphics[width=0.31\columnwidth]{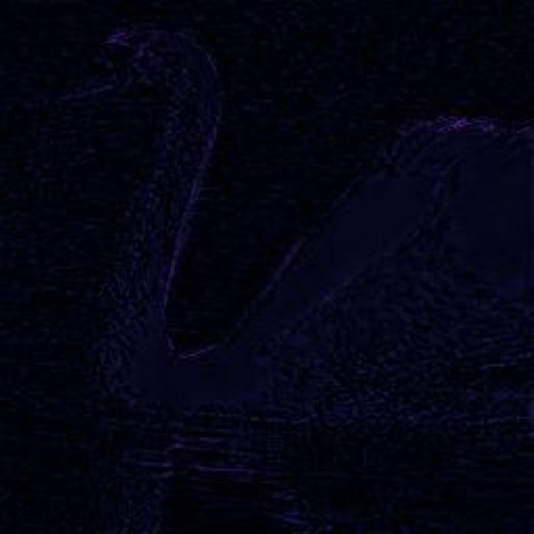}
    }
    \hfill
    \subfloat[][\parbox{0.31\columnwidth}{\centering (f) prNet-Large-Adversarial (+TTA)\\PSNR:35.21,SSIM:0.93}]{%
        \includegraphics[width=0.31\columnwidth]{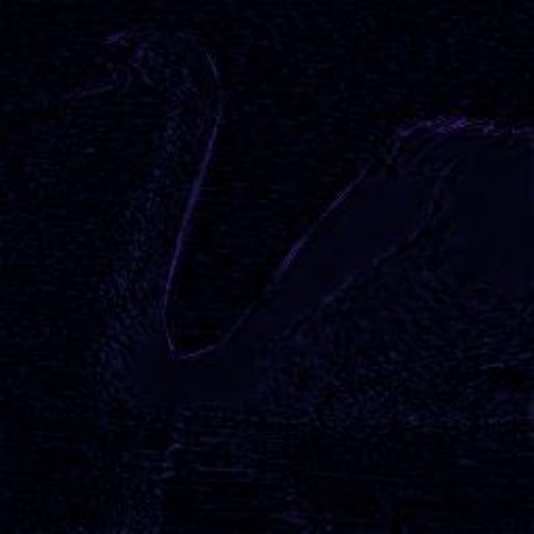}
    }
    \caption{Error maps of the reconstructions shown in Fig.~\ref{fig:error_map_prnettttintermediate}. Brighter colors indicate larger absolute error.}
\end{figure}

%%%%%%%%%%%%%%%%%%%%%%%%%%%%%%%%%%%%%%%%%

\begin{figure}[]
    \centering

    % First row
    \subfloat[][\parbox{0.31\columnwidth}{\centering (a) Ground truth}]{%
        \begin{minipage}[t]{0.31\columnwidth}
            \begin{tikzpicture}
                \node[anchor=south west,inner sep=0] (image) at (0,0) {
                    \includegraphics[width=\linewidth]{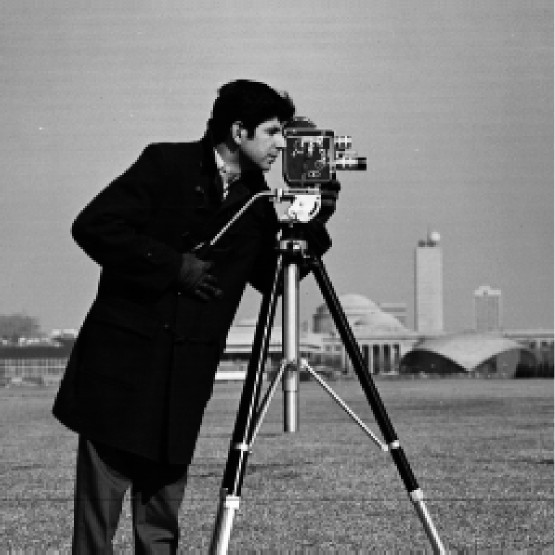}
                };
                \begin{scope}[x={(image.south east)}, y={(image.north west)}]
                    \draw[white, dashed, very thick] (0.35,0.56) rectangle (0.70,0.85);
                \end{scope}
            \end{tikzpicture}
            \vspace{2pt}
            \begin{adjustbox}{center,minipage=2.85\linewidth,trim={0.35\width, 1.1\height, 0.3\width, 0.3\height},clip}
                \includegraphics[width=\linewidth]{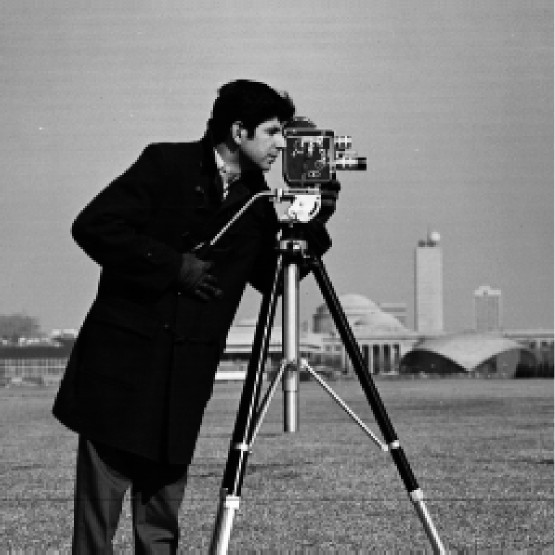}
            \end{adjustbox}
        \end{minipage}
    }
    \hfill
    \subfloat[][\parbox{0.31\columnwidth}{\centering (b) prDeep~\cite{pmlr-v80-metzler18a}\\PSNR:23.44,SSIM:0.66}]{%
        \begin{minipage}[t]{0.31\columnwidth}
            \includegraphics[width=\linewidth]{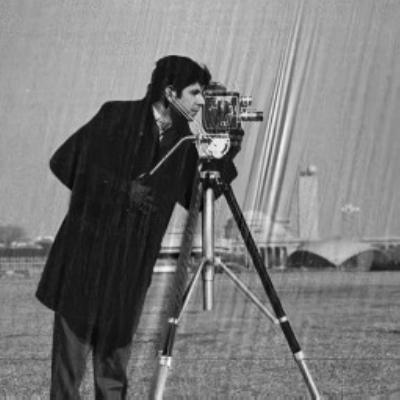}
            \vspace{2pt}
            \begin{adjustbox}{center,minipage=2.85\linewidth,trim={0.35\width, 1.1\height, 0.3\width, 0.3\height},clip}
                \includegraphics[width=\linewidth]{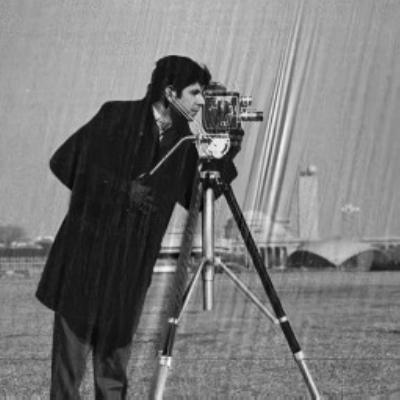}
            \end{adjustbox}
        \end{minipage}
    }
    \hfill
    \subfloat[][\parbox{0.31\columnwidth}{\centering (c) I2I-PR~\cite{kaya2026i2i} %DIR~\cite{Isil:19}
    \\PSNR:27.39,SSIM:0.70}]{%
        \begin{minipage}[t]{0.31\columnwidth}
            \includegraphics[width=\linewidth]{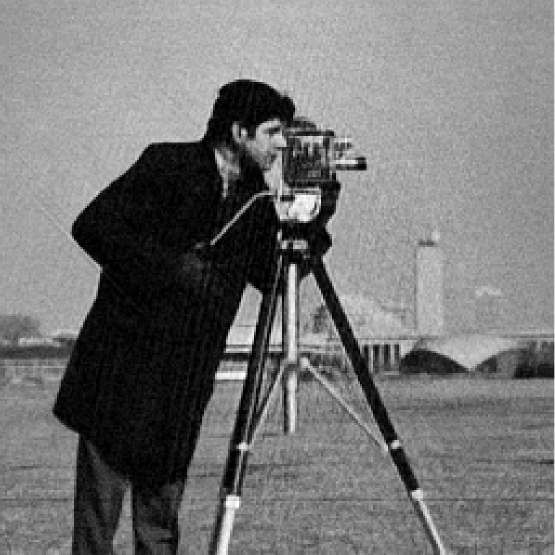}
            \vspace{2pt}
            \begin{adjustbox}{center,minipage=2.85\linewidth,trim={0.35\width, 1.1\height, 0.3\width, 0.3\height},clip}
                \includegraphics[width=\linewidth]{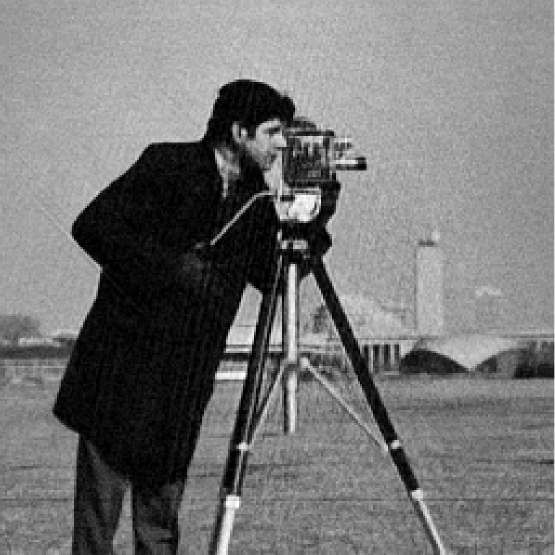}
            \end{adjustbox}
        \end{minipage}
    }

    % Second row
    \subfloat[][\parbox{0.31\columnwidth}{\centering (d) DiffFPR$^{\ddagger}$~\cite{li2024difffpr}\\PSNR:21.44,SSIM:0.49}]{%
        \begin{minipage}[t]{0.31\columnwidth}
            \includegraphics[width=\linewidth]{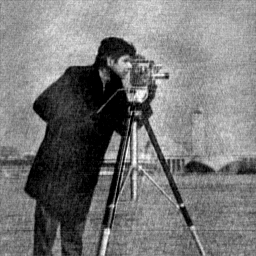}
            \vspace{2pt}
            \begin{adjustbox}{center,minipage=2.85\linewidth,trim={0.35\width, 1.1\height, 0.3\width, 0.3\height},clip}
                \includegraphics[width=\linewidth]{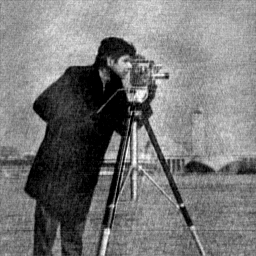}
            \end{adjustbox}
        \end{minipage}
    }
    \hfill
    \subfloat[][\parbox{0.31\columnwidth}{\centering (e) DAPS$^{\ddagger}$~\cite{zhang2025improving}\\PSNR:13.49,SSIM:0.42}]{%
        \begin{minipage}[t]{0.31\columnwidth}
            \includegraphics[width=\linewidth]{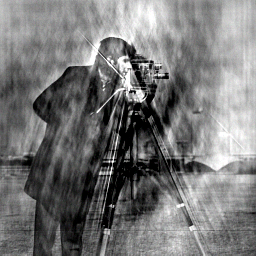}
            \vspace{2pt}
            \begin{adjustbox}{center,minipage=2.85\linewidth,trim={0.35\width, 1.1\height, 0.3\width, 0.3\height},clip}
                \includegraphics[width=\linewidth]{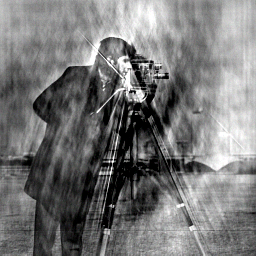}
            \end{adjustbox}
        \end{minipage}
    }
    \hfill
    \subfloat[][\parbox{0.31\columnwidth}{\centering (f) LatentDAPS$^{\ddagger}$~\cite{zhang2025improving}\\PSNR:12.78,SSIM:0.30}]{%
        \begin{minipage}[t]{0.31\columnwidth}
            \includegraphics[width=\linewidth]{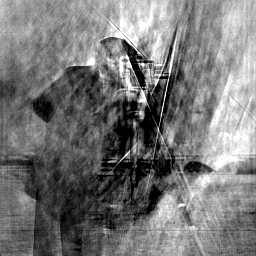}
            \vspace{2pt}
            \begin{adjustbox}{center,minipage=2.85\linewidth,trim={0.35\width, 1.1\height, 0.3\width, 0.3\height},clip}
                \includegraphics[width=\linewidth]{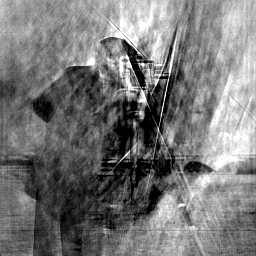}
            \end{adjustbox}
        \end{minipage}
    }

    % Second row
    \subfloat[][\parbox{0.31\columnwidth}{\centering (g) prNet-Small\\PSNR:27.12,SSIM:0.88}]{%
        \begin{minipage}[t]{0.31\columnwidth}
            \includegraphics[width=\linewidth]{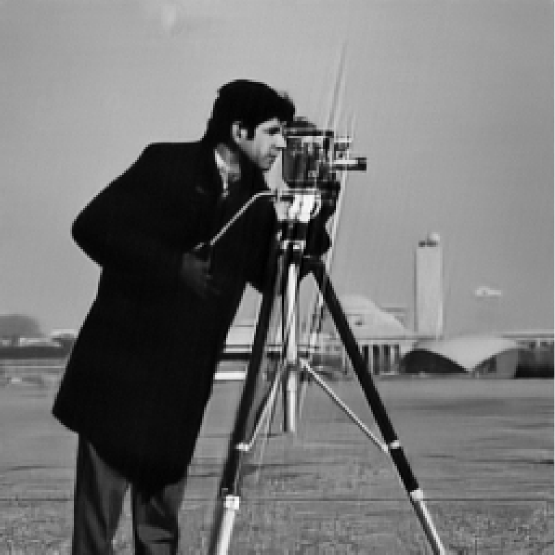}
            \vspace{2pt}
            \begin{adjustbox}{center,minipage=2.85\linewidth,trim={0.35\width, 1.1\height, 0.3\width, 0.3\height},clip}
                \includegraphics[width=\linewidth]{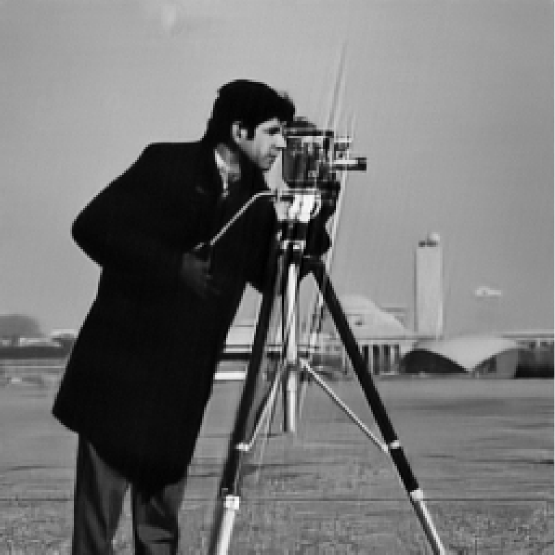}
            \end{adjustbox}
        \end{minipage}
    }
    \hfill
    \subfloat[][\parbox{0.31\columnwidth}{\centering (h) prNet-Large\\PSNR:30.87,SSIM:0.93}]{%
        \begin{minipage}[t]{0.31\columnwidth}
            \includegraphics[width=\linewidth]{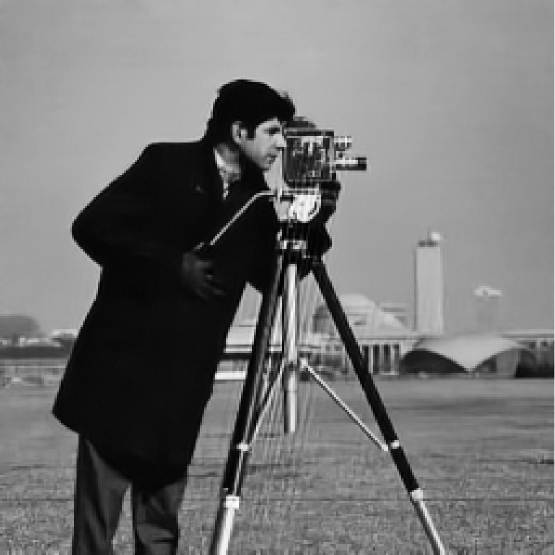}
            \vspace{2pt}
            \begin{adjustbox}{center,minipage=2.85\linewidth,trim={0.35\width, 1.1\height, 0.3\width, 0.3\height},clip}
                \includegraphics[width=\linewidth]{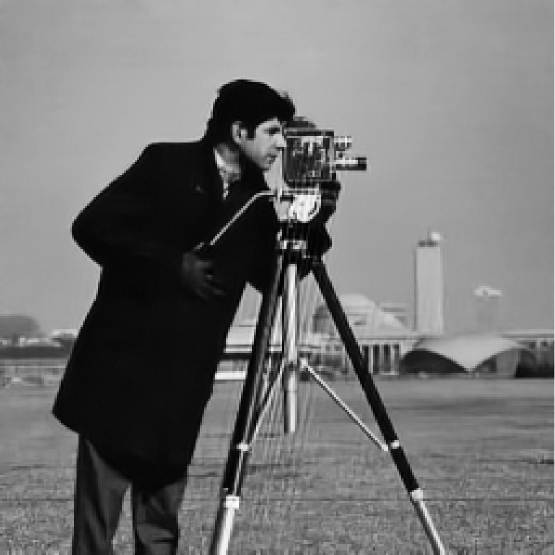}
            \end{adjustbox}
        \end{minipage}
    }
    \hfill
    \subfloat[][\parbox{0.31\columnwidth}{\centering (i) prNet-Large-Adversarial (+TTA)\\PSNR:31.23,SSIM:0.93}]{%
        \begin{minipage}[t]{0.31\columnwidth}
            \includegraphics[width=\linewidth]{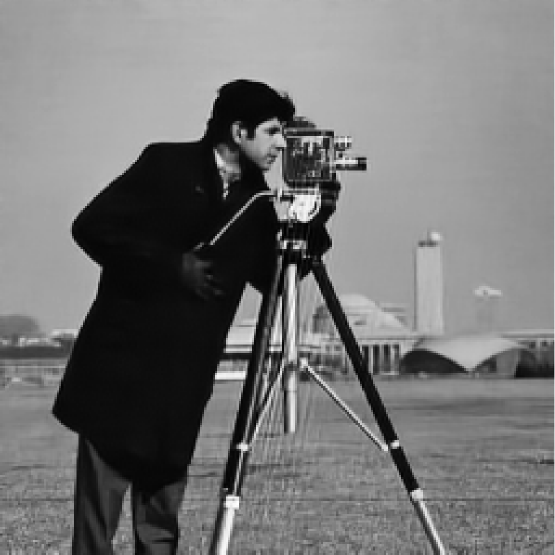}
            \vspace{2pt}
            \begin{adjustbox}{center,minipage=2.85\linewidth,trim={0.35\width, 1.1\height, 0.3\width, 0.3\height},clip}
                \includegraphics[width=\linewidth]{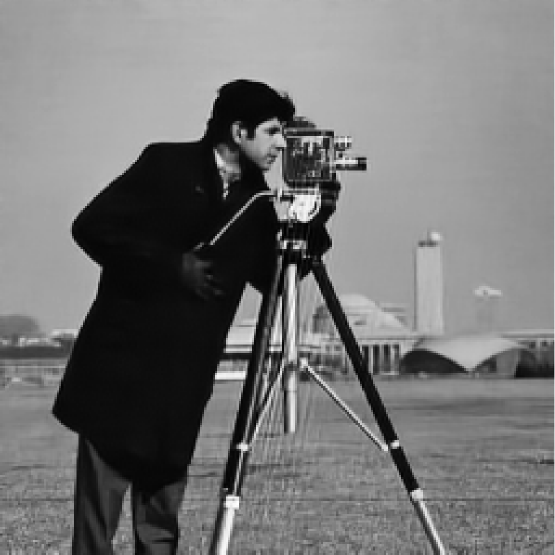}
            \end{adjustbox}
        \end{minipage}
    }

    \caption{The outputs of various algorithms for the ``Cameraman" test image subjected to $\alpha=3$ noise (SNR=31.61dB).}
    \label{fig:error_map_prnettttcameraman}
\end{figure}

\begin{figure}[]
    \centering

    % First row
    \subfloat[][\parbox{0.31\columnwidth}{\centering (a) Ground truth}]{%
        \begin{minipage}[t]{0.31\columnwidth}
            \begin{tikzpicture}
                \node[anchor=south west,inner sep=0] (image) at (0,0) {
                    \includegraphics[width=\linewidth]{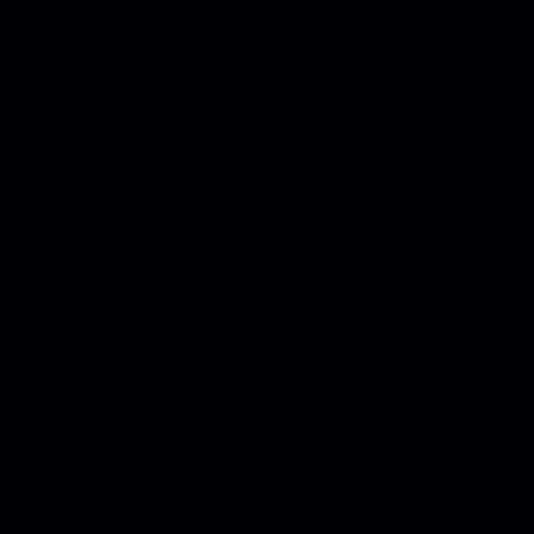}
                };
                \begin{scope}[x={(image.south east)}, y={(image.north west)}]
                    \draw[white, dashed, very thick] (0.35,0.56) rectangle (0.70,0.85);
                \end{scope}
            \end{tikzpicture}
            \vspace{2pt}
            \begin{adjustbox}{center,minipage=2.85\linewidth,trim={0.35\width, 1.1\height, 0.3\width, 0.3\height},clip}
                \includegraphics[width=\linewidth]{figures/figures_errormaps_aligned/fig_2/gt.pdf}
            \end{adjustbox}
        \end{minipage}
    }
    \hfill
    \subfloat[][\parbox{0.31\columnwidth}{\centering (b) prDeep~\cite{pmlr-v80-metzler18a}\\PSNR:23.44,SSIM:0.66}]{%
        \begin{minipage}[t]{0.31\columnwidth}
            \includegraphics[width=\linewidth]{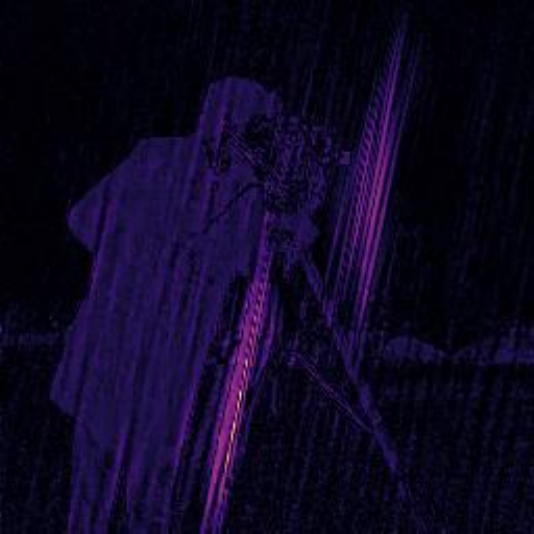}
            \vspace{2pt}
            \begin{adjustbox}{center,minipage=2.85\linewidth,trim={0.35\width, 1.1\height, 0.3\width, 0.3\height},clip}
                \includegraphics[width=\linewidth]{figures/figures_errormaps_aligned/fig_2/prdeep.pdf}
            \end{adjustbox}
        \end{minipage}
    }
    \hfill
    \subfloat[][\parbox{0.31\columnwidth}{\centering (c) I2I-PR~\cite{kaya2026i2i} %DIR~\cite{Isil:19}
    \\PSNR:27.39,SSIM:0.70}]{%
        \begin{minipage}[t]{0.31\columnwidth}
            \includegraphics[width=\linewidth]{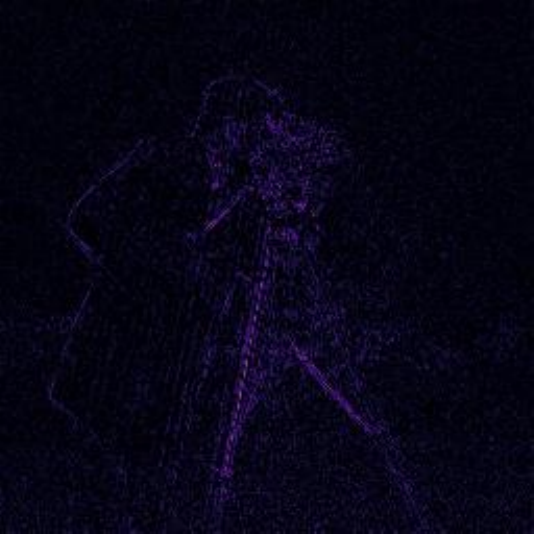}
            \vspace{2pt}
            \begin{adjustbox}{center,minipage=2.85\linewidth,trim={0.35\width, 1.1\height, 0.3\width, 0.3\height},clip}
                \includegraphics[width=\linewidth]{figures/figures_errormaps_aligned/fig_2/i2ipr.pdf}
            \end{adjustbox}
        \end{minipage}
    }

    % Second row
    \subfloat[][\parbox{0.31\columnwidth}{\centering (d) DiffFPR$^{\ddagger}$~\cite{li2024difffpr}\\PSNR:21.44,SSIM:0.49}]{%
        \begin{minipage}[t]{0.31\columnwidth}
            \includegraphics[width=\linewidth]{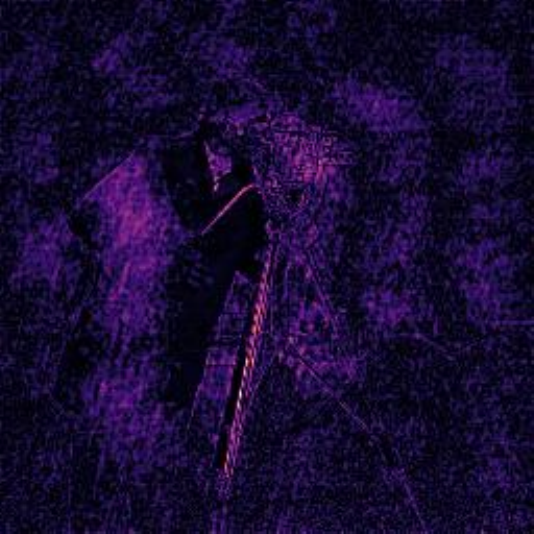}
            \vspace{2pt}
            \begin{adjustbox}{center,minipage=2.85\linewidth,trim={0.35\width, 1.1\height, 0.3\width, 0.3\height},clip}
                \includegraphics[width=\linewidth]{figures/figures_errormaps_aligned/fig_2/difffpr.pdf}
            \end{adjustbox}
        \end{minipage}
    }
    \hfill
    \subfloat[][\parbox{0.31\columnwidth}{\centering (e) DAPS$^{\ddagger}$~\cite{zhang2025improving}\\PSNR:13.49,SSIM:0.42}]{%
        \begin{minipage}[t]{0.31\columnwidth}
            \includegraphics[width=\linewidth]{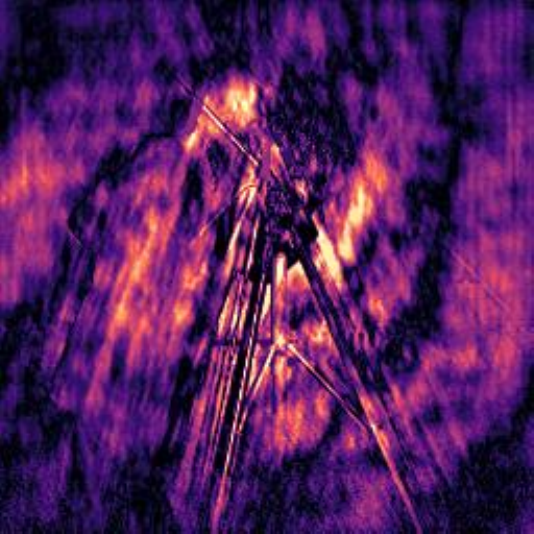}
            \vspace{2pt}
            \begin{adjustbox}{center,minipage=2.85\linewidth,trim={0.35\width, 1.1\height, 0.3\width, 0.3\height},clip}
                \includegraphics[width=\linewidth]{figures/figures_errormaps_aligned/fig_2/daps.pdf}
            \end{adjustbox}
        \end{minipage}
    }
    \hfill
    \subfloat[][\parbox{0.31\columnwidth}{\centering (f) LatentDAPS$^{\ddagger}$~\cite{zhang2025improving}\\PSNR:12.78,SSIM:0.30}]{%
        \begin{minipage}[t]{0.31\columnwidth}
            \includegraphics[width=\linewidth]{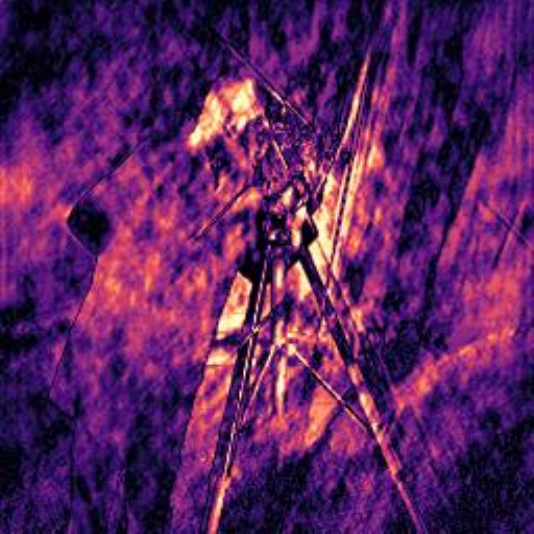}
            \vspace{2pt}
            \begin{adjustbox}{center,minipage=2.85\linewidth,trim={0.35\width, 1.1\height, 0.3\width, 0.3\height},clip}
                \includegraphics[width=\linewidth]{figures/figures_errormaps_aligned/fig_2/latentdaps.pdf}
            \end{adjustbox}
        \end{minipage}
    }

    % Second row
    \subfloat[][\parbox{0.31\columnwidth}{\centering (g) prNet-Small\\PSNR:27.12,SSIM:0.88}]{%
        \begin{minipage}[t]{0.31\columnwidth}
            \includegraphics[width=\linewidth]{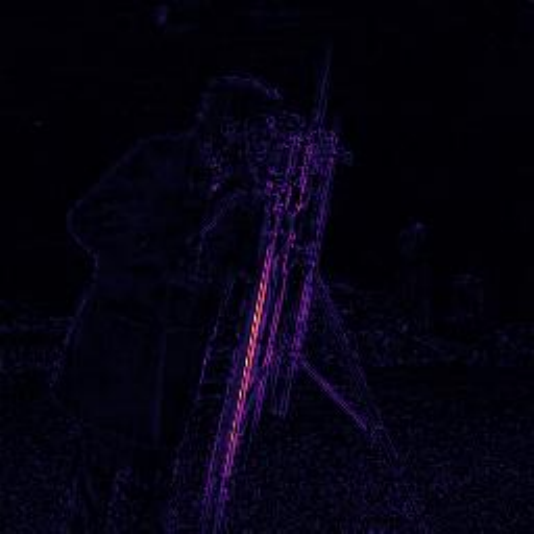}
            \vspace{2pt}
            \begin{adjustbox}{center,minipage=2.85\linewidth,trim={0.35\width, 1.1\height, 0.3\width, 0.3\height},clip}
                \includegraphics[width=\linewidth]{figures/figures_errormaps_aligned/fig_2/prnet_small.pdf}
            \end{adjustbox}
        \end{minipage}
    }
    \hfill
    \subfloat[][\parbox{0.31\columnwidth}{\centering (h) prNet-Large\\PSNR:30.87,SSIM:0.93}]{%
        \begin{minipage}[t]{0.31\columnwidth}
            \includegraphics[width=\linewidth]{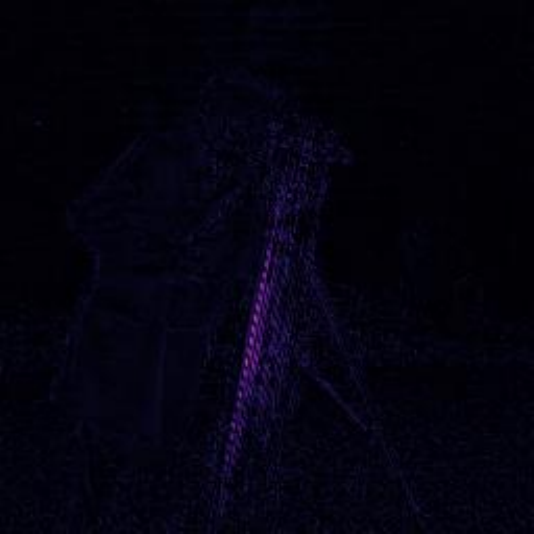}
            \vspace{2pt}
            \begin{adjustbox}{center,minipage=2.85\linewidth,trim={0.35\width, 1.1\height, 0.3\width, 0.3\height},clip}
                \includegraphics[width=\linewidth]{figures/figures_errormaps_aligned/fig_2/prnet_large.pdf}
            \end{adjustbox}
        \end{minipage}
    }
    \hfill
    \subfloat[][\parbox{0.31\columnwidth}{\centering (i) prNet-Large-Adversarial (+TTA)\\PSNR:31.23,SSIM:0.93}]{%
        \begin{minipage}[t]{0.31\columnwidth}
            \includegraphics[width=\linewidth]{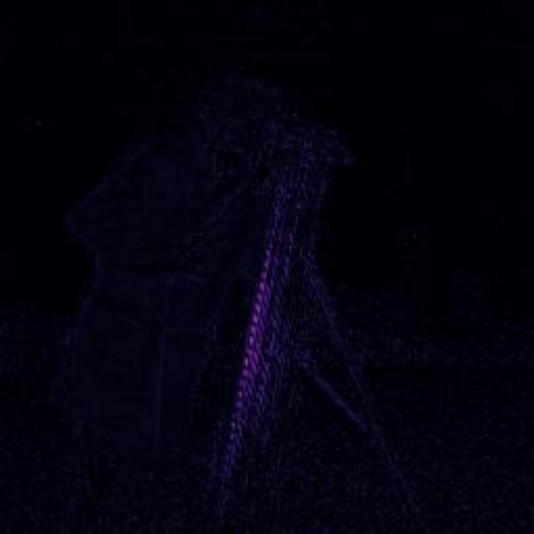}
            \vspace{2pt}
            \begin{adjustbox}{center,minipage=2.85\linewidth,trim={0.35\width, 1.1\height, 0.3\width, 0.3\height},clip}
                \includegraphics[width=\linewidth]{figures/figures_errormaps_aligned/fig_2/prnet_large_adversarial_tta.pdf}
            \end{adjustbox}
        \end{minipage}
    }

    \caption{Error maps of the reconstructions shown in Fig.~\ref{fig:error_map_prnettttcameraman}. Brighter colors indicate larger absolute error.}
\end{figure}

%%%%%%%%%%%%%%%%%%%%%%%%%%%%%%%%%%%%%%%%%

\begin{figure}[th]
    \centering

    % First row
    \subfloat[][\parbox{0.31\columnwidth}{\centering (a) Ground truth}]{%
        \begin{minipage}[t]{0.31\columnwidth}
            \begin{tikzpicture}
                \node[anchor=south west,inner sep=0] (image) at (0,0) {
                    \includegraphics[width=\linewidth]{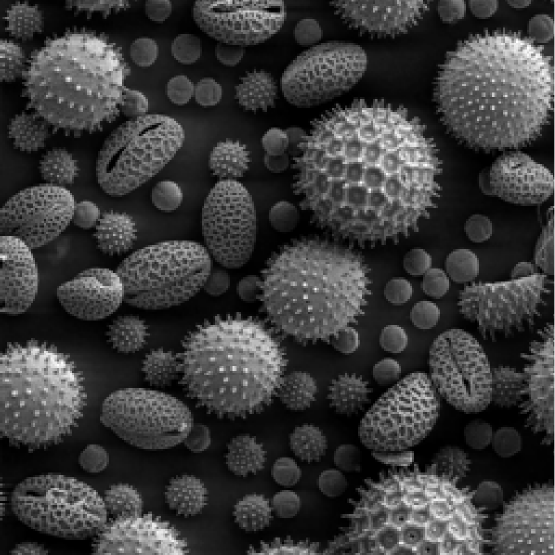}
                };
                \begin{scope}[x={(image.south east)}, y={(image.north west)}]
                    \draw[white, dashed, very thick] (0.35,0.35) rectangle (0.65,0.64);
                \end{scope}
            \end{tikzpicture}
            \vspace{2pt}
            \begin{adjustbox}{center,minipage=3.33\linewidth,trim={.35\width,.35\height,.25\width,.25\height},clip}
                \includegraphics[width=\linewidth]{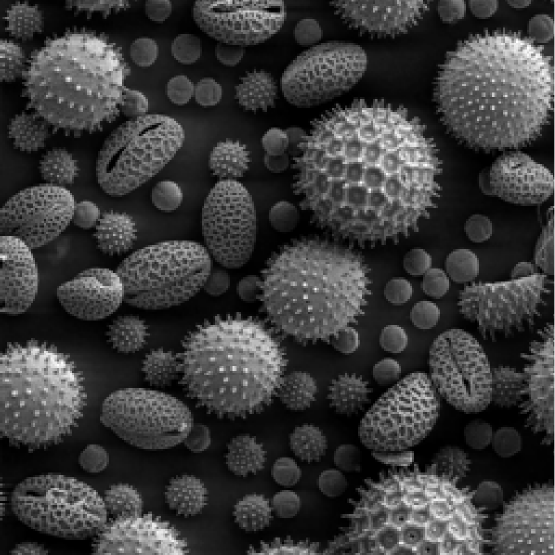}
            \end{adjustbox}
        \end{minipage}
    }
    \hfill
    \subfloat[][\parbox{0.31\columnwidth}{\centering (b) prDeep~\cite{pmlr-v80-metzler18a}\\PSNR:19.37,SSIM:0.47}]{%
        \begin{minipage}[t]{0.31\columnwidth}
            \includegraphics[width=\linewidth]{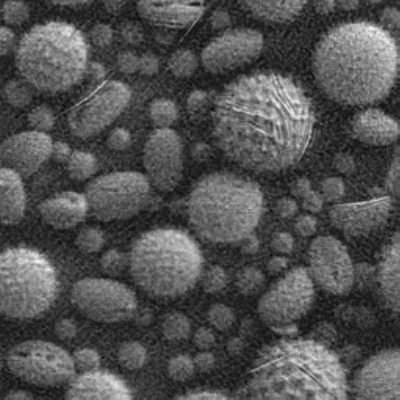}
            \vspace{2pt}
            \begin{adjustbox}{center,minipage=3.33\linewidth,trim={.35\width,.35\height,.25\width,.25\height},clip}
                \includegraphics[width=\linewidth]{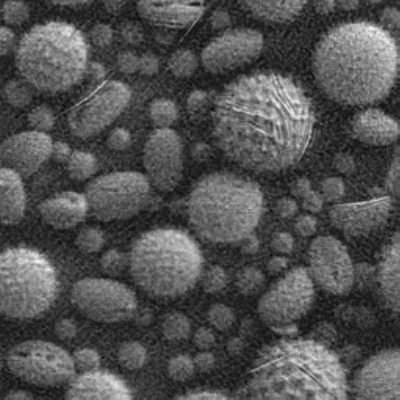}
            \end{adjustbox}
        \end{minipage}
    }
    \hfill
    \subfloat[][\parbox{0.31\columnwidth}{\centering (c) I2I-PR~\cite{kaya2026i2i} %DIR~\cite{Isil:19}
    \\PSNR:28.95,SSIM:0.86}]{%
        \begin{minipage}[t]{0.31\columnwidth}
            \includegraphics[width=\linewidth]{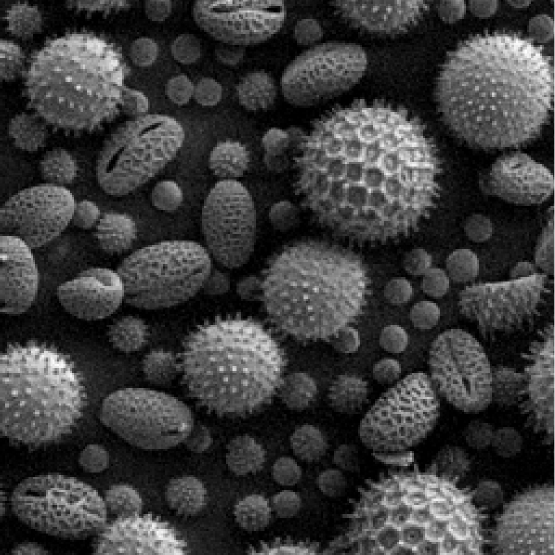}
            \vspace{2pt}
            \begin{adjustbox}{center,minipage=3.33\linewidth,trim={.35\width,.35\height,.25\width,.25\height},clip}
                \includegraphics[width=\linewidth]{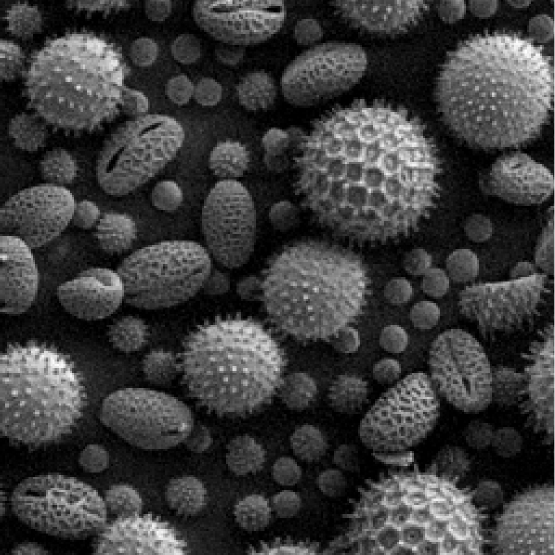}
            \end{adjustbox}
        \end{minipage}
    }

    % Second row
    \subfloat[][\parbox{0.31\columnwidth}{\centering (d) DiffFPR$^{\ddagger}$~\cite{li2024difffpr}\\PSNR:24.92,SSIM:0.73}]{%
        \begin{minipage}[t]{0.31\columnwidth}
            \includegraphics[width=\linewidth]{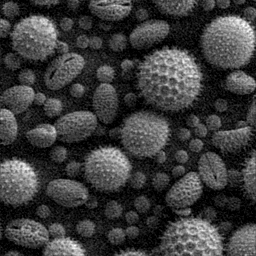}
            \vspace{2pt}
            \begin{adjustbox}{center,minipage=3.33\linewidth,trim={.35\width,.35\height,.25\width,.25\height},clip}
                \includegraphics[width=\linewidth]{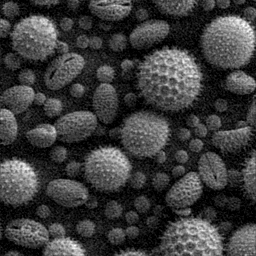}
            \end{adjustbox}
        \end{minipage}
    }
    \hfill
    \subfloat[][\parbox{0.31\columnwidth}{\centering (e) DAPS$^{\ddagger}$~\cite{zhang2025improving}\\PSNR:24.10,SSIM:0.85}]{%
        \begin{minipage}[t]{0.31\columnwidth}
            \includegraphics[width=\linewidth]{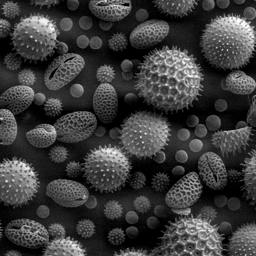}
            \vspace{2pt}
            \begin{adjustbox}{center,minipage=3.33\linewidth,trim={.35\width,.35\height,.25\width,.25\height},clip}
                \includegraphics[width=\linewidth]{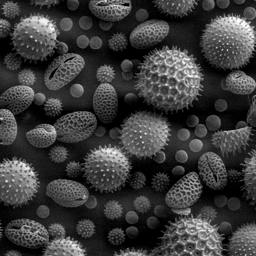}
            \end{adjustbox}
        \end{minipage}
    }
    \hfill
    \subfloat[][\parbox{0.31\columnwidth}{\centering (f) LatentDAPS$^{\ddagger}$~\cite{zhang2025improving}\\PSNR:21.53,SSIM:0.80}]{%
        \begin{minipage}[t]{0.31\columnwidth}
            \includegraphics[width=\linewidth]{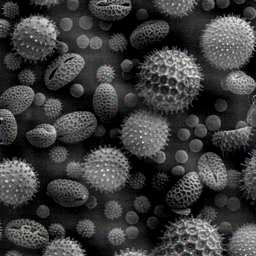}
            \vspace{2pt}
            \begin{adjustbox}{center,minipage=3.33\linewidth,trim={.35\width,.35\height,.25\width,.25\height},clip}
                \includegraphics[width=\linewidth]{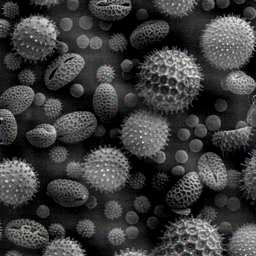}
            \end{adjustbox}
        \end{minipage}
    }

    % Second row
    \subfloat[][\parbox{0.31\columnwidth}{\centering (g) PnP-HIO~\cite{Isil:20}\\PSNR:26.28,SSIM:0.71}]{%
        \begin{minipage}[t]{0.31\columnwidth}
            \includegraphics[width=\linewidth]{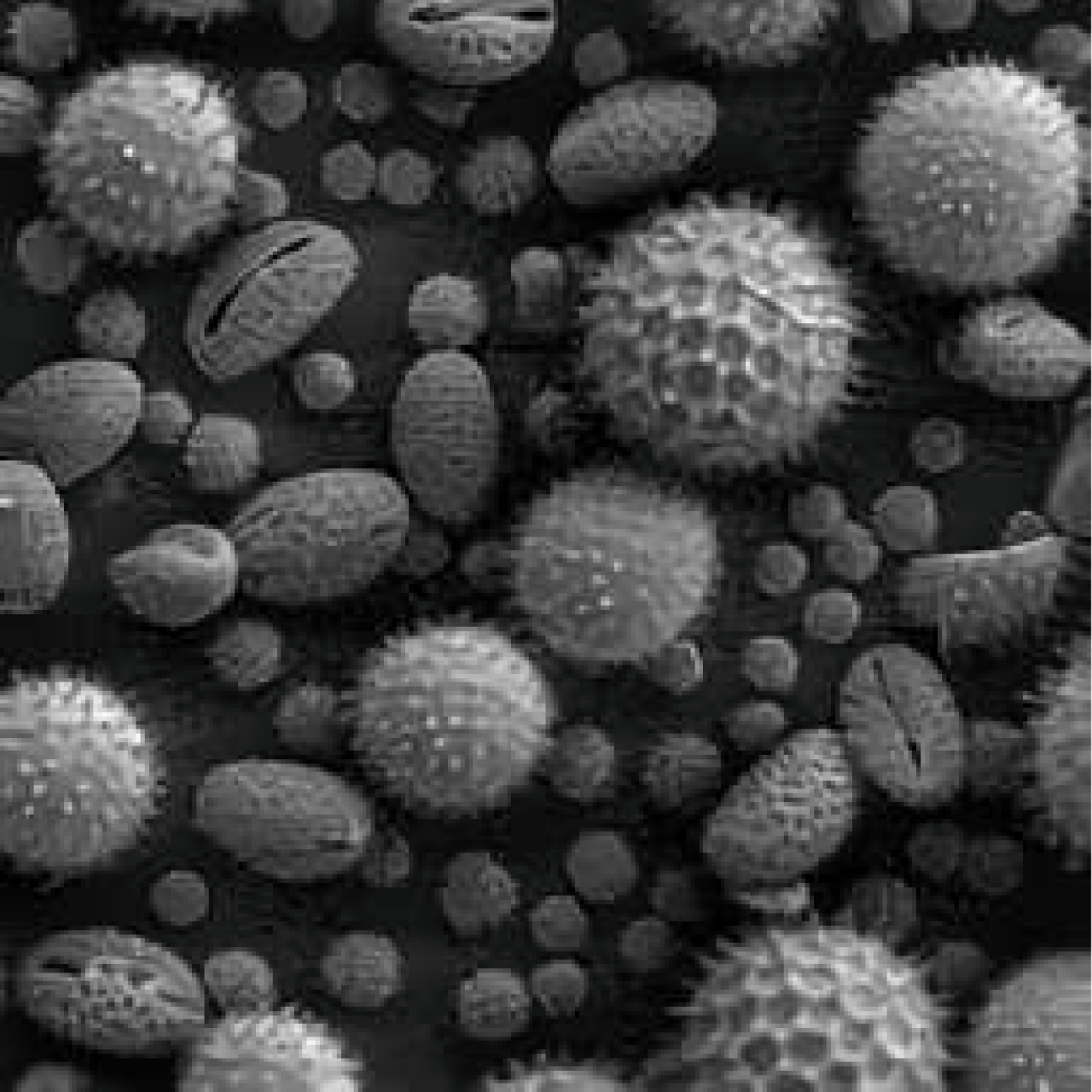}
            \vspace{2pt}
            \begin{adjustbox}{center,minipage=3.33\linewidth,trim={.35\width,.35\height,.25\width,.25\height},clip}
                \includegraphics[width=\linewidth]{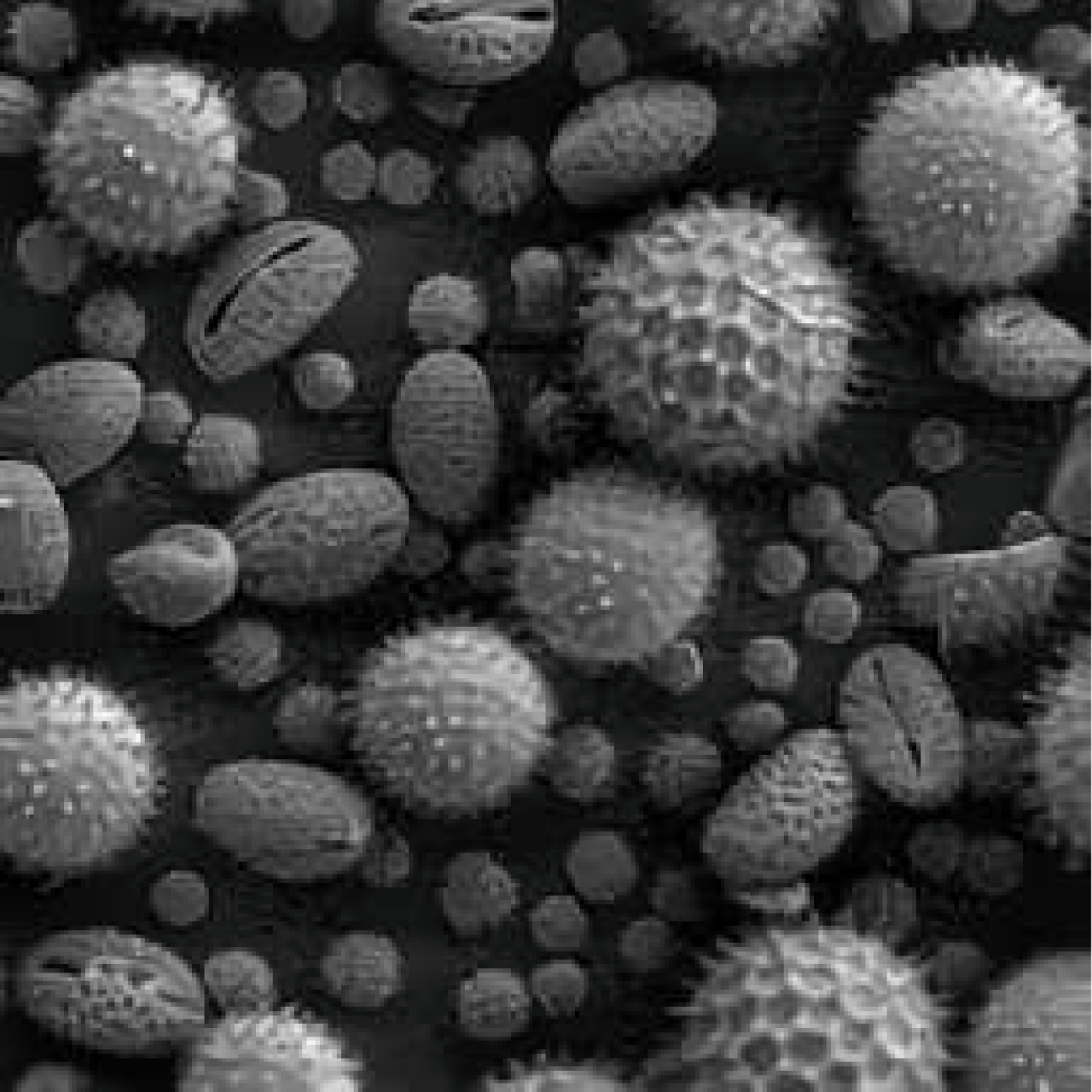}
            \end{adjustbox}
        \end{minipage}
    }
    \hfill
    \subfloat[][\parbox{0.31\columnwidth}{\centering (h) prNet-Small\\PSNR:28.67,SSIM:0.89}]{%
        \begin{minipage}[t]{0.31\columnwidth}
            \includegraphics[width=\linewidth]{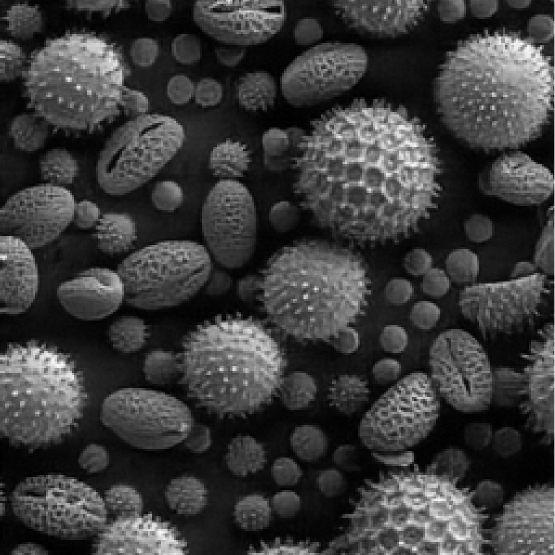}
            \vspace{2pt}
            \begin{adjustbox}{center,minipage=3.33\linewidth,trim={.35\width,.35\height,.25\width,.25\height},clip}
                \includegraphics[width=\linewidth]{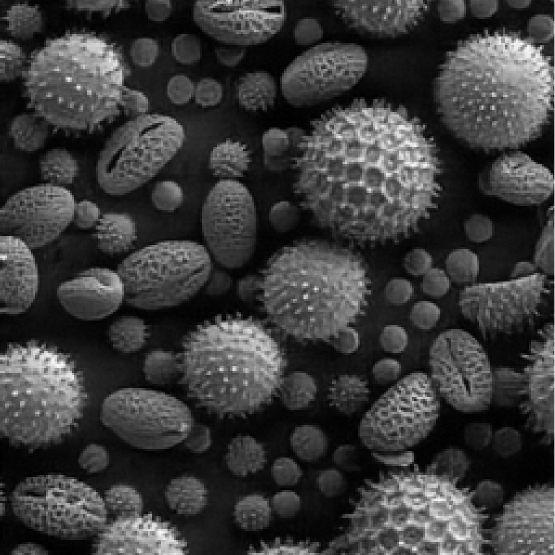}
            \end{adjustbox}
        \end{minipage}
    }
    \hfill
    \subfloat[][\parbox{0.31\columnwidth}{\centering (i) prNet-Large-Adversarial (+TTA)\\PSNR:30.88,SSIM:0.92}]{%
        \begin{minipage}[t]{0.31\columnwidth}
            \includegraphics[width=\linewidth]{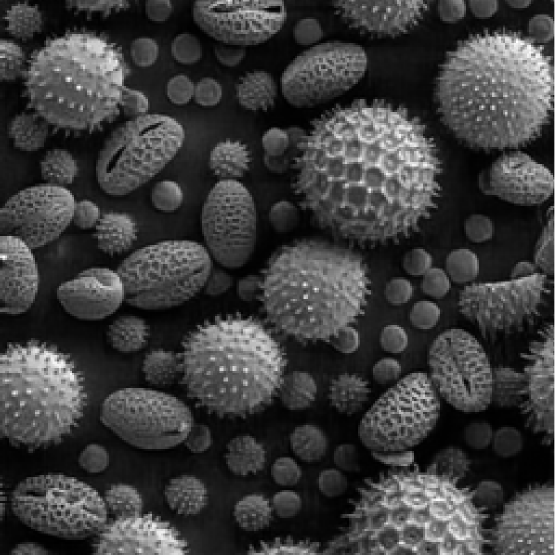}
            \vspace{2pt}
            \begin{adjustbox}{center,minipage=3.33\linewidth,trim={.35\width,.35\height,.25\width,.25\height},clip}
                \includegraphics[width=\linewidth]{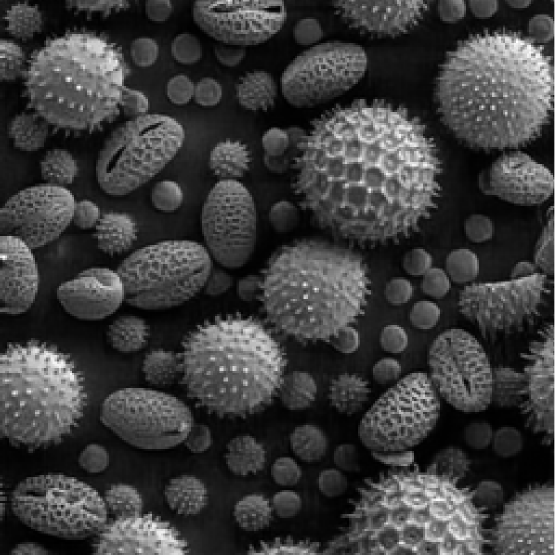}
            \end{adjustbox}
        \end{minipage}
    }

    \caption{The outputs of various algorithms for the out-of-domain ``Pollen" test image subjected to $\alpha=3$ noise (SNR=28.10dB).}
    \label{fig:error_map_prnettttpollen}
\end{figure}

\begin{figure}[th]
    \centering

    % First row
    \subfloat[][\parbox{0.31\columnwidth}{\centering (a) Ground truth}]{%
        \begin{minipage}[t]{0.31\columnwidth}
            \begin{tikzpicture}
                \node[anchor=south west,inner sep=0] (image) at (0,0) {
                    \includegraphics[width=\linewidth]{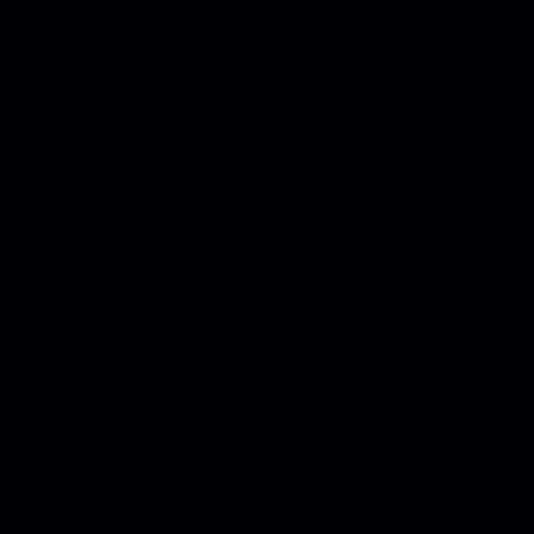}
                };
                \begin{scope}[x={(image.south east)}, y={(image.north west)}]
                    \draw[white, dashed, very thick] (0.35,0.35) rectangle (0.65,0.64);
                \end{scope}
            \end{tikzpicture}
            \vspace{2pt}
            \begin{adjustbox}{center,minipage=3.33\linewidth,trim={.35\width,.35\height,.25\width,.25\height},clip}
                \includegraphics[width=\linewidth]{figures/figures_errormaps_aligned/fig_4/gt.pdf}
            \end{adjustbox}
        \end{minipage}
    }
    \hfill
    \subfloat[][\parbox{0.31\columnwidth}{\centering (b) prDeep~\cite{pmlr-v80-metzler18a}\\PSNR:19.37,SSIM:0.47}]{%
        \begin{minipage}[t]{0.31\columnwidth}
            \includegraphics[width=\linewidth]{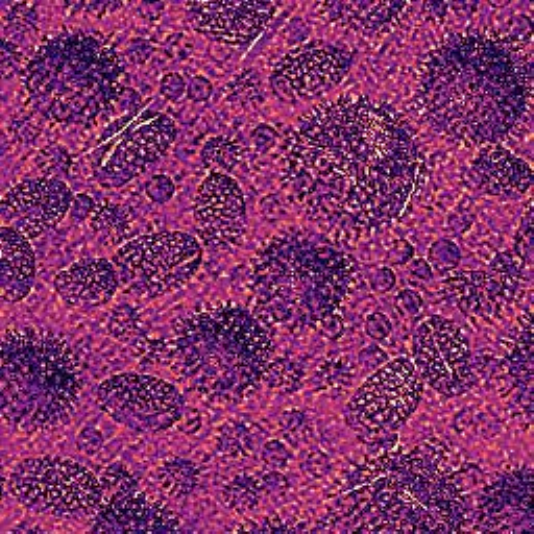}
            \vspace{2pt}
            \begin{adjustbox}{center,minipage=3.33\linewidth,trim={.35\width,.35\height,.25\width,.25\height},clip}
                \includegraphics[width=\linewidth]{figures/figures_errormaps_aligned/fig_4/prdeep.pdf}
            \end{adjustbox}
        \end{minipage}
    }
    \hfill
    \subfloat[][\parbox{0.31\columnwidth}{\centering (c) I2I-PR~\cite{kaya2026i2i} %DIR~\cite{Isil:19}
    \\PSNR:28.95,SSIM:0.86}]{%
        \begin{minipage}[t]{0.31\columnwidth}
            \includegraphics[width=\linewidth]{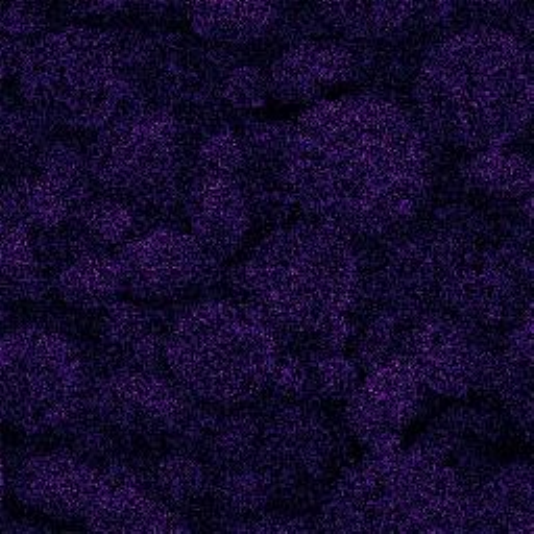}
            \vspace{2pt}
            \begin{adjustbox}{center,minipage=3.33\linewidth,trim={.35\width,.35\height,.25\width,.25\height},clip}
                \includegraphics[width=\linewidth]{figures/figures_errormaps_aligned/fig_4/i2ipr.pdf}
            \end{adjustbox}
        \end{minipage}
    }

    % Second row
    \subfloat[][\parbox{0.31\columnwidth}{\centering (d) DiffFPR$^{\ddagger}$~\cite{li2024difffpr}\\PSNR:24.92,SSIM:0.73}]{%
        \begin{minipage}[t]{0.31\columnwidth}
            \includegraphics[width=\linewidth]{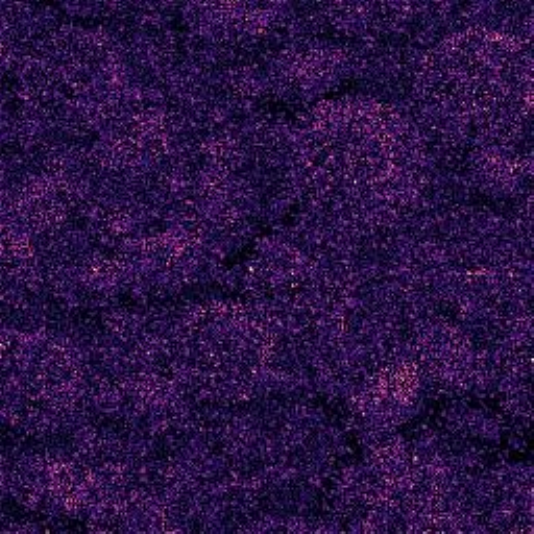}
            \vspace{2pt}
            \begin{adjustbox}{center,minipage=3.33\linewidth,trim={.35\width,.35\height,.25\width,.25\height},clip}
                \includegraphics[width=\linewidth]{figures/figures_errormaps_aligned/fig_4/difffpr.pdf}
            \end{adjustbox}
        \end{minipage}
    }
    \hfill
    \subfloat[][\parbox{0.31\columnwidth}{\centering (e) DAPS$^{\ddagger}$~\cite{zhang2025improving}\\PSNR:24.10,SSIM:0.85}]{%
        \begin{minipage}[t]{0.31\columnwidth}
            \includegraphics[width=\linewidth]{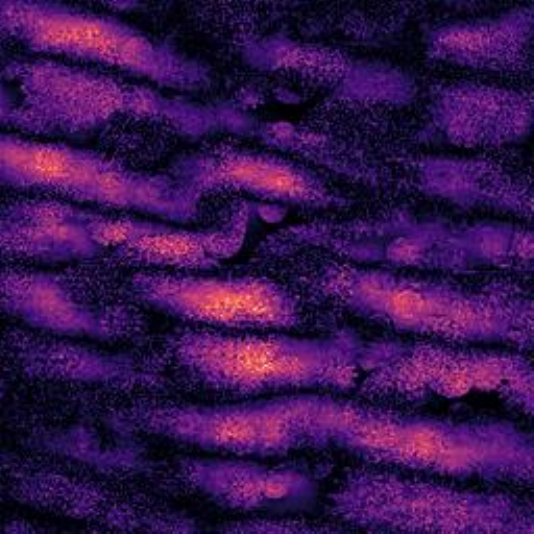}
            \vspace{2pt}
            \begin{adjustbox}{center,minipage=3.33\linewidth,trim={.35\width,.35\height,.25\width,.25\height},clip}
                \includegraphics[width=\linewidth]{figures/figures_errormaps_aligned/fig_4/daps.pdf}
            \end{adjustbox}
        \end{minipage}
    }
    \hfill
    \subfloat[][\parbox{0.31\columnwidth}{\centering (f) LatentDAPS$^{\ddagger}$~\cite{zhang2025improving}\\PSNR:21.53,SSIM:0.80}]{%
        \begin{minipage}[t]{0.31\columnwidth}
            \includegraphics[width=\linewidth]{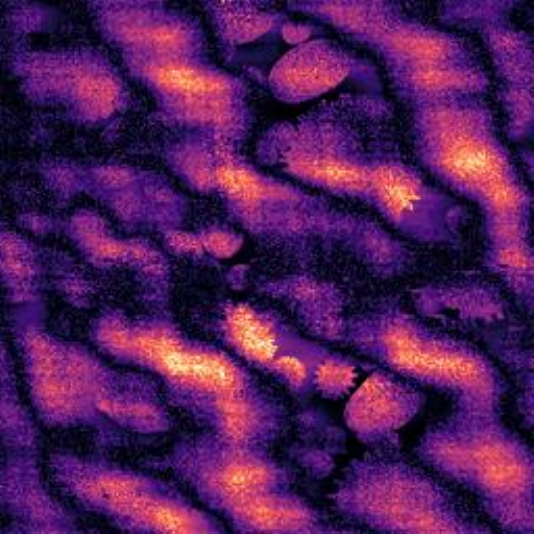}
            \vspace{2pt}
            \begin{adjustbox}{center,minipage=3.33\linewidth,trim={.35\width,.35\height,.25\width,.25\height},clip}
                \includegraphics[width=\linewidth]{figures/figures_errormaps_aligned/fig_4/latentdaps.pdf}
            \end{adjustbox}
        \end{minipage}
    }

    % Second row
    \subfloat[][\parbox{0.31\columnwidth}{\centering (g) PnP-HIO~\cite{Isil:20}\\PSNR:26.28,SSIM:0.71}]{%
        \begin{minipage}[t]{0.31\columnwidth}
            \includegraphics[width=\linewidth]{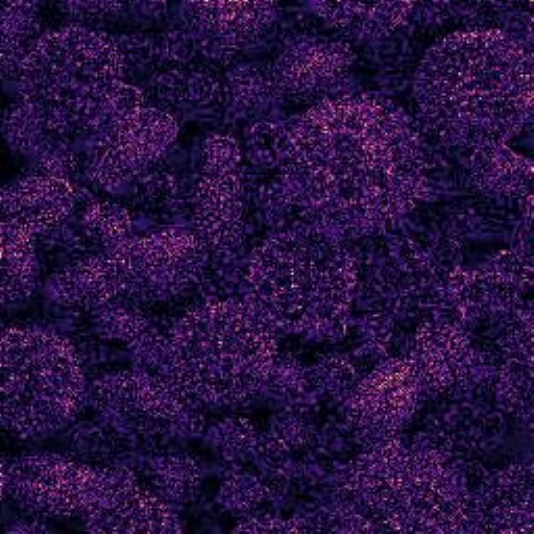}
            \vspace{2pt}
            \begin{adjustbox}{center,minipage=3.33\linewidth,trim={.35\width,.35\height,.25\width,.25\height},clip}
                \includegraphics[width=\linewidth]{figures/figures_errormaps_aligned/fig_4/pnp_hio.pdf}
            \end{adjustbox}
        \end{minipage}
    }
    \hfill
    \subfloat[][\parbox{0.31\columnwidth}{\centering (h) prNet-Small\\PSNR:28.67,SSIM:0.89}]{%
        \begin{minipage}[t]{0.31\columnwidth}
            \includegraphics[width=\linewidth]{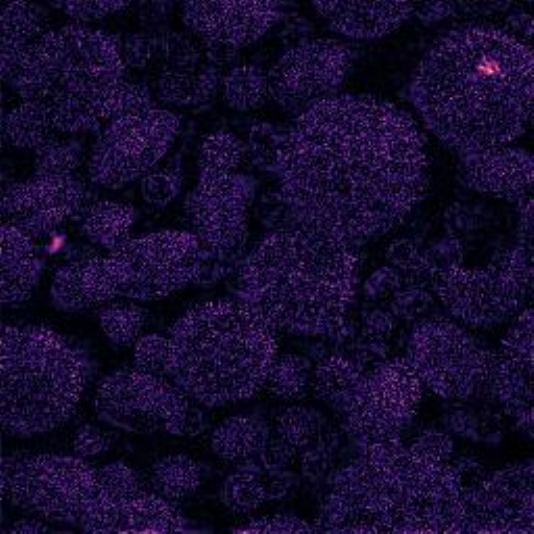}
            \vspace{2pt}
            \begin{adjustbox}{center,minipage=3.33\linewidth,trim={.35\width,.35\height,.25\width,.25\height},clip}
                \includegraphics[width=\linewidth]{figures/figures_errormaps_aligned/fig_4/prnet_small.pdf}
            \end{adjustbox}
        \end{minipage}
    }
    \hfill
    \subfloat[][\parbox{0.31\columnwidth}{\centering (i) prNet-Large-Adversarial (+TTA)\\PSNR:30.88,SSIM:0.92}]{%
        \begin{minipage}[t]{0.31\columnwidth}
            \includegraphics[width=\linewidth]{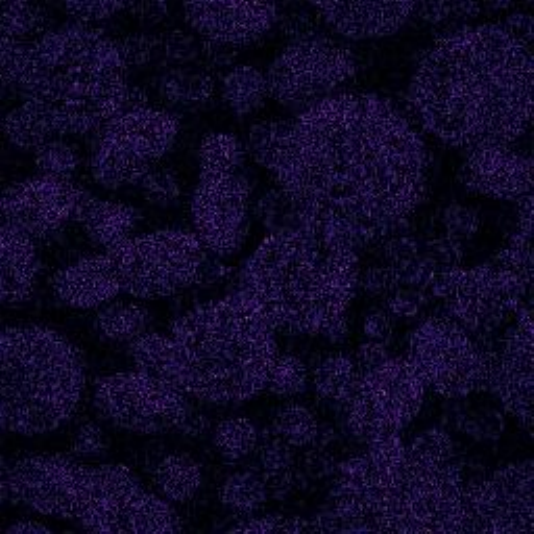}
            \vspace{2pt}
            \begin{adjustbox}{center,minipage=3.33\linewidth,trim={.35\width,.35\height,.25\width,.25\height},clip}
                \includegraphics[width=\linewidth]{figures/figures_errormaps_aligned/fig_4/prnet_large_adversarial_tta.pdf}
            \end{adjustbox}
        \end{minipage}
    }

    \caption{Error maps of the reconstructions shown in Fig.~\ref{fig:error_map_prnettttpollen}. Brighter colors indicate larger absolute error.}
\end{figure}

% \end{document}

\vfill

\end{document}